\documentclass[review, sort&compress]{elsarticle}

\usepackage[colorlinks,citecolor=blue,linktoc=all,linkcolor=cyan]{hyperref}
\usepackage{graphicx}
\usepackage[T1]{fontenc}          
\usepackage{dsfont}               
\usepackage{mathrsfs}             
\usepackage{slashed}              
\usepackage{amsmath}
\usepackage{amssymb}
\usepackage{amsbsy}
\usepackage{amsfonts}
\usepackage{color,xcolor}
\usepackage{cleveref}
\usepackage[normalem]{ulem}
\usepackage{caption}
\usepackage{subcaption}
\usepackage{floatflt,tcolorbox}

\usepackage{tikzpagenodes} 

\numberwithin{equation}{section}
\numberwithin{table}{section}
\numberwithin{figure}{section}

\DeclareMathAlphabet{\mathbbold}{U}{bbold}{m}{n}

\newcommand{\be}{\begin{equation}}
\newcommand{\ee}{\end{equation}}
\renewcommand\({\left(}
\renewcommand\){\right)}
\renewcommand\[{\left[}
\renewcommand\]{\right]}
\newcommand{\dd}{{\rm d}}

\newcommand{\e}{{\rm e}}
\newcommand\vp{\varphi}
\newcommand\eps{\epsilon}

\let\vec\mathbf
\let\vecS\boldsymbol

\def\L{\mathcal{L}}
\def\O{\mathcal{O}}

\def\D{\mathcal{D}}
\def\A{\mathcal{A}}
\def\C{\mathcal{C}}
\def\P{\mathcal{P}}

\renewcommand{\Im}{{\rm Im}}
\newcommand{\sgn}{\,{\rm sgn}}

\def\nn{\nonumber}
\def\Tr{{\rm Tr}}
\def\tr{{\rm tr}}

\newcommand{\bea}{\begin{eqnarray}}
\newcommand{\eea}{\end{eqnarray}}

\newcommand{\tb}{\bar \theta}

\graphicspath{{Figures/}}

\journal{Progress in Particle and Nuclear Physics}

\usepackage{titlesec}
\usepackage{sectsty}
\titleformat{\section}{\normalfont\Large\bfseries}{\thesection}{1em}{}
\titleformat{\subsection}{\normalfont\large\bfseries}{\thesubsection}{1em}{}
\titleformat{\subsubsection}{\normalfont\normalsize\bfseries}{\thesubsubsection}{1em}{}

\usepackage{fancyhdr}
\makeatletter
\fancypagestyle{pprintTitle}{
    \fancyhf{} 
    \fancyhead[R]{\raisebox{-0.8\baselineskip}[0pt][0pt]{\footnotesize CERN-TH-2025-161, Nikhef 2025-012}} 
}
\makeatother
\begin{document}

	\begin{frontmatter}
	
		\title{Bubble Trouble: a Review on Electroweak Baryogenesis}
        \author[cern]{Jorinde van de Vis\corref{mycorrespondingauthor}}
        \cortext[mycorrespondingauthor]{Corresponding author}
		\ead{jorinde.van.de.vis@cern.ch}
        \author[uva,nikhef]{Jordy de Vries}
		\author[nikhef,RU]{Marieke Postma}
		
\address[cern]{Theoretical Physics Department, CERN,
1 Esplanade des Particules, CH-1211 Geneva 23, Switzerland}		
		\address[uva]{Institute for Theoretical Physics, University
of Amsterdam,\\Science Park 904, 1098 XH Amsterdam, The Netherlands}
		\address[nikhef]{Nikhef, Science Park 105, 1098 XG Amsterdam, The Netherlands}
        \address[RU]{Institute for Mathematics, Astrophysics and Particle Physics, Radboud University, 6500 GL
Nijmegen, The Netherlands}
		
		\begin{abstract}
        The origin of the universal asymmetry between matter and antimatter remains a mystery.
        Electroweak baryogenesis is a well-motivated mechanism for generating the asymmetry dynamically, using interesting features of the Standard Model. In addition, it relies on beyond-the-Standard Model physics active around the electroweak scale: new physics coupling to the Higgs to make the electroweak phase transition first order, and a new mechanism of CP violation.
        The relatively low energy scale at which electroweak baryogenesis occurs makes certain aspects of the mechanism testable through collider experiments, electric dipole moment measurements, and gravitational wave observations. However, scenarios of electroweak baryogenesis are increasingly challenged by results from contemporary experiments. The developing experimental programs will play a crucial role in either falsifying or  detecting the new physics responsible for electroweak baryogenesis.
        To achieve this, it is essential to make precise predictions for the baryon asymmetry and the corresponding experimental signatures within specific scenarios. This review aims to provide a comprehensive overview of the rich physics involved in these predictions. Our goal is to offer a practical computational guide, with a focus on recent developments in the field. 
		\end{abstract}
		\vspace{5mm}
		
	\end{frontmatter}

	\newpage
	
	\thispagestyle{empty}
	\tableofcontents
	

	\newpage
	\section{Introduction}\label{intro}
		Everything around us – from rocks, to people, to stars – is made out of matter. Although usually taken for granted, this is baffling. Our state-of-the-art understanding of the world of elementary particles, captured by the Standard Model (SM), is that every matter particle comes with an antiparticle \cite{Dirac:1930ek}, which has the same mass and lifetime but opposite charge(s). When a particle and antiparticle meet they can annihilate and only leave energetic radiation behind. Why then, is there even any matter left in our universe, and how is it possible that we exist at all? What is the origin of the matter-antimatter asymmetry?

As normal matter mostly resides in baryons, these questions are usually rephrased in terms of the baryon asymmetry  of the universe (BAU). The  temperature fluctuations in the cosmic microwave background (CMB) and the light elements produced during Big Bang nucleosynthesis (BBN) depend sensitively on the matter content of the universe, on the amount of matter and radiation. This allows to quantify the baryon asymmetry $Y_b$ \cite{Cooke:2013cba,Planck:2015fie}:
\be
Y_b = \frac{n_b- \bar n_{ b}}{s} = \frac1{7.04}\frac{n_b- \bar n_{
    b}}{n_\gamma} = \left\{ \begin{array}{ll}
                              8.2-9.4 \times 10^{-11}, &  \quad ({\rm BBN}),\\
                              8.65 \pm 0.09 \times 10^{-11}, &\quad   ({\rm CMB}).
                            \end{array} \right.
                            \label{Yb}
\ee
Here $n_b, \, \bar n_{b},\, n_\gamma$ denote the number density of baryons, antibaryons and photons, and  $s =\frac{2\pi^2}{45} g_{* s} T^3$ is the entropy density, with $g_{*s}$ the effective number of degrees of freedom.
The systematics and parameter dependencies are different for these two cosmological probes, and they thus provide fairly independent measurements of $Y_b$, although both results assume standard ($\Lambda$CDM) cosmology. 
$Y_b$ is a useful quantity as it remains constant after the baryon-number violating interactions are turned off (and there is no significant entropy injection in the thermal plasma).
In our present day universe the density of antibaryons is negligible, and there is approximately one baryon for every 10 billion photons. This can be deduced from the cosmic ray spectrum: even though  positrons and antibaryons are observed in cosmic rays, the spectrum can be fully accounted for assuming all the antimatter originates from matter primaries. 

In an initially baryon-symmetric universe, there will be trace amounts of (anti)baryons left after the freeze-out of the baryon-antibaryon annihilation processes, but the corresponding densities fall short by orders of magnitude to explain the observed values. Moreover, no bright electromagnetic radiation from matter-antimatter annihilation has been observed,  indicating that there are no  antimatter pockets in our universe either. Explaining the observed baryon asymmetry as a consequence of antisymmetric initial conditions of the universe is not consistent with cosmological inflation. The enormous expansion of the universe during inflation effectively erases all initial conditions, including any pre-existing baryon asymmetry. 
Hence, the BAU asks for a dynamical {\it baryogenesis} mechanism. 

There are many baryogenesis scenarios on the market -- after all, there is just one number to explain, \cref{Yb} -- although some are better motivated than others.  
The two mechanisms for generating the BAU that have received the most attention in the literature are leptogenesis \cite{Fukugita:1986hr}, where the baryon asymmetry is produced during the production, evolution and/or decay of right-handed neutrinos; and electroweak baryogenesis \cite{Kuzmin:1985mm, Shaposhnikov:1986jp, Shaposhnikov:1987tw, Cohen:1990py, Cohen:1993nk}. Some other proposed scenarios are the Affleck-Dine mechanism \cite{Affleck:1984fy}, spontaneous baryogenesis \cite{Cohen:1987vi}, inflationary baryogenesis \cite{Anber:2015yca}, and mesogenesis \cite{Elor:2018twp}.
This review focuses on electroweak baryogenesis (EWBG) in which the baryon asymmetry is created during the electroweak phase transition. This scenario requires new, beyond-the-SM (BSM) physics not far from the electroweak (EW) scale. This physics can be readily probed by experiments, and in particular, by the Large Hadron Collider (LHC) (or future variants thereof),  electric dipole moment (EDM) searches, and gravitational wave observatories.

In this paper, we provide an extensive review of electroweak baryogenesis. Although one can explain the main idea in less than a page, and we will do so in \cref{sec:EWBG}, the actual quantitative calculation of the baryon asymmetry is rather involved. Our aim with this review is twofold. First, we want to give a state-of-the-art overview of the calculational approaches and the experimental bounds, with an emphasis on recent developments.  
Secondly, we hope that after reading this review, a researcher who is fairly new to the field will be familiar with all the steps in the calculation and, with the pointers to the literature, can compute the asymmetry for a given model. It is also good to stress what this review is not. We will not provide an exhaustive review of specific BSM models that can lead to successful electroweak baryogenesis. There are too many papers to discuss and they differ mainly in the particular BSM fields and interactions that are introduced. In this review we will concentrate on the underlying framework and methodology of electroweak baryogenesis itself.

\subsection{Electroweak baryogenesis in a nutshell}
To set the stage, we now give a brief overview of electroweak baryogenesis. 

\subsubsection{The Sakharov conditions}\label{sec:sakh}
Andrei Sakharov was the first to seriously think about a dynamical  baryogenesis mechanism, and in his seminal paper from 1967 \cite{Sakharov:1967dj} he identifies the necessary conditions. In many ways this work was far ahead of its time. In those days baryon number was thought to be conserved -- this was before the discovery/invention of chiral anomalies, grand unified theories, and black hole evaporation -- and cosmological inflation had not yet been invented.  
Sakharov's conditions are:
\begin{enumerate}
\item Baryon number violation. 
\item Charge (C) and charge-parity (CP) violation. 
\item Departure from thermal equilibrium. 
\end{enumerate}
The need for baryon number violation is obvious:  if all processes have the same baryon charge in the initial and in the final state, no asymmetry will be created.

A charge transformation flips all charges, and parity flips the spatial momenta.  Without C and CP violation the rate for left-chiral particle creation equals the rate for right-chiral antiparticle creation, and thus no net baryon asymmetry can be generated. A quick way to see this is to note that baryon number $B$ is odd under  C and CP, and a non-zero expectation value $\langle B \rangle$ requires that the Hamiltonian $H$ violates both C and CP.  Let's see this explicitly.  The SM is classically invariant under the global $U(1)_B$ transformation $q(x)  \to  e^{i\eps/3}q(x)$ and $\bar q(x)  \to  e^{-i\eps/3}\bar q(x)$ for the quarks and antiquarks.
Noether's theorem gives the associated baryon current and charge
\be
\partial^{\mu}J^B_{\mu}  = \partial^{\mu}
\sum_q \frac{1}{3}{\bar q}\gamma_{\mu} q = 0  \quad \Rightarrow \quad
{ B} = \int \dd^3x \, J^B_{0}(x) =\sum_q \frac{1}{3}\int \dd^3x \, {\bar q}\gamma_{0} q .
\ee
As reviewed in \cref{A:CP}, the quark fields transform under P and C as in \cref{CP_fermions}, from which it follows that $P{ B} P^{-1} = { B}$ and $
C{ B} C^{-1} = -{ B}$, and thus baryon number is indeed odd under C and CP.

In thermal equilibrium no preferred direction for time can be defined.
This can be proven as follows \cite{Bernreuther:2002uj}. The baryon expectation value in thermal equilibrium  is constant in time $\langle  B(t)\rangle \equiv  \tr[\e^{-H/T} B(t) ] =\tr[\e^{-H/T} \e^{iHt}{B(0)}\e^{-iHt} ] =\langle  B(0)\rangle$.
The baryon number
$B(0)$ is even under time translation T and odd under CP, and thus odd under $\Theta =$CPT.
In thermal equilibrium baryon number must then vanish, since
\be
\langle B(t) \rangle 
= \tr({\Theta}^{-1}\Theta
\e^{-H/T}
B(0))=
\tr(\e^{-H/T}\Theta B(0) {\Theta}^{-1}) = - \langle B(t)\rangle,
\ee
where we used that the Hamiltonian is CPT invariant.\footnote{If CPT is violated then a baryon asymmetry can be produced without a departure from thermal equilibrium \cite{Cohen:1987vi}. We will not pursue this in this review.}

The immediate question to ask is: can the Standard Model fulfill the Sakharov conditions?  Perhaps surprisingly, there is baryon number violation in the SM. Although the Lagrangian is classically invariant under both the global baryon $U(1)_B$ and lepton $U(1)_L$ symmetry, these symmetries are broken by quantum effects. As a consequence, non-perturbative  processes violate $B$ and $L$ by three units each. In today's universe these transitions are exponentially suppressed and negligibly small. However, in the hot environment of the primordial plasma, baryon number-violating {\it sphaleron} rates are large.  

The SM has more trouble with the other two Sakharov conditions. The weak gauge force only couples to left-handed fermions, which breaks P maximally. CP is violated by the phase in the CKM matrix. All three fermion generations should be involved in the process for this phase to be physical. CP violation is parameterized in a basis-invariant way by the Jarlskog invariant $J_{\rm CP} = {\rm det}[m_u^2,m_d^2]$, in terms of a commutator of the mass matrix for the three-generation up and down sector (see e.g. \cite{Cline:2006ts}).  Dividing by the powers of the electroweak scale to obtain a dimensionless measure of CP violation gives
\be
\frac{J_{\rm CP}}{(100\,{\rm GeV})^{12}} \sim 10^{-20},
\ee
which is too small for baryogenesis at or above the EW scale, where sphaleron transitions are active.\footnote{In cold electroweak baryogenesis the EW phase transition takes place at a temperature $T_{\rm PT}$ much below the EW scale; the CP violation  $J_{\rm CP}/T_{\rm PT}^{12}$ can then be large enough for baryogenesis, see e.g. Ref.~\cite{Tranberg:2010af}.}
The other potential SM source for CP violation,  the QCD $\theta$-angle, is likewise too small \cite{Kuzmin:1992up} as it is severely constrained by the non-observation of the electric dipole moment of the neutron and diamagnetic atoms. We can be short about the third Sakharov condition. Assuming the standard hot big bang cosmology, within the SM there are no sources of out-of-equilibrium dynamics that are significant enough for successful baryogenesis. In particular, the EW phase transition is a smooth cross over~ \cite{Kajantie:1995kf, Kajantie:1996mn, Kajantie:1996qd, Gurtler:1997hr, Csikor:1998eu, Aoki:1999fi}. 
To summarize, only one of the Sakharov conditions is (sufficiently) satisfied in the SM, and BSM physics is required to dynamically explain the BAU.


\subsubsection{Electroweak baryogenesis: a lightning review}
\label{sec:EWBG}

In electroweak baryogenesis \cite{Kuzmin:1985mm, Shaposhnikov:1986jp, Shaposhnikov:1987tw, Cohen:1990py, Cohen:1993nk} the baryon asymmetry is created during the electroweak phase transition (EWPT), in which the Higgs field obtains a non-zero vacuum expectation value (vev), which breaks the EW symmetry. A first-order transition gives the necessary non-equilibrium dynamics; new sources of CP-violation active during the phase transition are added to satisfy the second Sakharov condition. Electroweak sphalerons cause baryon number violation and address the first Sakharov condition. In this section we will describe a `vanilla' implementation of baryogenesis, i.e. a set-up in which only the SM Higgs field obtains a vev during the EWPT, and CP violation resides in the effective Yukawa coupling to the Higgs field.
Because of its relative simplicity, this serves well to highlight the necessary ingredients and describe the relevant dynamics. However, as we will discuss later in this review, the minimal set-up is under severe pressure from experimental constraints.

Quantum corrections to the Higgs potential are temperature dependent in the hot early universe, as virtual particles can exchange energy-momentum with the plasma. At high temperature, the dominant effect is a large temperature-dependent effective mass term for the Higgs, and the minimum of the Higgs potential is at the origin; the electroweak symmetry is restored. If at the critical temperature there is a barrier between the degenerate high- and low-temperature minima, the EWPT is first order and proceeds via the nucleation of bubbles; otherwise a second order phase transition or smooth cross-over takes place.  With the particle content of the SM the EWPT is a cross-over, and new physics in the Higgs sector is needed for a first-order phase transition (FOPT). The LHC data currently only weakly constrain the triple Higgs interaction, and large corrections are still allowed. 

For the new source(s) of CP violation to be important during the EWPT, they should live around the TeV scale. A large number of variants can be considered depending on the number and nature of the BSM fields. In this section, we will use an effective field theory approach where new sources of CP violation arise from higher-dimensional operators with a cutoff scale just above the EW scale.  As an explicit example, consider the dimension-six correction to the top quark Yukawa interaction (see e.g.~\cite{Bodeker:2004ws,Huber:2006ri,Balazs:2016yvi,deVries:2017ncy} for studies of EWBG with this operator)
\be
\L \supset -y_t \bar Q_L \tilde \vp t_R\(1+ c_t\vp^\dagger \vp\)+\mathrm{h.c.},
\label{CP_yukawa}
\ee
with $Q_L = (t_L\,\, d_L)^T$, the third-generation left-handed quark doublet, and $c_t$ a coupling constant with mass dimension minus 2. CP is violated in top-Higgs interactions if $\Im (c_t) \neq 0 $. The large value of the top Yukawa enhances the baryon yield. 
The Higgs doublet in unitary gauge is $\vp =\frac1{\sqrt{2}} \binom{0}{\phi+ h}$, with $\phi$ the background field -- in today's universe $\phi_0 \equiv v =246\,$GeV -- and $h$ the excitation around it. In a constant background, the phase in the effective top mass $m_t = \frac{y_t}{\sqrt{2}} \phi (1+ c_t \phi^2)$ can be removed by a chiral rotation of the top quark fields.  However, in the bubble wall background the Higgs field value $\phi(x)$ is space-time dependent, and the phase cannot be removed.  Hence, the effective top mass in the bubble wall background is complex, in particular inside the bubble wall region where the vev changes, and the corresponding CP violation is physical. Note that even in the vacuum the $\bar t t h$-coupling violates CP, which can be probed in EDM experiments or, to lower accuracy, at the LHC.

\begin{figure}[t!]
    \centering
    \includegraphics[width=0.6\textwidth]{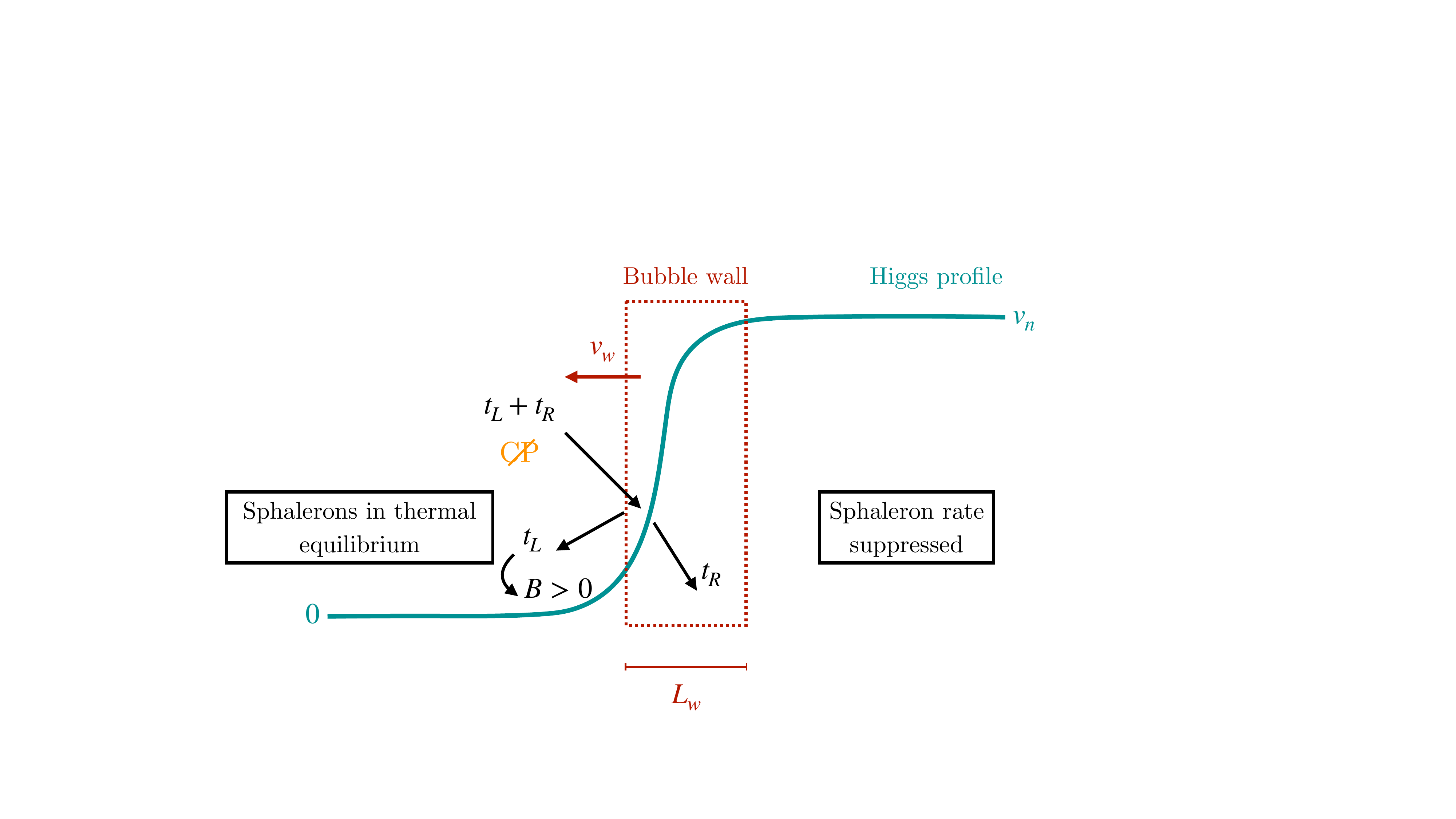}
    \caption{EWBG in a nutshell. Figure inspired by Ref.~\cite{Morrissey:2012db}.}
    \label{fig:nutshell}
\end{figure}

Having discussed how the Sakharov conditions are incorporated in EWBG, let's now describe the dynamics of the  scenario. See \cref{fig:nutshell} for an illustration.
At high temperature the EW sphalerons are in thermal equilibrium and any pre-existing $B$-$L$ symmetry is erased. The universe cools, and at EW scale temperatures the first-order EWPT takes place. Bubbles of true vacuum are nucleated, and expand into the surrounding plasma. 
The plasma particles scatter off the bubble wall, and if this interaction violates CP, 
the distributions of particles and antiparticles -- throughout this review we will use `antiparticle' for the CP conjugate state -- in front of and behind the wall will be different. 
For example, the presence of the CP-violating top Yukawa coupling \cref{CP_yukawa} can lead to an overdensity of left-handed anti-top quarks  over left-handed top quarks in front of the bubble wall\footnote{Here we define the left-handed antiquark as the CP conjugate of the left-handed quark, which is an SU(2) doublet. As CP conjugation flips chirality, in the literature one also encounters the nomenclature in which the right-handed antiquark is the doublet state. 
} (and a compensating overdensity of right-handed tops over anti-tops, as the interaction does not violate baryon number).  The EW sphaleron only involves the left-handed fermions. As there are more left-handed antiparticles outside the bubble, this will bias the sphaleron rate which destroys antiparticles (in favor of particles as it violates baryon number) over the inverse rate, and a net baryon asymmetry is created.

Left to itself, all the fast plasma interactions would bring the system back to equilibrium, and erase the asymmetry.  However, before the asymmetry is washed out the baryons are swept up by the expanding bubbles.  If the PT is sufficiently strong, the sphaleron rate is suppressed inside the bubble, and the baryons remain. At the end of the PT, when bubbles collide and coalesce, there will be a net baryon number created.

This, in a nutshell, is electroweak baryogenesis. The key ingredients are a strong first-order EWPT and a CP-violating interaction involving the scalar field that undergoes bubble formation. The mechanism is attractive as it involves interesting phenomena that only require rather mild modifications of the SM. However, the devil is in the details.  The main goal of this review is to fill in these details.

\subsection{The organization of this paper and its place in the  landscape of reviews}

Much progress has been made in recent years regarding various  aspects of electroweak baryogenesis. In particular, with respect to older reviews \cite{Trodden:1998ym,Riotto:1998bt,Riotto:1999yt,Cline:2006ts,Morrissey:2012db,Konstandin:2013caa} tremendous experimental progress has been made. Most dramatic are the discovery of the Higgs boson, the non-detection of supersymmetry at the LHC, the first detection of gravitational waves, and roughly 1.000 times more precise EDM experiments. That being said, the aforementioned reviews still cover the essential aspects of the topic and are well worth studying. 

The theoretical description of electroweak baryogenesis has also improved significantly (although, as this review shows, major open questions remain). In particular, it has been shown that a popular method to compute the CP-violating source term that drives the generation of the chiral asymmetry in front of the bubble wall, the so-called vev-insertion approximation, has been derived incorrectly. Relatively recent reviews \cite{Morrissey:2012db, White:2016nbo} are partially based on this approximation and thus require updating. Since these and other \cite{Konstandin:2013caa, Garbrecht:2018mrp} reviews, major progress has been made in the computation of both the phase transition dynamics and the bubble wall velocity. It has been shown that a CP asymmetry can still be generated at relativistic wall velocities, refuting previous expectations. 
This is interesting as the gravitational wave signal is generically larger for faster moving bubbles. 
These developments have been spurred by the ever increasing constraints on CP violation from improved EDM experiments.

This review is organized as follows. We start in the next section with the dynamics of the first-order electroweak phase transition. We review the finite-temperature effective potential and bubble nucleation, highlighting the state-of-the art approaches and the improved understanding of theoretical uncertainties. We provide examples of models that do give a first-order phase transition, and discuss if it is possible to describe the FOPT in a model-independent way using the Standard Model Effective Field Theory. 
\Cref{sec:CP} continues with the creation of the CP asymmetry. 
The system is driven out of equilibrium by (CP-violating) {\it source} terms that turn on as the bubble passes by. We review the semi-classical, flavor, and (the vanishing of) the vev-insertion-approximation source terms. 
We revise how transport equations can be derived that describe the evolution of the quanta in the thermal plasma.
In \cref{sec:bubbles} we review how to solve the transport equations. In particular, we discuss how the bubble wall velocity  and the CP-violating number densities can be computed from the CP conserving and violating parts of the Boltzmann equations respectively.  Recent results for supersonic bubbles are addressed.
In \cref{sec:sphalerons}, we discuss the relevant transport equations for the conversion of a CP-asymmetry into baryon number and the state-of-the-art values for the symmetric and broken phase sphaleron rates 
EWBG requires new physics at the EW scale, which can be probed by experiment. 

The bounds on the amount of CP violation have improved tremendously over the last decades by the increasingly precise determination of the (absence of the) electric dipole moments of nucleons, atoms, and molecules. Much progress has also been made on determining the background of gravitational waves generated by a FOPT. We will discuss all these developments
in \cref{sec:experiment}, where we review the experimental probes. The experimental bounds increasingly constrain models of EWBG, forcing to look beyond the vanilla set-up; we highlight some of these model implementations in \cref{sec:models}. Finally, we end with a summary and outlook in \cref{sec:sum}.  To set notation, we list our conventions for the SM Lagrangian and CP transformations in \cref{A:SM,A:CP}, and introduce the closed-time path (CTP) formalism in \cref{A:CTP}. 

Some (sub)sections of this review necessarily will be technical. To guide the reader, we end the more technical \cref{sec:FOPT,sec:CP,sec:bubbles,sec:sphalerons} with a summary for the practitioner with explicit references to the most important formulae.  

\subsection{Conventions}
\label{sec:conventions}

We use the mostly-negative sign convention for the metric  $(+,-,-,-)$.
For large bubbles the bubble wall is well approximated by a planar surface. We follow the commonly used convention that the bubble wall is moving along the negative $z$-direction with velocity $v_w$, that is, the symmetric phase is at $z<0$. We tried to keep notation and conventions consistent throughout the review.  An exception is for the rates:
The nucleation and sphaleron rate $\Gamma_n$ and $\Gamma_{\rm sph}$ in \cref{sec:FOPT,sec:sphalerons} are defined as the rate per unit time and per unit volume, whereas the other rates appearing in this review -- also denoted by $\Gamma$ -- are the rates per unit time.  A list of often used abbreviations is given in the table below.\\[1cm]

\begin{center}
 \label{tab:conventions}
\begin{tabular}{ |l|l| }
 \hline
{\bf acronyms} &\\
\hline
 (B)SM & (Beyond the) Standard Model \\  \hline
 BAU & baryon asymmetry of the universe \\   \hline
 (FO)PT & (first-order) phase transtion \\ \hline
 EW & electroweak\\ \hline
 EWBG& electroweak baryogenesis\\ \hline
 EDM & electric dipole moment \\  \hline
 VIA & vev-insertion-approximation\\ 
 \hline
 KB equations & Kadanoff-Baym equations \\ \hline
 CPV & CP-violating/CP violation \\
 \hline
 CPC & CP-conserving/ CP conservation \\
 \hline
\end{tabular}
\end{center}

	\newpage
	\section{First-order phase transition}\label{sec:FOPT}
	   A first-order phase transition (FOPT) is one of the essential ingredients in EW baryogenesis. 
For the measured value of the Higgs mass, $m_h = 125 \, {\rm GeV}$, the electroweak phase transition in the SM is a cross-over \cite{Kajantie:1995kf, Kajantie:1996mn, Kajantie:1996qd, Gurtler:1997hr, Csikor:1998eu, Aoki:1999fi}, and new states coupling to the Higgs are necessary to change the order of the PT. We will start with a very brief review of the finite-temperature effective potential, and the computation of the nucleation temperature and the bubble profile.
We will then roughly demonstrate why the phase transition in the SM is not first order. We will provide examples of models that \emph{do} give a FOPT and discuss if it is possible to describe the FOPT in a model-independent way using the Standard Model Effective Field Theory. Several more detailed reviews have appeared that focus on phase transitions  
\cite{Quiros:2007zz,Mazumdar:2018dfl,Hindmarsh:2020hop}. Here we focus on the aspects of the problem most directly connected to EWBG, and highlight recent developments.

\subsection{Finite-temperature effective potential}\label{sec:Potential}

The effective potential 
is computed by adding quantum corrections to the tree-level expression. As the particle distributions describing the plasma in the early universe  are temperature dependent, the loop corrections and the resulting effective potential become temperature dependent  as well. This gives
the possibility of cosmological phase transitions.
There are several formalisms to compute the finite-temperature effective potential, such as the imaginary-time, or Matsubara formalism \cite{Matsubara:1955ws}, and the real-time formalism \cite{Niemi:1983nf, Landsman:1986uw}; see \cite{Quiros:1999jp, Laine:2016hma} for reviews. For the content of this section, they all lead to the same conclusions.

Let's assume for simplicity that the Higgs field is the only scalar field in the effective field theory at the EW scale; the description of the effective potential in this subsection is straightforwardly generalized to a multi-scalar set-up.
At the one-loop level, the scalar potential for the background Higgs  field $\phi$ consists of the following contributions:
\begin{equation}
	V_{T}(\phi,T) = V_{\rm tree} (\phi) + V_{\rm CW}(\phi) 
    + V_{T, {\rm 1-loop}}(\phi, T),\label{eq:VT}
\end{equation}
where $V_{\rm tree}$ denotes the zero-temperature tree level potential. $V_{\rm CW}(\phi)$ is the zero-temperature one-loop (Coleman-Weinberg) correction \cite{Coleman:1973jx}, given by
\begin{equation}
    V_{\rm CW} = \frac{1}{64 \pi^2} \sum_i n_i m_i^4 \left[\log{\frac{m_i^2}{\mu^2}} -C_i  \right],
    \label{V_CW}
\end{equation}
in the $\overline{\rm MS}$ renormalization scheme, with $C_i = 3/2$ for scalars and fermions and $C_i=5/6$ for vector bosons, and with $\mu$ the renormalization scale. The divergent part of the one-loop potential has been canceled by the counter-term potential $V_{\rm ct}$, leaving the finite CW-potential given above. 
The sum $i$ runs over particle species, $n_i$ denotes the number of degrees of freedom (a sign has been absorbed in the definition such that $n_i$ is positive for bosons and negative for fermions), and $m_i$ the mass eigenvalues. In the SM, the most significant contributions come from the $W$ and $Z$ bosons, the top quark and the Higgs field itself.
In principle, the RG-scale $\mu$ can be chosen freely, as long as the masses and couplings are evaluated at that scale.
A typical choice is $\mu = 4\pi e^{-\gamma_E}T$, with $\gamma_E$ the Euler-Mascheroni constant, which cancels the logarithmic term appearing in the (high-temperature expansion of) the finite-temperature potential, {\it cf}.~\eqref{eq:JB/FHighT}.
However, as discussed below, finite-temperature perturbation theory suffers from slow convergence, aggravated by the large loop corrections that are typically needed for a FOPT, which is signaled by residual dependence on the value of $\mu$.

The one-loop finite-temperature contribution is \cite{Dolan:1973qd}
\begin{equation}
    V_{T, {\rm 1-loop}}(\phi, T) = \frac{T^4}{2\pi^2} \left[\sum_{i \subset \rm bosons} n_i \,J_B[m_i(\phi)^2/T^2] +\sum_{i \subset \rm fermions} n_i \,J_F[m_i(\phi)^2/T^2] \right],
    \label{V_T}
\end{equation}
with 
\begin{equation}
    J_{B,F}(y^2) = \int_0^\infty \dd k \,k^2 \log{[1 \mp \exp{[-\sqrt{k^2 + y^2}]}]},
    \label{eq:J_BF}
\end{equation}
with the minus sign for bosons and the plus sign for fermions. In the high-temperature regime $T^2 \gg m_i^2$, the functions $J_{B,F}$ can be expanded as follows 
\begin{align}
    J_B(m_i^2/T^2) &= -\frac{\pi^4}{45} + \frac{\pi^2}{12} \frac{m_i^2}{T^2} -\frac{\pi}{6} \frac{m_i^3}{T^3} - \frac{1}{32}\frac{m_i^4}{T^4} \left(\log{\frac{m_i^2}{16\pi^2 T^2}} - \frac 3 2 + 2\gamma_E \right) \cdots\,\nn , \\
    J_F(m_i^2/T^2) &= \frac{7\pi^4}{360} - \frac{\pi^2}{24} \frac{m_i^2}{T^2} - \frac{1}{32}\frac{m_i^4}{T^4} \left(\log{\frac{m_i^2}{\pi^2 T^2}} - \frac 3 2 + 2\gamma_E \right) \cdots \, ,\label{eq:JB/FHighT}
\end{align}
where the ellipsis denotes higher order terms in $m_i^2/T^2$ and $\gamma_E$ is the Euler-Mascheroni constant.
From the one-loop, leading order (in $m_i^2/T^2$) term, we can already see how the temperature-corrections affect the shape of the potential: the $m_i^2/T^2$-terms can give a positive contribution to the Higgs mass parameter, making the symmetric minimum the preferred vacuum state at high temperature. 

It should be stressed that the one-loop potential~\cref{eq:VT} typically does not give an accurate description of the thermodynamics. The reason is that IR-modes (Matsubara zero-modes) receive Bose-enhancement, which jeopardizes the convergence of the loop expansion. The most well-known manifestation of this is the IR-divergence in the diagrams corresponding to the cubic terms in \cref{eq:JB/FHighT} \cite{Dolan:1973qd,Kirzhnits:1976ts}. This IR-divergence is caused by loop contributions from massless modes (see e.g. \cite{Kapusta:2006pm, Laine:2016hma}), and it gets cured by taking the thermal correction to these masses into account \cite{Parwani:1991gq,Carrington:1991hz,Arnold:1992rz}. This procedure is usually referred to as daisy resummation, and it corresponds at high temperatures to adding the following term to the effective potential (this corresponds to the Arnold-Espinosa resummation scheme)
\begin{equation}
	V_{\rm daisy} = -\frac{T}{12 \pi} \sum_i n_i \left[(m_{i,{\rm th}}(\phi,T))^{3/2} - m_i^2(\phi)^{3/2} \right],\label{eq:daisy}
\end{equation}
where the sum runs over all scalars and the transverse polarizations of the gauge bosons. Diagrammatically, this contribution arises from an infinite number of diagrams of a one-loop zero-mode diagram with nonzero-mode one-loop diagrams attached around; hence the name {\it daisy} resummation.
Further $m^2_{i,{\rm th}}$ corresponds to the thermally corrected masses, obtained from the one-loop effective action in the high temperature expansion ($T^2 \gg m_i^2)$; these inputs can be taken from the literature, or alternatively, calculated e.g. using the {\tt DRalgo} package  \cite{Ekstedt:2022bff}.

To cubic order in the masses, simply replacing $m_{i}(\phi) \to m_{i,{\rm th}}(\phi,T)$ in $V_T$ gives the same effect in the high-temperature expansion as adding $V_{\rm daisy}$, and this is the prescription often used in the literature. Although the use of $V_T(m_{i,{\rm th}})$ can strictly be motivated only at high temperatures, the procedure is generically extended beyond this regime as well (using the integral expression for $J_{B,F}$ given in \cref{eq:J_BF} rather than the expansions \cref{eq:JB/FHighT}). Moreover, there is confusion in the literature as to whether to substitute the thermally corrected masses for the spin-0 fields only in the thermal potential (e.g. \cite{Jiang:2022btc,Blinov:2015vma,Gil:2012ya}), in the thermal and CW potential
(e.g \cite{Cline:2011mm,Branchina:2025jou}), or in the thermal and renormalized one-loop potential (e.g \cite{Benincasa:2022elt}). 
The first approach corresponds to Arnold-Espinosa or daisy resummation as introduced above. This is the leading order of a (dimensionally reduced) EFT description of the phase transition. 
In this sense, the approach is better justified than the other approaches; but as we will discuss in \cref{sec:VTaccuracy}, an accurate description of the phase transition requires going \emph{beyond} leading order.

\subsubsection{Obtaining the nucleation temperature and bubble profile}
\begin{figure}
    \centering
     \includegraphics[scale=0.55,trim=0 0 0 0]{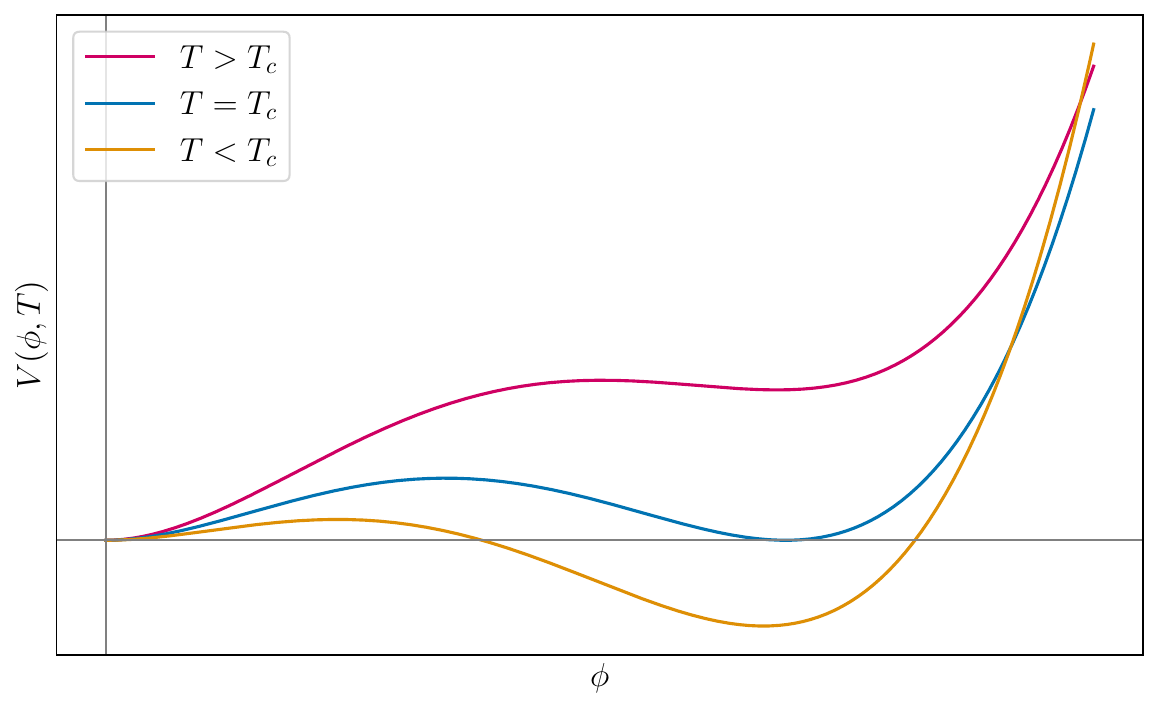}
    \caption{A sketch of the finite temperature effective potential without the field-independent ($\propto T^4$) contribution. }
    \label{fig:VT}
\end{figure}
From the point of view of EWBG, the main quantities of interest regarding the PT are the nucleation temperature, the bubble wall profile, the strength of the PT, and the bubble wall velocity. The latter will be discussed in section~\ref{sec:Bubblevw}. 
A typical evolution of the effective potential is demonstrated in figure~\ref{fig:VT}. The magenta line shows the potential at high temperature, where the origin is the vacuum state. As the temperature decreases, a second minimum forms. The two minima are degenerate at the critical temperature $T_c$, indicated by the blue line; the orange line gives the potential at below the critical temperature $T<T_c$. 

Once the low-temperature vacuum state becomes energetically favorable, bubbles of broken vacuum will start to form. At high temperature  $T R_b \gg 1$ with $R_b$ the size of the critical bubble at the time of nucleation -- this can be checked a posteriori -- the dynamics is dominated by the Matsubara zero-modes and the system becomes effectively 3-dimensional \cite{Linde:1981zj}. The bubbles are $O(3)$ symmetric solutions of the 3D-Euclidean action, that is, they are solutions of the 3D  `bounce' equation of motion \cite{Coleman:1977py,Linde:1981zj}
\begin{equation}
    \frac{d^2 \phi_b}{dr^2} + \frac{2}{r} \frac{d\phi_b}{dr} = \frac{dV_T(\phi_b)}{d\phi_b},\label{eq:Bounce}
\end{equation}
with boundary conditions $d\phi_b/dr = 0$ for $r=0$ and $\phi_b \rightarrow 0$ for $r \rightarrow \infty$. This equation is typically solved using a shooting algorithm, and several packages are available to solve the bounce equation \cite{Wainwright:2011kj, Masoumi:2016wot, Basler:2018cwe, Athron:2019nbd, Sato:2019wpo, Guada:2020xnz}. 
The temperature-dependent nucleation rate per unit volume $\Gamma_n$ is then given by 
\begin{equation}
    \Gamma_n(T) = A \e^{-S_3(T)/T},\label{eq:rate}
\end{equation}
where $S_3(T)$ is the 3-dimensional Euclidean action evaluated at the bounce configuration, which has mass dimension one.
The prefactor $A$ splits into a dynamical part and a statistical part \cite{Langer:1967ax, Langer:1969bc, Gould:2021ccf}. The former was estimated in~\cite{Langer:1969bc, Hanggi:1990zz, Berera:2019uyp}. The latter corresponds to the fluctuation determinant appearing in the one-loop effective action.
Often, the prefactor $A$ is assumed to be of lesser importance than the exponential contribution and simply estimated as $T^4$ based on dimensional analysis. See \cref{sec:VTaccuracy} for a discussion about this estimate. 

Even though bubbles of the new phase can already be formed at the critical temperature, initially the bubble nucleation rate $\Gamma_{n} (T)$ is smaller than the expansion rate of the universe, $\Gamma_{n} < H^4$, where $H$ is the Hubble parameter.
We compute the baryon asymmetry at the \emph{nucleation temperature} $T_n$, which is the temperature where the average number of bubbles in each Hubble patch,
$N$, equals unity \cite{Linde:1980tt, Linde:1981zj}
\begin{equation}
    N(T_n) =  \int_{T_n}^{T_c} \frac{\dd T}{T} \frac{\Gamma_n}{H^4}=1\,.\label{eq:NuclCrit}
\end{equation} 
The nucleation condition can be approximated by the assumption that the integral is dominated by the contribution from $T\sim T_n$, yielding the following approximation \cite{Caprini:2019egz}
\begin{equation}
    \frac{S_3(T_n)}{T_n} = 141 +\log{\(\frac{A}{T_n^4}\)} -4\log{\(\frac{T_n}{100 \, {\rm GeV}}\)} - \log{\(\frac{\beta/H}{100} \)},   \label{eq:NuclCritApprox}
\end{equation}
where $\beta/H$ is given by 
\begin{equation}
    \frac{\beta}{H} = - \frac{\dd\log{\Gamma_n}}{\dd\log{T}},\label{eq:betaoH}
\end{equation}
and for the EWPT the prefactor is estimated as $\log(A/T^4) \sim -14$ \cite{Carrington:1993ng}. $T_n$ determines the onset of the phase transition. In fact, a more relevant 
(but slightly more difficult to compute, see \cite{Guth:1981uk, Enqvist:1991xw}) temperature is the percolation temperature, which is defined as the temperature where a significant volume fraction ($\mathcal O (0.3)$) has converted to the true vacuum. A simplified criterion for percolation is given by \cite{Caprini:2019egz}
\begin{equation}
    \frac{S_3(T_p)}{T_p} = 131 +\log{\(\frac{A}{T_p^4}\)} -4\log{\(\frac{T_p}{100 \, {\rm GeV}}\)} - \log{\(\frac{\beta/H}{100} \)} + 3 \log{(v_w)}.
\end{equation}
For models with only a moderate amount of supercooling $T_n \sim T_p$ and we do not make a strict distinction between the two temperatures in the rest of this review. The action $S_3$ and rate $\Gamma_n$ are evaluated for
the bubble profile $\phi_b$  that is the solution of eq.~\eqref{eq:Bounce} at $T_n$.
Once $T_n$ is known, the ratio $v_n/T_n$ can also be computed, where $v_n$ is the value of the Higgs field in the broken phase minimum. As we will see in section~\cref{sec:broken_sphaleron}, a heuristic way to estimate  whether the phase transition is strong enough to avoid a disastrous washout of the baryon asymmetry inside the bubbles is to satisfy $v_n/T_n \gtrsim 1$.
Note that this condition is gauge-dependent \cite{Patel:2011th, Garny:2012cg}. We will discuss this in more detail in \cref{sec:broken_sphaleron}.

For large bubbles the curvature of the bubble can be neglected, and the bubble wall -- the region over which the classical Higgs field changes -- is well approximated by a planar wall.  The bubble profile is often fitted to a kink solution \cite{John:1998ip} in the plasma frame (the frame where the plasma is at rest far away from the bubble wall) of the form
\begin{equation}
	\phi_b(z,t) = \frac{v_n}{2} \left(1 + \tanh \frac{z + v_w t}{L_w} \right)\,,
    \label{bubble_profile}
\end{equation}
and chosen here to be moving along the minus-$z$ direction.  
\hyperlink{box2}{Box 2} in \cref{sec:Bubblevw} reviews how the profile can be written in a frame-invariant form.
The parameters $v_n$, $T_n$, and (to some extent) the bubble wall width $L_w$ can be extracted from the fit to the bounce solution.
Strictly speaking, the bounce solution does not reach $v_n$ in the broken phase, but a smaller value. After the transition, the field value at the center of the bubble will quickly relax to $v_n$.
Using the exact bounce profile therefore carries the risk of underestimating the asymmetry.
The most realistic profile is given by the one which reaches $v_n$ in the broken phase, and includes hydrodynamics and friction effects.
In \cref{sec:Bubblevw} we review how to calculate this profile and the bubble wall velocity $v_w$.

\subsubsection{Accuracy  and gauge-(in)dependence of the thermal parameters}\label{sec:VTaccuracy}

Unfortunately, the potential obtained after daisy-resummation often does not provide sufficient accuracy \cite{Croon:2020cgk, Niemi:2021qvp, Gould:2023jbz, Kierkla:2023von}. This is, for example, signaled by a strong dependence of the thermodynamics on the choice of RG-scale \cite{Ghisoiu:2015uza, Gould:2021oba}. This problem is particularly severe for the single-step FOPT  with the potential barrier generated by radiative corrections, as this requires Higgs couplings to the new degrees of freedom close to the perturbativity bound. It is a model-dependent question whether a two-step phase transition, possibly in a multi-scalar model, fares better, but \cite{Gould:2021oba} shows that the problem also arises in the singlet extension with a two-step PT.

A systematic approach to include higher-order thermal contributions is provided by the method of High Temperature Dimensional Reduction \cite{Ginsparg:1980ef, Appelquist:1981vg, Braaten:1995cm, Kajantie:1995dw}.
In this framework, applicable in thermal equilibrium, one uses that the temperature can be viewed as a compact imaginary time direction, and the fields can be expanded as a tower of modes, called the Matsubara modes.
After integrating out all the heavy modes, a (3-dimensional) effective theory is derived for the light modes that are relevant for the phase transition, and the effects of the heavy fields are captured by the masses and couplings of the effective theory -- the IR-divergences described around \cref{eq:daisy} are resummed in the effective parameters.
This process has been automated in the package {\tt DRalgo}~\cite{Ekstedt:2022bff}.
Within the effective theory, the effective potential can be computed up to several orders in the relevant coupling expansion.
In pure QCD this was e.g. done to $\mathcal O(g_s^6 \ln(1/g_s))$,
corresponding to four loop orders \cite{Kajantie:2002wa, Kajantie:2003ax} (this is not so relevant to EWBG). 
For an SU(2) theory coupled to a scalar doublet, which resembles the SM but lacks the SU(3) and U(1) gauge groups, the effective potential was computed to $\mathcal O (g^{11/2})$, corresponding to three loops \cite{Ekstedt:2024etx}.
The often-employed daisy-resummed effective potential, discussed in \cref{sec:Potential}, typically corresponds to the leading-order effective potential (e.g. $\mathcal O(g^3)$, but this depends on the relevant power counting)
in the effective theory, with truncated matching relations (see e.g. \cite{Kierkla:2023von} for an explicit comparison). 

Usually, a high-temperature expansion is employed in the construction of the effective theory, which converges when the fields contributing to the effective potential have masses smaller than the temperature.
Higher order terms in this expansion show up as higher-dimensional operators in the effective theory, and the validity of the high-temperature regime was explored in \cite{Laine:2017hdk, Bernardo:2025vkz, Chala:2024xll}, showing that the expansion could break down for strong phase transitions.
In \cite{Navarrete:2025yxy} a framework for including higher loop effects without high-temperature expansion was developed. 
This is a promising development, but
so far, it has only been applied to a toy model.

Even in the dimensionally reduced effective theory, the accuracy that can be obtained perturbatively is limited by gauge modes that do \emph{not} develop a thermal mass of the order $gT$ (like the Debye mass) and remain IR-singular \cite{Linde:1980ts, Gross:1980br}. This non-perturbativity implies that phase transitions should really be studied on the lattice, as was e.g. done for the Standard Model and various extensions in~\cite{Kajantie:1996mn, Farakos:1994xh, Jansen:1995yg, Rummukainen:1996sx,Rummukainen:1998as, Kainulainen:2019kyp, Niemi:2020hto, Niemi:2024axp}. 
It has been demonstrated for a range of theories and observables that perturbation theory can obtain results close to the lattice \cite{Kajantie:2002wa, Ekstedt:2024etx,
Niemi:2020hto,
Niemi:2024axp,
Ghiglieri:2021bom, Ekstedt:2022zro,  Gould:2023ovu}, at sufficiently high loop order. An important caveat is that these successful comparisons all concerned \emph{equilibrium} quantities such as the critical temperature, and not the nucleation temperature. 
Summing up, perturbation theory can be a useful tool to study the order of the phase transition, but an accurate prediction of the phase transition parameters requires going beyond the one-loop effective potential and in some cases a lattice computation. 

The nucleation rate of eq.~(\ref{eq:rate}) also requires further discussion. First, using the loop-corrected effective potential in the computation of the bounce in eq.~(\ref{eq:Bounce}) and then plugging the corresponding bounce effective action into eq.~(\ref{eq:rate}) leads to a double-counting of the thermal fluctuations, as discussed in \cite{Croon:2020cgk, Weinberg:1992ds, Buchmuller:1993bq, Gleiser:1993hf, Bodeker:1993kj, Berges:1996ib, Surig:1997ne, Strumia:1998nf, Garbrecht:2015yza}. This problem is resolved by recognizing and using the hierarchy of scales between the UV modes providing the barrier, and the IR modes relevant for the phase transition, see \cite{Langer:1969bc,Gould:2021ccf, Croon:2020cgk, Weinberg:1992ds, Buchmuller:1993bq, Bodeker:1993kj, Berges:1996ib, Strumia:1998nf, 
Langer:1974cpa,
Hirvonen:2022jba}. 
The statistical part of prefactor $A$ in eq.~(\ref{eq:rate}) is then determined by the functional determinant of (IR) fluctuations around the bounce solution \cite{Callan:1977pt, Baacke:1993ne, Baacke:1995bw} obtained from the leading-order effective action. A computation of the fluctuation determinant can lead to significant corrections to the nucleation temperature \cite{ Strumia:1998nf, 
Baacke:1995bw,Ekstedt:2021kyx, Ekstedt:2022ceo, Ekstedt:2023sqc,Kierkla:2025qyz}, as compared to a simple estimate based on dimensional analysis. The challenge of computing the fluctuation determinant is alleviated by the software package {\tt BubbleDet} \cite{Ekstedt:2023sqc}, which can be used for single field phase transitions. 

A clear symptom of an inconsistent treatment of the thermodynamics is when physical observables, such as the critical and nucleation temperatures, become gauge dependent. In the daisy-resummed effective potential gauge dependence stems from the Goldstone boson contributions (see e.g. \cite{Croon:2020cgk} for an explicit demonstration in $R_\xi$-gauge), which remains uncancelled.
A common approach to deal with this unwanted gauge-dependence is to simply exclude the gauge-dependent contributions, sacrificing the accuracy of the effective potential \cite{Schicho:2022wty}.
A more effective approach is to obtain the thermodynamical parameters in a strict perturbative expansion (sometimes called $\hbar$-expansion) in the 3D EFT \cite{Ekstedt:2022zro, Gould:2023ovu,Laine:1994zq, Baacke:1999sc, Lofgren:2021ogg, Hirvonen:2021zej} in the appropriate expansion parameter, e.g. the ratio of the scalar quartic coupling to the gauge coupling (see \cref{sec:SM3dEFT}). Such an expansion has the added benefit that the effective potential and derived quantities are real. 
It can then be demonstrated that the results are gauge-independent order by order in perturbation theory.

The observations of the limited accuracy of the daisy-resummed effective potential and the prefactor $A$ in the decay rate were made in the context of the phase transition parameters itself, or the corresponding GW signal \cite{Croon:2020cgk, Niemi:2021qvp, Gould:2023jbz, Kierkla:2023von}.
It is very probable that these uncertainties also affect the prediction of the baryon asymmetry, as the outcome depends on the bubble profile and the nucleation temperature. We are not aware of any works in which this uncertainty is investigated.

\subsection{Electroweak phase transition in the Standard Model}\label{sec:PTSM}
Let us briefly review the EWPT in the Standard Model and demonstrate the absence of a FOPT. For illustration, we will use the effective potential of eq.~\eqref{eq:VT}, with the high-temperature expansion of eq.~\eqref{eq:JB/FHighT}. 
As argued in the previous subsection,
an accurate description of the EWPT requires including higher-order corrections or even a lattice computation. That being said, in our simplified description we can already see that the phase transition will not be strongly first order for a Higgs mass of $125\, {\rm GeV}$, and therefore not suitable for EW baryogenesis.

We write the tree-level, zero-temperature Higgs potential for the background field as
\begin{equation}
    V_{\rm tree}(\phi) = -\frac{\mu_h^2}{2} \phi^2 + \frac{\lambda}{4} \phi^4,
\end{equation}
with positive $\mu_h^2$ and $\lambda$.
The dominant corrections to the one-loop potential come from the EW gauge bosons and the top quark. The corresponding masses, given in \cref{W_mass,mass_yukawa}, enter the Coleman-Weinberg potential and thermal potential \cref{V_CW,V_T}.
To study the effects of the thermal corrections, we apply the high-temperature expansion of the thermal functions, eq.~\eqref{eq:JB/FHighT}. The 
potential becomes
\begin{align}
	V_{\rm SM}(\phi , T)  \sim &-\frac{\mu_h^2}{2}\phi^2 + \frac{T^2}{4}\left(m_W^2 (\phi) + \frac{1}{2}m_Z^2(\phi) - m_t^2(\phi) \right)  
    -\frac{T}{4 \pi}\left( 2(m_W(\phi) ^2)^{3/2}
    +(m_Z(\phi)^2 )^{3/2}\right)\nonumber \\
    &+\frac{\lambda}{4}\phi^4 + \frac{1}{64\pi^2}\left(\sum_{i=W,Z} n_i m_i^4 (\phi)\left( \log{\left[\frac{(4\pi e^{\gamma_E}T)^2}{\mu^2}\right]}-\frac{3}{2}-C_i \right) + 12 m_t^4(\phi) \log{\left[\frac{(\pi e^{\gamma_E}T)^2}{\mu^2}\right]} \right)\cdots.\label{eq:VSMHT}
\end{align}
The ellipsis denote higher-order terms in $m_i^2/T^2$ and the field-independent $T^4$-term, which does not contribute to the PT dynamics. The thermal corrections have two main effects: they shift the mass of the Higgs field via the terms proportional to $T^2$, such that it becomes positive at large temperature, and they provide a cubic term $\sim \phi^3$. In the last line, we have combined the CW contribution with the logarithmic term of the one-loop thermal function. Together, these terms (mildly) affect the quartic coupling.

Let us neglect the loop correction to the quartic coupling and introduce the following notation (see e.g. \cite{Cline:2006ts,Anderson:1991zb, Dine:1992wr})
\begin{equation}
	V_{\rm SM}(\phi, T) \sim \frac{1}{2} m_{T}^2\phi^2 - ET \phi^3 + \frac{\lambda}{4} \phi^4, \qquad
    {\rm with} \;\; E =  \frac{1}{4 \pi v^3} (2 m_W^3 + m_Z^3), \qquad m_T^2 = -\mu_h^2 + \frac{T^2}{16}(3g^2 + {g'}^2 + 2 y_t^2).
\end{equation}
We can determine the critical temperature assuming the phase transition to be first order, by solving simultaneously $V(\phi, T_c) =V(0, T_c) = 0$ and $dV(\phi, T_c)/d\phi=0$ at ${\phi = v_c} $.
This gives 
\begin{equation}
	\frac{v_c}{T_c} = \frac{2 E}{\lambda} = \frac{4E v^2}{m_h^2}\,,
    \label{FOPT_estimate}
\end{equation}
with $v$ and $m_h$ the zero-temperature vev and mass respectively. Assuming that the nucleation temperature is close to the critical temperature, a strong FOPT $v_n/T_n \approx v_c/T_c \gtrsim 1$ requires
\begin{equation}
	m_h\lesssim \sqrt{4E v^2} \sim 48 \, {\rm GeV},
\end{equation}
where we used the values of the couplings at the $Z$-mass scale.
This is much below the experimentally observed value of the Higgs mass. This conclusion is confirmed by more accurate lattice calculations, which find that for a Higgs mass larger than $m_h \gtrsim 67\,$GeV the transition is a crossover \cite{Kajantie:1996mn,Csikor:1998eu, Kajantie:1995dw}.

\subsubsection{A standard-model like 3-dimensional theory}\label{sec:SM3dEFT}
As discussed in \cref{sec:VTaccuracy}, a more accurate description of thermodynamics can be obtained in a dimensionally reduced effective theory. For the Standard Model phase transition, the appropriate three-dimensional effective action is
\begin{equation}
    \hat S_3 = \int d^3x \left( \frac{1}{4} F_{ij}^a F_{ij}^a+(D_i \phi_3)^\dagger (D_i \phi_3) + m_3^2 \phi_3^\dagger \phi_3 + \lambda_3 (\phi_3^\dagger \phi_3)^2\right),\label{eq:SM3dEFT}
\end{equation}
with $F_{ij} = \partial_i A_j -\partial_j A_i - g_3[A_i,A_j]$ and $D_i = \partial_i + ig_3 A_i$. Here, $A_i$ denotes the spatial part of the SU(2) gauge field. The temporal gauge field  has been integrated out, as well as all the fermions and the non-zero Matsubara modes. The field $\phi_3$ denotes the Higgs field. 
The Higgs field has been rescaled to $\phi_3 = \phi/\sqrt{T}$, such that $\phi_3$ has units of $\sqrt T$, and $\hat S_3$ is dimensionless. We have introduced the hat to distinguish it from the $S_3$ appearing in \cref{eq:rate} (which is the effective theory from integrating out the non-zero Matsubara modes, and which was defined as an object of mass dimension one).
The coefficients $g_3$, $m_3$ and $\lambda_3$ can be obtained from the matching to the UV-theory. In principle, the theory also contains the dimensionally reduced SU(3) and U(1) gauge groups. We can neglect the SU(3) group because it only couples to the Higgs via loop corrections; also the inclusion of the U(1) only has a negligible effect \cite{Kajantie:1996mn}.
Expressions for $g_3$, $\lambda_3$, and $m_3$ in the Standard Model can be found in \cite{Kajantie:1995dw}\footnote{Ref.~\cite{Laine:2017hdk} pointed out a mistake in the expression for the top quark self energy.}.

A very attractive feature of the theory described by \cref{eq:SM3dEFT} is that it does not only describe the Standard Model, but also extensions with sufficiently heavy scalar fields such that they can be integrated out. In contrast to the pure Standard Model, these models may actually feature a FOPT. Examples of models that map into the EFT of \cref{eq:SM3dEFT} are (parts of the parameter space of) the MSSM \cite{Losada:1996ju}, the two-Higgs doublet model \cite{Losada:1996ju, Andersen:1998br, Andersen:2017ika, Gorda:2018hvi}, and real singlet \cite{Brauner:2016fla} and real triplet \cite{Niemi:2018asa} extensions.
We can define two dimensionless parameters
\begin{equation}
    x = \frac{\lambda_3}{g_3^2}, \qquad y = \frac{m_3^2}{g_3^4},\label{eq:xy}
\end{equation}
that fully determine the theory. $x$ is typically a weak function of the temperature, while $y$ varies more strongly. See for example \cite{DOnofrio:2014rug} for a graph showing the temperature-dependence of $x$ and $y$ in the SM. By studying the thermodynamics as a function of $x$ and $y$, it is possible to study a larger set of models and parameter points. The theory described by \cref{eq:SM3dEFT} has been studied in perturbation theory in \cite{Kajantie:1995dw,Ekstedt:2024etx, Ekstedt:2022zro} and on the lattice in \cite{Kajantie:1995kf,Gurtler:1997hr,Rummukainen:1998as,Laine:1998jb, Moore:2000jw, Gould:2022ran}.
Lattice results show that a FOPT can exist for $x<0.0983$. In the SM, $x = 0.29$ \cite{Kajantie:1996mn}, confirming the conclusion of \cref{sec:PTSM} that there is no FOPT in the SM.

In practice, the lattice results of \cite{Moore:2000jw, Gould:2022ran} can be used to determine the percolation temperature as follows: compute the coefficients of the effective theory \cref{eq:SM3dEFT}, for example using {\tt DRalgo} \cite{Ekstedt:2022bff}, and determine $x(T)$ and $y(T)$.
The temperature dependence follows from the matching relations, and determines a trajectory in the $x,y$-plane. 
The results of \cite{Moore:2000jw, Gould:2022ran} yield a line in the $x,y$ plane, denoted by $y_p(x_p)$, corresponding to percolation. 
The temperature at which the model-specific $y(x)$ trajectory intersects the percolation line $y_p(x_p)$ corresponds to the percolation temperature.

\subsection{Extensions of the SM with a FOPT}\label{sec:BSMFOPT}
Let us now give some examples of models that can feature a sufficiently strong electroweak phase transition. EWBG requires, in addition, new sources of CP violation to fulfill the second Sakharov condition, and we will return to the full model-building challenge in \cref{sec:models}.
A much more extensive overview of FOPT models, not per se related to EWBG, can be found for example in Ref.~\cite{Caprini:2019egz}.

\subsubsection{Gauge singlet}\label{sec:xSM}
One of the simplest extensions of the SM with a FOPT is the extension by a (real) SM gauge singlet $S$. The Lagrangian is 
\begin{equation}
	\L=\mathcal L_{SM} + \frac 1 2 \partial_\mu S \partial^\mu S  -\( \frac 1 2 a_1 |\vp|^2 S +\frac 1 2 a_2 |\vp|^2 S^2 + \frac{1}{2} b_2 S^2 + \frac{1}{3} b_3 S^3 + \frac 1 4 b_4 S^4\),\label{eq:SMsinglet}
\end{equation}
where we have defined the field $S$ such that the zero-temperature tadpole term vanishes. A special case that is often studied is the $Z_2$-symmetric set-up, with $a_1= b_3=0$, as the terms odd under $S \to -S$ are forbidden by the symmetry. In this model, a FOPT can arise in two different ways:
\begin{itemize}
	\item Radiative corrections from loops with $S$-particles contribute to the cubic term in the Higgs potential, similarly to the $W$ and $Z$ in \eqref{eq:VSMHT}. This will increase  the value of $E$ in \cref{FOPT_estimate}, making the phase transition first order \cite{Espinosa:1993bs, Benson:1993qx}. In this scenario, the singlet does not obtain a vev.
	\item There is a two-step phase transition. In the first step, the singlet obtains a non-zero vev, in a transition that is typically weakly first order or cross-over. There can now be a (tree-level) barrier between the minimum where only the singlet has a vev and the global minimum where only the Higgs has a vev \cite{Choi:1993cv,Espinosa:2011ax,Cline:2012hg,Kurup:2017dzf,Huang:2018aja,Lewicki:2024xan, Niemi:2024vzw}, allowing for the 2nd step to be a FOPT. 
    As both fields change field value during this transition, the strength is no longer given by the potential along the Higgs direction, avoiding the bound in \cref{FOPT_estimate}. 
\end{itemize}
Variations of this model are the Standard Model extended by a complex singlet \cite{Espinosa:1993bs, Barger:2008jx}, or multiple singlets \cite{Espinosa:2008kw}.

\subsubsection{Scalar electroweak multiplets}
\label{sec:2HDM}

A FOPT can also arise in extensions of the SM with new electroweak multiplets, such as one \cite{Cline:2011mm, Huet:1995mm, Fromme:2006cm, Dorsch:2016nrg} or several additional Higgs doublets \cite{Ahriche:2015mea}, isospin triplets \cite{Niemi:2020hto, FileviezPerez:2008bj, Patel:2012pi, Inoue:2015pza, Chala:2018opy}, or even higher dimensional electroweak multiplets \cite{Chala:2018ari}. We will focus on the two Higgs doublet model (2HDM) here \cite{Branco:2011iw}, as it is the scenario that has been studied most extensively in the context of EWBG. Denoting the two doublets by $\vp_1, \vp_2$, the potential is given by
\begin{align}
	V(\vp_1, \vp_2) = &\mu_1^2 |\vp_1|^2 + \mu_2^2 |\vp_2|^2 - \mu^2[\vp_1^\dagger \vp_2 + {\rm h.c.}] + |\vp_1|^4 + \lambda_2 |\vp_2|^4 \nonumber \\ &+\lambda_2 |\vp_1|^2 |\vp_2|^2 + \lambda_4 |\vp_1^\dagger \vp_2|^2 + \frac{\lambda_5}{2}\left[(\vp_1^\dagger \vp_2)^2 + {\rm h.c.} \right].
\end{align}
We have included a soft $Z_2$-breaking term proportional to $\mu^2$, but have forbidden hard $Z_2$-breaking terms, see \cite{Ginzburg:2008efb,Ginzburg:2009dp} for further discussion.
At zero temperature, the vacuum state is
\begin{equation}
	\vp_1|_{T = 0} = \frac{1}{\sqrt 2}\begin{pmatrix} 0 \\ v_1\end{pmatrix}, \qquad \vp_2|_{T = 0} = \frac{1}{\sqrt 2}\begin{pmatrix} 0 \\ v_2\end{pmatrix}\,,
\end{equation}
where the two vevs satisfy $(v_1^2 + v_2^2)^{1/2} = 246 \, {\rm GeV}$ and $\beta = \tan v_2/v_1$. Out of the eight fields contained in the two complex doublets, three get `eaten' by the $W$ and $Z$ bosons, and five scalar degrees of freedom remain. After rotating to a basis where the mass matrix of the neutral scalars is diagonal (the corresponding angle is referred to as $\alpha$), one identifies two neutral scalars, typically denoted as $h$ and $H$, of which one corresponds to the $125 \, {\rm GeV}$ scalar particle measured at the LHC. The other physical fields are a charged scalar field $H^{\pm}$ and a pseudoscalar $A$. The angles $\alpha$ and $\beta$ enter in the effective couplings between the mass eigenstates $h$, $H$, and the SM fields. 

In the so-called Type I 2HDM, only the $\vp_2$-doublet couples to fermions, which can be enforced by a discrete $Z_2$ symmetry under which only the $\vp_1$ doublet is odd; in Type II models $\vp_1$ couples to down-type fermions and $\vp_2$ to up-type (see \cite{Branco:2011iw} for other possibilities). The top quark is the only SM fermion which has a significant effect on the PT, and since it couples to $\vp_2$ in both scenarios, the phase transition dynamics does not depend on the type of 2HDM. In the so-called alignment limit, $\cos(\beta - \alpha) \rightarrow 0$, the couplings of $h$ to the SM particles approach those in the SM, and LHC constraints indicate that deviations from this limit must be small \cite{ATLAS:2014kua}. 
A special case of the 2HDM is the so-called Inert Doublet Model \cite{Deshpande:1977rw, Ma:2006km, Barbieri:2006dq}, where $\mu^2 = 0$ and one of the doublets does not couple at all to fermions and does not obtain a vev.

Despite several stringent experimental constraints (see e.g. \cite{Haller:2018nnx}), the 2HDM can still  accommodate a FOPT, as explored in e.g. \cite{Cline:2011mm, Huet:1995mm, Fromme:2006cm,Dorsch:2016nrg, Kakizaki:2015wua,  Basler:2016obg, Dorsch:2017nza, Basler:2017uxn}.  
Just as in the SM + singlet case, the phase transition can proceed in one or two steps \cite{Blinov:2015sna, Bernon:2017jgv, Aoki:2021oez}. Large couplings are needed for a FOPT in models that satisfy all experimental constraints.
The consequences for the applicability of perturbation theory was discussed in \cite{Laine:2017hdk}, and further explored in lattice studies in \cite{Kainulainen:2019kyp, Andersen:2017ika}.


\subsubsection{Conformal models}\label{sec:conformal}
Conformal extensions of the SM giving rise to a FOPT have received considerable interest, 
as they can provide a solution to the hierarchy problem~\cite{Foot:2007iy, Iso:2012jn}, harbor a dark matter candidate \cite{Hambye:2013dgv, Khoze:2013uia, Karam:2016rsz},
and may feature a first-order phase transition with promising prospects for generating a strong GW signal \cite{Hambye:2013dgv,Konstandin:2011dr, Jaeckel:2016jlh, Hashino:2016rvx, Prokopec:2018tnq, Kierkla:2022odc}.
The classically conformal version of the SM corresponds to the SM Lagrangian without an explicit mass term for the Higgs. 
Electroweak symmetry breaking and generation of the masses for the SM particles occurs via the Coleman-Weinberg mechanism (also referred to as radiative symmetry breaking or dimensional transmutation). However, without any additional particle content, the required Higgs and top quark masses are excluded by their observed values \cite{Coleman:1973jx}. 
Consequently, many different extensions of the conformal SM have been considered, 
e.g. with extended scalar sectors \cite{Foot:2007iy,Meissner:2006zh,Alexander-Nunneley:2010tyr, Brdar:2019qut} or new dark gauge sectors, e.g. \cite{Hempfling:1996ht, Chang:2007ki,Khoze:2013oga, Khoze:2014xha}.
The models have in common that the tree-level Lagrangian contains only dimension-4 operators and 
the particle masses are generated via the Coleman-Weinberg mechanism.

The conformal  symmetry-breaking phase transition can be  first order, and was studied in e.g.~\cite{Kierkla:2023von,Kierkla:2025qyz, Chataignier:2018kay}.
Due to the absence of a (negative) bare mass term, the effective potential features a barrier originating from thermal masses up to low temperatures, {\it cf}. \eqref{eq:JB/FHighT}.
Consequently, these phase transitions are strongly supercooled, i.e. the nucleation temperature is orders of magnitude below the critical temperature, resulting in a very strong gravitational wave signal.
In the context of EWBG, the strong supercooling poses a challenge as the bubbles typically accelerate to wall velocities close to the speed of light, 
which makes it difficult to generate the baryon asymmetry (see the discussion in \cref{sec:supersonic}). Moreover, the field undergoing the FOPT is typically \emph{not} the Higgs field, but rather an additional BSM scalar field, which means the EW symmetry is broken in a subsequent phase transition. Hence, extra ingredients are required to implement EWBG in these set-ups.  To our knowledge, no scenarios exist that combine classical conformal symmetry and EWBG.
  
\subsubsection{Composite Higgs models}\label{sec:composite}
The previous subsection considered the scenario where a new dark sector (and extra symmetries) are added to the SM. A somewhat related class of models are the composite Higgs models (see \cite{Panico:2015jxa, Witzel:2019jbe} for reviews), where the Higgs arises as a bound state of a (nearly) conformal strongly coupled sector which is confined at low temperatures and deconfined at high temperatures. 
Composite Higgs models were developed to address the hierarchy problem of the Higgs mass. 
The Higgs field remains light compared to the scale of new physics because it is a pseudo-Nambu-Goldstone (PNGB) of a broken global flavor symmetry.
The minimal coset space in which the PNGBs can live that is consistent with EW precision tests is SO(5)/SO(4), but larger cosets such as SO(6)/SO(5) and SO(7)/SO(6) are possible as well. In the non-minimal case, the Higgs is accompanied by additional light PNGBs.

The EWPT can become first order in the following ways. 
The confining phase transition is naturally first order, and in the regime of a light dilaton (the pseudo-Goldstone boson of an approximate scale invariance of the composite
sector), the EWPT occurs simultaneously with the confining PT \cite{Bruggisser:2018mus, Bruggisser:2018mrt, Bruggisser:2022rdm}.
In this scenario the PT can be very strong, {\it cf}. \cref{sec:conformal}, and therefore potentially observable in gravitational wave detectors \cite{Caprini:2019egz, Konstandin:2011dr}, as further discussed in \cref{sec:GWs}.
Alternatively, the EWPT can occur after the confining phase transition. Other light states that arise in the non-minimal case could make the EWPT first order. From the point of view of the scalar sector, this scenario could look very similar to e.g. the SM coupled to a scalar (with $Z_2$ symmetry), as discussed in \cref{sec:xSM} (this scenario could also generate CPV operators involving the new scalar field \cite{Espinosa:2011eu, Chala:2016ykx}). 
A third option that has been considered in the literature is the case where the effects of the UV-physics are fully captured in effective operators with only the SM content~\cite{Grinstein:2008qi}. We will discuss the consistency of such a picture in the following section.

Due to the strong dynamics, the confinement phase transition can not be studied with the methods discussed in \cref{sec:Potential}. 
Instead, results have been obtained by using holography and studying a 5-dimensional weakly coupled gravity theory (the Randall-Sundrum model \cite{Randall:1999ee}) \cite{Creminelli:2001th, Randall:2006py, Nardini:2007me}, or alternatively, in the large-$N$ limit \cite{Bruggisser:2018mrt}.

\subsubsection{FOPT in the Standard Model Effective Field Theory}
\label{sec:SMEFT_FOPT}

It would be great if the various models for a FOPT could all be described within a single framework. The SM effective field theory (SMEFT) \cite{Buchmuller:1985jz,Grzadkowski:2010es} could provide such a model-independent description, at least for single step transitions. SMEFT assumes that the new physics degrees of freedom are heavy and can be integrated out, and  their effects at low energies is captured by a tower of gauge-invariant higher-dimensional operators containing just SM fields
\be
{\cal L}_{\rm SMEFT} = {\cal L}_{\rm SM}+ \sum_i \frac{c_i \O_i}{\Lambda^{d_i-4}}\,.
\ee
Here $\Lambda$ is the break-down scale of the EFT, $c_i$ dimensionless Wilson coefficients and $d_i$ the mass dimension of the operator $\O_i$. Under the assumption that $v\ll \Lambda$, the most significant deviations from SM predictions arise from the lowest-dimensional EFT operators.

For the EWPT the operators containing the Higgs field are relevant.  In particular,  the dim-6 $\O_\vp = (\vp^\dagger \vp)^3/\Lambda^2$ operator adds to the effective potential, where we have absorbed the Wilson coefficient in the cutoff scale. In addition, there are dim-6 operators containing derivatives of the Higgs field; these are generically subdominant during the EWPT \cite{Postma:2020toi}, and for simplicity we neglect them in this discussion. The potential for the classical background Higgs becomes
\be
V_{\rm SMEFT} =\frac12 a_2 \phi^2 +\frac14 a_4 \phi^4 + \frac16 a_6 \phi^6
= -\frac14(m_{h}^2 -2 a_6 v^4) \phi^2 +\frac14\(\frac{m_{h}^2}{2v^2} -2 a_6 v^2\)\phi^4 +\frac18 a_6 \phi^6\,,
\label{V_phys}
\ee
where in the second equation we have substituted $a_2,a_4$ with the physical Higgs mass $m_h^2 =125\,$GeV and Higgs vev $v=246\,$GeV. The potential is only a function of the cutoff scale through $a_6 =3/(4\Lambda^{2})$.  
A FOPT requires a barrier in the finite temperature potential.
Neglecting the cubic thermal corrections this appears as the quartic term becomes negative, which happens for   $\Lambda \sim \sqrt{3}v^2/m_h \approx 800\,$GeV, while the quadratic and dim-6 terms are positive. Slightly larger cutoffs are possible when the cubic thermal corrections are included, while a lower bound $\Lambda \sim \sqrt{3}v^2/m_h \gtrsim 600\,$GeV  follows from the requirement that the quadratic $a_2$ needs to be positive at zero temperature. These estimates agree well with numerical results such as in \cite{Grojean:2004xa}, who find a FOPT is possible for cutoffs in the $600$-$1000\,$GeV range.

The SMEFT description of the EWPT is very tractable, with 
only one free parameter, and it has been widely used, see e.g. Refs.~\cite{Bodeker:2004ws,Balazs:2016yvi,Croon:2020cgk, Chala:2018ari,Grojean:2004xa,Delaunay:2007wb,Ellis:2019flb}.
The cutoff scale is higher than the EW scale, although not by much. 
As was pointed out in Refs.~\cite{deVries:2017ncy,Postma:2020toi, Damgaard:2015con,deVries:2018tgs}, despite appearances, there is no hierarchy of scales. One cannot simply compare the cutoff and the EW scale, as there are small coefficients involved, to wit the effective  Higgs quartic coupling.  This is also immediately clear from the above discussion. The broken phase minimum arises from balancing the negative quartic and positive dim-6 term in the potential; almost by construction, then, there is no hierarchy between the SM potential and the dim-6 operator.  Another way to reach the same conclusion: within the SM the EWPT is first order for a Higgs mass $m_h \lesssim 67\,$GeV, much smaller than the measured mass $m_h \approx 125\,$GeV. Hence, a FOPT would require $\mathcal O(1)$ corrections to the SM Higgs potential, which is not amenable to treatment in SMEFT.
This would imply the SMEFT description of a FOPT, and the resulting values of the nucleation temperature and the bubble wall width, is not reliable \cite{Postma:2020toi}.

We end this section with some comments on recent work.
Refs.~\cite{Camargo-Molina:2021zgz,Camargo-Molina:2024sde} discuss a potential way of embedding the FOPT into SMEFT: the dim-6 operator $\mathcal O_\phi$ is used to obtain the observed Higgs mass for a Higgs self-coupling $\lambda$ that is much smaller than in the SM. In their power counting scheme $O_\phi$ is nevertheless subleading around the critical temperature, and can be neglected for the PT dynamics. Effectively this then leads to a smaller value of $x$ as defined in \cref{eq:xy} and thus the possibility of a strong FOPT. However, it is not clear whether the power counting works and $O_\phi$ can indeed be neglected during the phase transition, nor whether this scenario can be embedded in a UV-complete scenario \cite{Postma:2020toi} and fit in with the analysis in this subsection. For example, the claim that larger cutoffs are possible in the limit that the quartic term approaches zero \cite{Camargo-Molina:2021zgz,Camargo-Molina:2024sde}  seems to contradict with setting $a_4=0$ in \cref{V_phys}, which fixes the cutoff scale to $\Lambda \sim 800\,$GeV for $a_6>0$. 

Even more recent, Ref.~\cite{Chala:2025xlk} analyses the (radiative) effect of dim-6 SMEFT operators in the dimensionally reduced EFT, fitting it to the $(x,y)$ parameters defined in  \cref{eq:xy}.  They find that turning on the dim-6 correction to the top Yukawa coupling, together with either the dim-6 correction to the Higgs potential $\mathcal O_\phi$ or to the Higgs kinetic term, can induce a FOPT for cutoff scales around the TeV scale. As expected on general grounds, and borne out by the explicit expressions and results, an $\mathcal O(1)$ correction to the SM Lagrangian is needed to obtain a FOPT. To check the validity of the EFT approach, it should be checked whether the impact of dim-8 operators on the phase transition dynamics \cite{Postma:2020toi, Damgaard:2015con}, and on the BAU calculation \cite{deVries:2017ncy,deVries:2018tgs}, is small.

\subsection{Summary of \cref{sec:FOPT}}

Successful EWBG requires an extension of the SM with a sufficiently strong first-order phase transition. Some examples of models are listed in \cref{sec:BSMFOPT}. 
For the computation of the BAU, we need the nucleation temperature, the bubble profile, and the bubble wall velocity. The latter is discussed in \cref{sec:Bubblevw} and the first two are determined as follows:
\begin{itemize}
    \item Construct the finite-temperature effective potential. To obtain accurate results, it is typically necessary to go \emph{beyond} the one-loop approximation, which is most easily done in the 3D EFT approach. The package {\tt DRalgo} \cite{Ekstedt:2022bff} can provide the effective potential  to two-loop order.
      \item Determine the nucleation temperature by finding the temperature at which the number of bubbles per Hubble volume is one. That is, solve \cref{eq:NuclCrit}. A simpler, approximate nucleation criterion is given in \cref{eq:NuclCritApprox}. 
     \item 
    The nucleation rate is given by \cref{eq:rate}; the different inputs are determined as follows:
    \begin{itemize}
    \item Solve the bounce equation of motion \cref{eq:Bounce} in the effective potential and determine the corresponding action $S_3$. This can be done with several tools, for instance {\tt CosmoTransitions} \cite{Wainwright:2011kj}, {\tt AnyBubble} \cite{Masoumi:2017trx}, {\tt SimpleBounce} \cite{Sato:2019wpo}, {\tt FindBounce} \cite{Guada:2020xnz}, {\tt BSMPT3} \cite{Basler:2024aaf}, and {\tt PhaseTracer2} \cite{Athron:2024xrh}.
    \item The pre-factor $A$ can be determined with {\tt BubbleDet} \cite{Ekstedt:2023sqc}; or alternatively approximated on dimensional grounds by $T^4$. 
    \end{itemize}
    \item An estimate of the theoretical uncertainty in $T_n$ can be obtained by varying the RG-scale between e.g. $\pi T/2$ and $2\pi T$. A large dependence on the RG-scale indicates that higher order contributions would have a significant effect. 
    \item The solution to the bounce equation of motion, or an analytic fit to it, \cref{bubble_profile}
    are often used as the bubble profile in the baryogenesis computations. 
    The most realistic profile is the one that corresponds to the solution for the wall velocity. This profile reaches $v_n$ in the broken phase, and includes hydrodynamics and friction effects (see  \cref{sec:Bubblevw}).
    \item It is important that the FOPT is sufficiently strong to avoid baryon number washout. An approximate criterion is $v_n/T_n \gtrsim 1$; this will be further discussed in \ref{sec:broken_sphaleron}.
    \item 
    In models that map onto the EFT of \cref{eq:SM3dEFT} one can determine the parameters $x,y$ by a matching procedure and find the percolation temperature from the intersection of $y(x)$ with the $y_p(x_p)$ curve of  lattice results \cite{Moore:2000jw, Gould:2022ran}. 
\end{itemize}
The nucleation temperature and bubble profile will play an important role in the generation of the CP asymmetry in front of the bubble wall that ultimately sources the baryon asymmetry of the universe. The next sections focus on the generation of this CP asymmetry.
	
	\newpage
	\section{Generating the CP-asymmetry}\label{sec:CP}

This section reviews the derivation of the quantum Boltzmann equations that track the densities of the plasma quanta during the phase transition, with a focus on the derivation of the source term that drives the system out of equilibrium; in  \cref{sec:bubbles} we address how to actually solve the set of equations. 

Early derivations of the transport equations were mostly based on phenomenological approaches (e.g. in \cite{Shaposhnikov:1987tw,Cohen:1993nk,Cohen:1991iu,Joyce:1994fu,Cline:2000nw,Cline:1997vk,Fromme:2006wx}), but in later work, this has been supplanted by a first-principles approach based on quantum field theory. The closed-time path (CTP), or in-in formalism, provides a path integral description of an out-of-equilibrium system at finite temperature. The starting point is the Schwinger-Dyson equation for the Green's functions, which can be derived from the 2-particle-irreducible effective action. It can be split into a hermitian and anti-hermitian part, usually referred to as the kinetic and constraint equations. 
These equations are brought to a solvable form by a gradient expansion in the inverse wall width. 
The set of (truncated) constraint equations provides the spectral information, whereas the kinetic equations determine the dynamical evolution of the system. Finally, integrating the kinetic equations over energy gives a set of Boltzmann equations for the distribution functions of the various plasma species. A quick review of the CTP formalism is given in \cref{A:CTP}, which also serves to introduce our notation and conventions; more details and derivations can be found in e.g. \cite{Garbrecht:2018mrp, Riotto:1998zb, Prokopec:2003pj,Kainulainen:2001cn,Berges:2004yj, Cirigliano:2011di,Cirigliano:2009yt,Kainulainen:2021oqs}.   
  
For completeness, we should also mention the recent approach to fermion-bubble interactions based on a real-time lattice simulation in 1+1 dimensions \cite{Carena:2024peb}; it would be very interesting if these simulations can be extended to 3+1 dimensions, as they can shed light on the validity and limitations of the analytic approaches.

As the EW sphaleron transitions are slow compared to the other relevant plasma interactions, to a good approximation a two-step approach is valid. In the first step the CP asymmetry is generated, which subsequently is transformed into a baryon asymmetry -- this latter step is discussed in \cref{sec:sphalerons}.  The generation of the CP asymmetry is difficult to model in a controlled approximation. Much work has been devoted to the case of thick bubble walls. The background can then be treated as slowly evolving compared to the typical momentum $k$ of the plasma excitations.
This is implemented by writing the Green's function in terms of the relative coordinate $r$ and collective coordinate $x$, performing a Fourier transform with respect to the former $G(r,x) \to G(k,x)$ (known as the Wigner transform) as in \cref{Wigner}, followed by a gradient expansion in derivatives acting on the background.  The validity of the gradient expansion requires
\be
\partial_x \sim L_w^{-1}\ll k \sim T\,,
\label{gradient} 
\ee
which corresponds to thick bubble walls $L_w T\gg 1$, with $L_w$ the bubble wall thickness, \emph{cf} \cref{bubble_profile}, and $T$ the plasma temperature.
For the constraint and kinetic equations in Wigner space, collectively known as the Kadanoff-Baym (KB) equations, the gradient expansion corresponds to an expansion of the diamond operator \cref{A:diamond}.
Even though this is a controlled expansion, the theoretical uncertainties in the baryon yield are large \cite{Cline:2017jvp, Cline:2020jre, Cline:2021dkf}, and in fact, not always well known. One of the main sources of uncertainty is in the derivation of the source terms in the transport equations, which turn on as the bubble passes by and drive the system out of equilibrium.

Several CPV source terms have been identified, each with a corresponding computational method. The Higgs vacuum expectation value (vev) varies from zero in the false vacuum outside the bubble to some finite value inside, leading to a spacetime dependent mass for all fields with a coupling to the Higgs.\footnote{A different source term, which directly enters the equation for the baryon asymmetry, is generated if the CP asymmetry is in effective $(\vp^\dagger \vp)\tilde F F$-interactions \cite{Dine:1990fj}; we will not discuss this possibility in this section.} The CP-violating {\it semi-classical source} or {\it WKB source} describes the difference in force for particle and antiparticle modes, which arises from the change in mass when crossing the bubble wall \cite{Cline:2000nw,Prokopec:2003pj, Kainulainen:2001cn,Joyce:1994zt,Kainulainen:2002th, Prokopec:2004ic}.  In a multi-flavor set-up, such as in the SM with quarks and leptons, the mass and flavor eigenstates may be  misaligned,  with the mixing angles varying across the bubble wall as well. The {\it flavor source} relies on the CP violation in the mixing matrix \cite{Cirigliano:2011di, Konstandin:2004gy, Konstandin:2005cd}.  The {\it vev-insertion approximation (VIA)} treats the spacetime-dependent part of the mass (proportional to the Higgs vev) as a small perturbation \cite{Riotto:1995hh,Riotto:1997vy}. 
In addition to the computational approach, this source distinguishes itself from the others by including thermal corrections, thermal masses, and thermal widths.
The physical interpretation would be that  thermal corrections are non-diagonal in the mass basis leading again to non-trivial flavor dynamics.
For example, the weak interactions are chiral, and the thermal corrections for left- and right-handed fermions differ. The chiralities can effectively be considered as different `flavors'. However, as discussed in detail below, the VIA source vanishes at leading order in the derivative expansion and higher-order corrections have not been systematically computed.

In this section we will review the semi-classical, flavor and (vanishing of) the VIA source term in \cref{sec:semiclassical,sec:via,sec:flavor}.  Model implementations and alternative ways for generating the asymmetry are discussed in \cref{sec:models}.  We start with a short summary of the Kadanoff-Baym equations for the systems of interest.


\subsection{Kadanoff-Baym equations}
A brief overview of the CTP formalism, and the derivation of the Kadanoff-Baym equations is given in \cref{A:CTP}. Here, we list the results needed for the derivation of the different source terms in the next sections. We stick to the notation in the literature, and for bosons we use $\theta$ for the mixing angle between flavors, and $\sigma$ denotes the CP violating phase. For fermions the CP violating phase instead is denoted by $\theta$. Although this notation can lead to confusion -- we do not hope so -- it allows for easier comparison with results in the literature.

\paragraph{Bosons.}
For illustration, we consider a toy set-up with two real scalar fields $\chi_f=( \chi_L \; \chi_R)^T$ with Lagrangian:
\be
\L = \frac12(\partial_\mu \chi_f)^\dagger (\partial^\mu \chi_f)
-\frac12\chi_f^\dagger M_0^2 \chi_f + \L_{\rm int} (\chi_f) 
\label{M_boson_VIA}
\qquad
M_0^2(\phi_b) = 
\(
\begin{array}{cc}
  m_{LL}^2 & m_{LR}^2(\phi_b) \\
 m_{RL}^2(\phi_b)  & m_{RR}^2
\end{array}
\),
\ee
where we have assumed that the off-diagonal mass terms appearing in $M_0^2$ originate from Higgs interactions, and are thus space-time dependent in the bubble wall background, $\phi_b=\phi_b(x^\mu)$.
The interactions with the thermal bath are diagonal for the interaction or flavor eigenstates, which we have denoted by $L,R$ (the labels are chosen to mimic chiral fermions).  The diagonal mass terms then get thermal mass corrections,  which are given by the hermitian self-energy $\Pi^h_f$. We absorb the thermal corrections in the mass matrix via $M_f^2 \equiv M_0^2(\phi_b) + \Pi^h_f(T)$; the KB equations only depend on $M_f^2$, see \cref{A:constraint_B,A:kinetic_B}, and to simplify notation we will work with $M_f^2$ from now on. In the flavor basis the interactions with the plasma in $\L_{\rm int} $ (including baryon-number-violating operators) are diagonal, whereas in the mass basis the mass matrix  is diagonal.

The two-point functions \cref{Greens} contain all the needed information about the system. Of particular interest are the Wightman functions in \cref{Wightman_def}. After integrating them over energy, we project out the distribution functions for particles and antiparticles, see \cref{project_boson}. The 
constraint and kinetic equation for the Wightman functions $\Delta^\lambda$ in the flavor basis, indicated by the subscript $f$, are
\begin{align}
  (k^2-\frac14 \partial_x^2) \Delta_f^\lambda &=\frac12\e^{-i\diamond}\bigg( \{M^2_{f}, \Delta_f^\lambda\}
+\frac{1}{2}  \{ \Pi^\lambda_f, \Delta_f^h\} +\C_{f,-}\bigg),
 \label{constraint_B}\\
2ik\cdot\partial_x \cdot \Delta_f^\lambda &= \e^{-i\diamond}\bigg( [M^2_{f}, \Delta_f^\lambda] + [ \Pi_f^\lambda, \Delta_f^h] +\C_{f,+}\bigg),
 \label{kinetic_B}
\end{align}
with $\lambda =\{>,\,<\}$, and the collision term $\C_{f,\pm}$ is defined in \cref{calC}. In a constant background, the constraint equation reduces to the Klein-Gordon equation, while the kinetic equation vanishes -- this explains the nomenclature.
The flavor basis KB equations \cref{constraint_B,kinetic_B} are used in the discussion of the VIA source in \cref{sec:via}, following the approach in the literature.

The flavor source, on the other hand, is commonly analyzed in the mass basis, in which the mass matrix is diagonal. Let us parameterize the complex off-diagonal masses in $M_f^2$ as $m_{LR}^2 = (m_{RL}^2)^*= |m|^2\e^{-i \sigma}$. The mass eigenstates are $\chi_m = U^\dagger \chi_f$, with $U$ a unitary matrix that diagonalizes the mass matrix $U^\dag M_f^2 U = {\rm diag}(m_1^2, m_2^2) = M_m^2$. We split $U = U_p U_r$
where $U_p $ removes the phase from the flavor-basis mass matrix, and $U_r$ diagonalizes the resulting real mass matrix:
\be
U_p ={\rm diag}( e^{-i\sigma/2},e^{i\sigma/2}),\qquad
U_r=\begin{pmatrix}
 \cos \theta  & -\sin \theta  \\
  \sin \theta  & \cos \theta
\end{pmatrix}.
\label{U}
\ee
The mass eigenstates are
\be
m_{1,2}^2 = \frac12\bigg((m_{L}^2+m_R^2) \pm \sqrt{(\delta m^2 )^2 +4 |m|^4}\bigg),
\quad
\tan 2 \theta = \frac{2 |m|^2}{\delta m^2 },
\label{m12}
\ee
with $\delta m^2 = (m_L^2-m_R^2)$.  
The mass ordering is such that $m_L^2 \geq m_R^2$.

The Lagrangian in the mass basis is \cite{Cirigliano:2011di, Cirigliano:2009yt}
\be
\L = -\chi_m^\dagger \[\partial^2 + M^2_m + \Sigma^2 +2\Sigma \cdot \partial +(\partial \Sigma) \]\chi_m +\L_{{\rm int}}(\chi_m),
\label{L_mass}
\ee
with $\Sigma_\mu = U^\dagger \partial_\mu U
=  U_r^\dagger \partial_\mu U_r + U_r^\dagger (U_p^\dagger \partial_\mu U_p) U_r $ or
\be
\Sigma_\mu =
 \begin{pmatrix} 0 & -1 \\ 1 & 0 \end{pmatrix}
 \partial_\mu \theta
+
\begin{pmatrix} \frac 1 2 (s_\theta^2-c_\theta^2 )  & c_\theta s_\theta   \\ c_\theta s_\theta   & -\frac 1 2(s_\theta^2-c_\theta^2 ) \end{pmatrix} i\partial_\mu \sigma\equiv \Sigma^\textsc{CP} + \Sigma^\textsc{CPV} .
\label{Sigma}
\ee
With our parameterization of the rotation matrix there is no explicit phase factor in $\Sigma$, and we can set the phase factor in the CP transformation \cref{phi_CP} to unity; in \cite{Cirigliano:2011di, Cirigliano:2009yt} instead, there is a phase factor in $\Sigma$ which necessitates to include a non-trivial  phase in the definition of the CP transformation. 
The Lagrangian is CP invariant if $\Sigma =-\Sigma^T$, which gives $\Sigma_{11} =\Sigma_{22}=0$; CPV requires  $\Sigma^\textsc{CPV} \propto \partial_\mu \sigma \neq 0$. CP violation needs a space-time dependent phase, as any constant phase can be removed by field redefinitions.

The KB-equations in the mass basis are
\begin{align}
  (k^2-\frac14 \partial_x^2) \Delta_m^\lambda &=\frac12\e^{-i\diamond}\bigg( \{M^2_m+ \Sigma^2-2i k\cdot \Sigma, \Delta_m^\lambda\}
+[ \Sigma,\partial \Delta_m^\lambda] +\{ \Pi^\lambda,\Delta_m^h\} +\C_{m,-}\bigg), \label{eq:KBflavorConst}
 \\
2ik\cdot\partial_x  \Delta_m^\lambda &= \e^{-i\diamond}\bigg( [M^2_m + \Sigma^2-2i k\cdot \Sigma, \Delta_m^\lambda] +\{ \Sigma,\partial \Delta_m^\lambda\} +[ \Pi_m^\lambda,\Delta_m^h]+\C_{m,+}\bigg)\,.
\label{eq:KBflavorKin}
\end{align}
These will be used as the starting point for the discussion of the flavor source in \cref{sec:flavor}.
The $[\Sigma,\partial G^\lambda_m]$-term (and the analogous term with anti-commutator) is absent in \cite{Cirigliano:2011di}. This does not affect the final result, as the term drops out at the approximation order that \cite{Cirigliano:2011di} assumes.

\paragraph{Fermions.}
We will consider a single fermionic species, with Lagrangian
\be
     \L = i\bar{\psi} \slashed{\partial} \psi - \bar{\psi}( m_h +i  m_a \gamma^5) \psi  + \L_{\rm int}.
\label{L_fermion}
\ee
We can rewrite the mass in terms of Weyl spinors $\bar{\psi}( m_h +i  m_a \gamma^5) \psi= \bar{\psi}_L M \psi_R
            - \bar{\psi}_R M ^*\psi_L $, with
\be
      M = m_h+ i m_a = m \e^{i\theta},
\label{mass_fermion}
\ee
where the parameterization in terms of $(m,\theta)$ is introduced for future reference.
In the bubble wall background $M =M (\phi_b(x^\mu))$ is spacetime dependent.
The Kadanoff-Baym equation for fermions is
\begin{align}
    \( \slashed{k} + \frac{i}{2} \slashed{\partial}  - ( m_h +i  m_a \gamma^5) e^{-\frac{i}{2}\stackrel{\!\!\leftarrow}{\partial_x}\cdot\,\partial_{k}}\)S^\lambda = e^{-i\diamond} \left(  \{\slashed{\Pi}^h \} \{ S^\lambda \}+ \{\slashed{\Pi}^\lambda \} \{ S^h \}  - \frac 1 2 \left( \{\slashed{\Pi}^> \} \{S^< \}  -\{\slashed{\Pi}^< \} \{S^> \}  \right) \right).\label{eq:KBfermions}
\end{align}
Just as for bosons,  we absorb the thermal mass contribution from $\Pi^h$ into the mass $m_h$. \Cref{eq:KBfermions} is the starting point for the derivation of the semi-classical source for fermions in the next subsection. Taking the hermitian and anti-hermitian parts, we obtain the constraint and kinetic equations; this is done in the next subsection using an explicit parameterization \cref{decomp_semi_0} for the Wightman functions.


\subsection{Semi-classical force}
\label{sec:semiclassical}

For a single species with dispersion relation $p^2=m^2$, the classical 4-force can be identified as $F^\mu = \partial_{x^\mu} m$, which can be seen from \cite{Garbrecht:2018mrp}
\be
p^\mu F_\mu = p^\mu \frac{\dd}{\dd \tau} p_\mu = \frac12 \frac{\dd}{\dd \tau} p^2 = m \frac{\dd m }{\dd x^\mu}\frac{\dd x^\mu }{\dd \tau} = p^\mu \frac{\dd m }{\dd x^\mu}\ ,
\label{dm_force} 
\ee
with $p^\mu = m \dd x^\mu/\dd \tau$ and $\dd \tau^2 =\dd x_\mu \dd x^\mu$, {\it cf} \cref{eq:forcevw}. The classical force is CP conserving, and arises for fermions and bosons alike at first order in gradients. In the bubble wall background, the dispersion relation gets modified by derivative corrections. As a result, the force term picks up (CP-violating) corrections at higher order in the gradient expansion, which are referred to as the semi-classical or WKB force. 

Let us see how the (semi-)classical force arises from the KB equations. For a single complex boson the leading-order gradient correction in the constraint \cref{eq:KBflavorConst} is $\sim \diamond^2 \{M_m^2,\Delta^\lambda\}$, which gives $\O(\partial_x^2)$ corrections to the dispersion relation. 
The source follows from inserting the thermal-equilibrium value of the Wightman function into the kinetic equation  \cref{eq:KBflavorKin}; in particular, the (semi-)classical source term is generated by the mass commutator term expanded 
to first order in the gradients:
\be
\{ \partial_x M_m^2, \partial_k \Delta^\lambda\}.
\label{origin_semi}
\ee
The classical source arises from the 0th order gradient solution to the constraint equation 
and is $\O(\partial_x)$, while the $\O(\partial_x^2)$ corrections to the dispersion relation gives the semi-classical source term that is $\O(\partial_x^3)$.  For a charged scalar the sources conserve CP at all orders, as the boson mass is just a real number. For a singlet scalar, the phase in the mass term can be absorbed in a redefinition of the fields (but if there are e.g. complex quartic couplings in the theory, there will be CPV in the interactions).
CPV can be introduced in a multi-flavor system with a complex mass matrix. It is a bit arbitrary how to distinguish between flavor and classical source terms, as in the flavor frame they both arise from the same mass matrix.  In the mass frame \cref{L_mass}, CPV resides in the $\Sigma$-term whereas the mass matrix is real and diagonal. The effects of $\Sigma$ are here classified as the flavor source and discussed in \cref{sec:flavor}. The field-dependence of the mass eigenvalues corrects the classical force in a CPC way, in the same way as for the single boson in \cref{origin_semi}, at  $\O(\partial_x^3)$.

We can repeat the exercise for the fermionic system.  There are two important differences with respect to the bosonic system. First, the constraint equation, and thus the dispersion relation, get corrected already at first order in the gradient expansion. And second, the single-flavor system can be CP violating if the  mass is complex valued. 
The source in the kinetic equation comes from the hermitean part of the mass term in \cref{eq:KBfermions} expanded to first order in gradients, i.e. from the hermitean part of
\be
 -\frac{i}{2}   \partial_z( m_h -i  m_a \gamma^5) \cdot\,\partial_{k}S^\lambda .
\ee
The classical source term arises from inserting the 0th order solution for $S^\lambda$, similarly to the bosonic system.   But now the $\O(\partial_x)$ corrections to the dispersion relation yields the CPV semiclassical source at $\O(\partial_x^2)$ in the gradient expansion.
 We will derive this source in the next subsection. 
This source term is distinctive from the flavor source because the latter stems from mixing between different flavors and is thus only present in a multi-flavor system.

\begin{tcolorbox}[float,hypertarget=box1,before skip=0.5cm,after skip=1cm,colback=blue!5!white,colframe=blue!75!black,toptitle=0.1cm,bottomtitle=0.1cm, title= Box 1: Conventions: distribution functions] 
As already mentioned in the introduction, we refer to the CP conjugate particle as the antiparticle. Here we make this statement precise, introducing our notation in the process. As follows from charge conjugation symmetry, the particles and antiparticles correspond to positive and negative energy solutions of the dispersion relation. 
The corresponding distribution functions describing their phase space densities are found by integrating over the Wightman functions, see \cref{project_boson} for bosons and \cref{project_semi} for fermions. From the way Wightman functions transform under CP, \cref{CP_boson_delW,S_CP}, we can the identify the particle and antiparticle distributions.

To be explicit we define the local distributions $f(k_0,\vec k,x) = \mathfrak{n}(|k_0|,\vec k,x)$, with $k_0$ the solution of the dispersion relation; here we suppressed the species (and for fermions) spin indices.  In thermal equilibrium, indicated by a `0' subscript, $\mathfrak{n}_0(k_0,\vec k)$ reduces to the Bose-Einstein (Fermi-Dirac) distribution for bosons (fermions).  We use the (admittedly ugly) $\mathfrak{n} $ to distinguish the distribution function from the number density defined below. The distribution function for antiparticles is $\bar f(\omega_0,\vec k,x)$, defined in terms of positive energies, which is related to the CP conjugate via  $f_{\rm CP}( \omega_0,k_z,x)=\bar f^T(\omega_0,\vec k,x)$. The transpose acts in flavor space and is only relevant in a system with several species, as discussed in \cref{{sec:flavor}}. The Wightman functions can be written both in terms of $\mathfrak{n}$ and in terms of $f$ and $\bar f$, see e.g. \cref{Gnf}, and both forms appear in the literature. There is also no consistent notation for what we denote by $f$ and $\mathfrak{n}$, and sometimes their definitions are interchanged, e.g. in \cite{Bellac:2011kqa}.

Since both particle and antiparticle densities are defined for positive energy, the distinction between $f$ and $\mathfrak{n}$ becomes moot: we only use $\mathfrak{n}$ in the definition of the Wightman functions, while the notation $f$ and $\bar f$ is used in the Boltzmann equations.  It is important to note that $\omega_0$ should be seen as a label, in that in the Boltzmann equations (obtained from integrating the Wightman functions over $k_0$) the  $\partial_z$ and $\partial_{\vec k}$ derivatives only act on the explicit arguments in $f(\omega_0,\vec k,x)$. 

The thermal equilibrium distribution is  
\be
 f^0(k^\mu)=(\e^{k^\mu u_\mu/T} \mp 1)^{-1} ,
 \label{f0}
 \ee
 with  $k_0 =\omega_0 > 0$ and $u_\mu$ the plasma velocity; in the wall frame $u_\mu=\gamma_w(1,v_w)$. The minus and plus sign is for bosons and fermions, respectively. The number density is $n(x) = g\int \frac{\dd^3k}{(2\pi)^3} f(k_0,\vec k,x)$, with $g$ the number of degrees of freedom. The number density of particles minus antiparticles -- the relevant CP-violating quantity for the baryogenesis calculation --  is denoted by $\rho(x)= n(x)-\bar n(x)$.
\end{tcolorbox}

\subsubsection{Semi-classical source for a single fermion}

We will derive the (semi-)classical force for fermions from the KB equations following 
\cite{Kainulainen:2001cn,Kainulainen:2002th}
and the reviews in \cite{Prokopec:2003pj, Garbrecht:2018mrp,Kainulainen:2021oqs}. Alternatively, the source term can be derived using WKB-methods
~\cite{Joyce:1994fu,Cline:2000kb,Kainulainen:2002th}, which can be incorporated in kinetic theory \cite{Cline:2000kb,Huber:2000ih}. 
The latter approach derives the CP-violating source directly from the Dirac equation, whereas the former uses the CTP-formalism, which allows for a more straightforward inclusion of e.g. flavor and thermal effects.

We work in the bubble wall frame, in which the bubble is at rest. Approximating the bubble wall by a planar interface, the mass is only a function of the spatial coordinate  orthogonal to the wall, which we choose to be $z$, and the fermion mass in \cref{mass_fermion} is $M= M(z)=m(z)\e^{i\theta(z)}$. For a space-time dependent angle $\theta'(z)\neq 0$, the phase cannot be rotated away by field redefinitions, and CP is violated. In this section, the prime denotes a derivative with respect to $z$.

The strategy to derive the semi-classical source is as follows.  The KB equations are most easily solved in the frame in which the momenta parallel to the bubble wall vanish, and the system becomes effectively 1+1 dimensional; at any point in the calculation one can boost back to the wall frame to find the results for general momenta.  To deal with the Dirac structure, the Wightman functions are expanded on a basis of
independent gamma-matrix structures that commute with the spin operator. Taking appropriate traces of the KB equation then yields a set of equations for the expansion coefficients. This set can be split into anti-hermitean  and hermitean parts, which are the constraint and kinetic equations respectively. The former is solved first to find the dispersion relation and the corresponding expression for the Wightman functions in terms of distribution functions. To find the semiclassical source we need to include first-order gradient corrections to the Green's functions, which makes the expressions differ from the more familiar thermal equilibrium expressions. The Wightman functions are then substituted into the kinetic equation to find their evolution. Finally, we integrate the kinetic equation over energy to project out the Boltzmann equations for the thermal distribution functions. The source term is identified as the term that does not vanish for thermal equilibrium distribution functions. 

This is rather a mouthful so let's show how this works in practice. We start with the KB equations for fermions in \cref{eq:KBfermions}.
As the focus is on finding the source, we neglect thermal corrections in $\Pi^\lambda$ and the collision term to get
\be 
{\cal D} S^\lambda \equiv\(  \slashed{k}+\frac{i}2\slashed{\partial} - ( m_h +i  m_a \gamma^5) e^{-\frac{i}{2}\stackrel{\!\!\leftarrow}{\partial_z}\cdot\,\partial_{k_z}}\) 
S^\lambda = 0\,.
\label{KB_fermion_semi}
\ee
In the frame where the momenta parallel to the wall  $\vec k_\parallel =(k_1,k_2) $ vanish \cite{Prokopec:2003pj}, the system becomes effectively $1+1$-dimensional. We boost $k^\mu=(k_0, \vec k_\parallel, k_z) \to\tilde k^\mu= (\tilde k_0, 0, k_z)$, with boost parameters
\be
\vec v_\parallel =\frac{\vec k_\parallel}{k_0}\,, \quad \gamma_\parallel = \frac{k}{\tilde k_0}\,,\quad
\tilde k_0 =\sgn(k_0) \sqrt{k_0^2 -|\vec k_\parallel|^2}\,.
\label{boost}
\ee
Because of the planar wall approximation, the Dirac operator in the boosted frame $\tilde{\cal D}$ is obtained by replacing $k^\mu \to \tilde k^\mu$ and $\partial^\mu \to \tilde \partial^\mu$ in $\cal D$.  $\tilde{\cal D}$ commutes with the spin operator $[\tilde \D, \tilde S_z]=0$, with $\tilde S_z = \gamma^0\gamma^3\gamma^5$.  The Wightman function factorizes into eigenstates of the spin projector $\tilde P_s = \frac12(\mathbbold{1}+ s \tilde S_z)$ with $s =\pm 1$ eigenvalues: 
$\tilde S^\lambda =\sum_{s=\pm1} \tilde S^\lambda_s$ with $\tilde P_s \tilde S_s^\lambda = s \tilde S_s^\lambda$.
We can further decompose $\tilde S^\lambda_s$ into a set of matrices that commute with $\tilde P_s$
\be
\tilde S^\lambda_s = i \tilde P_s\( s \gamma^3 \gamma^5 g_0^{\lambda s} -s \gamma^3 g_3^{\lambda s} + \mathbbold{1} g_1^{\lambda s} -i \gamma^5 g_2^{\lambda s}\)\,.
\label{decomp_semi_0}
\ee
Using hermiticity $(i\gamma^0 \tilde S^\lambda)^\dagger = i\gamma^0 \tilde S^\lambda$ it follows that $g_a^{\lambda s} (\tilde k,\tilde z)$ with $a=0,..,3$ are hermitian functions of the $\tilde z$-coordinate and momentum. 
Multiplying the KB equation \cref{KB_fermion_semi} by $P_s\{\mathbbold{1},s\gamma^3\gamma^5,-is\gamma^3,-\gamma^5\}$, inserting the decomposition of the propagator \cref{decomp_semi_0}, and taking the trace gives four equations for the component fields. These can be split into their hermitian and anti-hermitian parts corresponding to the kinetic and constraint equations respectively. 
Expanding $\e^{-i \diamond} (m g^s) = m g^s-\frac{i}{2}  \partial_x m \cdot \partial_{k}g^s+..$, 
the constraint equations to first order in derivatives are
\begin{align}
\tilde k_0 g^{\lambda s}_0 - sk_z g^{\lambda s}_3 - m_h g^{\lambda s}_1 - m_a g^{\lambda s}_2 & = 0\,,
\nn \\
\tilde k_0 g^{\lambda s}_3 - sk_z g^{\lambda s}_0 + \frac12 m_h' \partial_{k_z} g^{\lambda s}_2
                          - \frac12 m_a' \partial_{k_z} g^{s}_1 &= 0\,,
\nn \\
\tilde k_0 g^{\lambda s}_1 + \frac{s}{2} \partial_z g^{\lambda s}_2
               - m_h g^{\lambda s}_0 + \frac12 m_a' \partial_{k_z} g^{\lambda s}_3 &= 0\,,
\nn \\
\tilde k_0 g^{\lambda s}_2 - \frac{s}{2} \partial_z g^{\lambda s}_1
               - m_a g^{\lambda s}_0 - \frac12 m_h' \partial_{k_z} g^{\lambda s}_3 &= 0\,.
\label{constraint_semi}
\end{align}
Here we have counted $\partial_z^n m_i$ (with $i=h,a$) as well as $\partial_z^n g_a^{\lambda s}$ as $n$'th order in derivatives; only the former corresponds to the diamond expansion. This power counting can fail in a multi-flavor system because there are more mass scales in the problem and the spatial derivative of the propagator can bring down a factor of the mass difference rather than the inverse wall width.
The constraint equations relate the various $g^{\lambda s}_a$, and only one of these functions is independent, which we take as
 $g_0^{\lambda s}$.
The bottom three equations can be solved for $g_i^{\lambda s}$  with $i=1,2,3$ in terms of $g_0^{\lambda s}$ and $\partial_{k_z} g^ s_0$ up to first order in the gradient expansion. Plugging the solutions into the first equation, all $\partial_{k_z} g ^{\lambda s}_0$ terms cancel, and we get
the algebraic result
\be
\Omega_s^2 g^{\lambda s}_0 \equiv\( \tilde k_0^2 -k_z^2 - m^2  + \frac{s}{\tilde k_0}m^2 \theta'\) g^{\lambda s}_0=0\,,
\label{constraint_g0_semi}
\ee
where we switched to the modulus $m$ and phase $\theta$ notation for the mass \cref{mass_fermion}.
This is solved by 
\be
g^{\lambda s}_0 = 2\pi |\tilde k_0|  \delta(\Omega_s^2) \mathfrak{g}^\lambda_s.
\label{spectral_semi_0}
\ee 
Here $\mathfrak{ g}^\lambda_s =\{\mathfrak{n}_s,\,-(1-\mathfrak{n}_s)\}$ for $\lambda=\{<,\,>\}$
with $\mathfrak{n}_s(k,z)$ the distribution function. Our conventions for the distribution functions can  be found in \hyperlink{box1}{Box 1}.  
The normalization is fixed by the sum rule \cref{sumrule_F}, which imposes $\int {\dd \tilde k_0}\,g_0^{\lambda s} =2\pi$ \cite{Prokopec:2003pj, Kainulainen:2001cn}.

Now we can boost back to the frame with non-zero momentum parallel to the wall $\tilde k^\mu \to k^\mu$, using the inverse boost of \cref{boost}. The Wightman function becomes
\be
S^\lambda_s = i P_s\( s \gamma^3 \gamma^5 g_0^{\lambda s} -s \gamma^3 g_3^{\lambda s} + \mathbbold{1} g_1^{\lambda s} -i \gamma^5 g_2^{\lambda s}\),
\label{decomp_semi}
\ee
with now $P_s =\frac12(\mathbbold{1}+s S_z)$ and  $S_z =\tilde k_0^{-1} ( k_0 \gamma^0 -\vec k_\parallel \cdot \vec \gamma) \gamma^3 \gamma^5$. Further $g_i(\tilde k, \tilde z) = g_i(k, z)$. Hence the wall-frame Wightman solution is still of the form \cref{spectral_semi_0}, where we can simply replace $\tilde k_0$ with the definition \cref{boost} to express the results in terms of $k_0$.  Explicitly, this gives\footnote{$\Omega_{s}^2=0$ yields a 3rd order polynomial in $\tilde k_0$ and has a third mass shell solution. Following the literature, we have neglected this pole as it corresponds to unphysical solutions, which are harmless as their contribution to the sum rule \cref{sumrule_F} vanishes \cite{Kainulainen:2001cn}.}
\be
g^{\lambda s}_0 = 2\pi |\tilde k_0|  \delta(\Omega_s^2) \mathfrak{g}^\lambda_s
=2\pi \sum_\pm Z_{s\pm} \frac{\delta(k_0 \mp \omega_{s\pm}) }{2 \omega_{s\pm} }
                 \, |\tilde k_0| \mathfrak{g}^\lambda_s\,.
\label{spectral_semi_1}
\ee 
The energies of the quasi-particles $\omega_{s\pm}$ correspond to the roots of $\Omega_{s}^2=0$, where the indices $\pm$ refer to the sign of $k_0$.  Solving the root equation perturbatively in the gradient expansion to first order gives the dispersion relation
\be
     \omega_{s\pm} = \omega \mp s \frac{m^2\theta'}{2\omega \omega_{z}}\,, \quad
      Z_{s\pm} = 1\pm s\frac{m^2\theta'}{2\omega^3_{z}}\,,
\label{energy_semi}
\ee
with $\omega= \sqrt{ k_z^2 +|\vec k_\parallel|^2+ m^2}$ and $\omega_{z}= \sqrt{ k_z^2 + m^2}$. At this order in the derivative expansion, the constraint equation does not contain $z$-derivatives of $g^{\lambda s}_0$ and therefore the quasi-particle picture holds.

The transformation of the Wightman function under CP conjugation \cref{S_CP} translates to
\be 
g_0^{<s}(k,z) \stackrel{{\cal CP}}{\to} -g^{>-s}_0 (-\bar k, \bar z) ,
\ee
where we remind ourselves of the definition $\bar k  = (k^0, -\vec k_\parallel,-k_z)$ and $\bar z =-z$.
Integrating over frequencies projects out the phase space distributions and their CP conjugate: 
\begin{align}
\int_0^\infty \frac{\dd k^0}{2\pi} 2\gamma_\parallel g_0^{<s} (k^0,k_z,z)&= Z_{s+}\mathfrak{n}_s( \omega_{s+},k_z,z) \equiv Z_{s+} f_s(\omega_{s+},k_z,z), \nn \\
\int_0^\infty \frac{\dd k^0}{2\pi} 2\gamma_\parallel (-g^{>-s} (-k^0,k_z,\bar z)) &=Z_{-s-} (1-\mathfrak{n}_{-s}(- \omega_{-s-},k_z,\bar z))\equiv Z_{s-} \bar f_s(\omega_{s-},k_z, z),
\label{project_semi}
\end{align}
with $f_s(k_0)=\mathfrak{n}_s(|k_0|)$.
The integration over $g^{<s}$ on the first line  picks out the positive pole $ \omega_{s+} $, while  the integration over $g^{>-s}$ on the second line instead picks out the negative pole $ \omega_{(-s)-} $.  The CP conjugate $\bar f_s$ antiparticle density is defined for positive energies and the same spin as the particle, and going to the final expression of \cref{project_semi} both the sign of the energy and of the spin are flipped.

The force is derived from the kinetic equation for $g^{\lambda s}_0$. We can obtain this equation from \cref{KB_fermion_semi} in the same way as the constraint equation \cref{constraint_semi}, but by taking the hermitian rather than anti-hermitian part:
\begin{align}
   0&=  s\partial_z g^{s}_3
            - m_h' \partial_{k_z}g^{\lambda s}_1
            - m_a' \partial_{k_z}g^{\lambda s}_2 \nn \\
& =\gamma_\parallel\[  \frac{k_z}{k_0}\partial_z-\frac1{2k_0}
       \(   {m^2}^{\,\prime}
                - \frac{s}{\tilde k_0}(m^2\theta^{\,\prime})^{\,\prime}
                         \)  \partial_{k_z}\]g^{\lambda s}_0 (k_0,k_z,z),
\label{boltzmann_semi}
\end{align}
up to second order in gradients, and where we used  that the background only depends on the $z$-coordinate in the bubble wall frame.  
On the 2nd line we used the solutions $g_i^{\lambda s}$ from the constraint equation. Integrating over positive $k_0$ gives the Boltzmann equation for the particle and antiparticle densities respectively
\begin{align}
Z_{s+} \frac{k_z}{\omega_{s+}} \[  \partial_z-\frac1{2 k_z}
       \(   {m^2}^{\,\prime}
                - \frac{s}{\omega_{z}}(m^2\theta^{\,\prime})^{\,\prime}
                         \)  \partial_{k_z}\]f_s(\omega_{s+},k_z,z) &=0,
\nn\\
Z_{s-}\frac{k_z}{\omega_{s-}}\[  \partial_z-\frac1{2 k_z}
       \(   {m^2}^{\,\prime}
                + \frac{s}{\omega_{z}}(m^2\theta^{\,\prime})^{\,\prime}
                         \)  \partial_{k_z}\]\bar f_s(\omega_{s-},k_z, z)  &=0.
\label{Znormalization}
\end{align}
The factor $v^{\rm gr}_{s\pm} = Z_{s\pm} k_z/\omega_{s\pm} = \partial \omega_{s\pm}/\partial k_z$  can be identified as the group velocity \cite{Kainulainen:2021oqs}. 
However, another common approach is to define a `velocity' $v_{s\pm} =k_z/\omega_{s\pm} $, and absorb the wavefunction renormalization in the collision terms, once these are reinstated on the right-hand-side of the equations; this is what we will do in the following. When comparing to the literature, care should be taken in what choice is made, and how the collision terms are defined. 
The term between round brackets can be identified as the force term. To separate the CP-conserving and -violating parts, we do a gradient expansion of $\omega_{s\pm}$.
This gives 
\begin{align}
\[  v_{s\pm}\partial_z+ ( F_s^\textsc{cp}\pm F_s^\textsc{cpv}) \partial_{k_z}\]f_s(\omega_{s\pm},k_z,z) &=0,
\label{Boltz_fermion}
\end{align}
where it is understood that $f_{s+}=f_s$ and $f_{s-}=\bar f_s $ correspond to the particle and antiparticle distributions respectively. 
The CP and CPV force (which are along the $z$-axis) are
\be
 F_s^\textsc{cp}  =-\frac{{m^2}^{\,\prime}}{2\omega},
\quad F_s^\textsc{cpv}=  s \( \frac{(m^2 \theta')'}{2 \omega \omega_z} - \frac{m^2 \theta' {m^2}^{\,\prime}}{4\omega^3 \omega_{z}}\).
\label{clas_force}
\ee
The CPV force has the opposite sign for particles and antiparticles.
The gradient-expanded velocity is
\be
v_{s\pm} =
\frac{k_z}{\omega_s} = \frac{k_z}{\omega} \pm s \frac{m^2 \theta'}{2\omega^2\omega_{z}}.
\ee

Treating the bubble as a perturbation on top of the local thermal equilibrium background, we expand the distribution function 
around the Fermi-Dirac distribution in the wall frame \cite{Fromme:2006wx,Kainulainen:2021oqs,Cline:2020jre}:
\be
f_{s\pm} (\omega_{s\pm},k_z,z) = f^0_{s\pm}(\omega_{s\pm},k_z)+ \delta f^\textsc{cp}_s(\omega_{s\pm},k_z,z) \pm \delta f^\textsc{cpv}_s(\omega_{s\pm},k_z,z).
\ee
Here $\omega_{s\pm}$ contains the first order gradient correction, and thus $f_{0s}(\omega_{s\pm}, k_z)$ tracks the local equilibrium distribution along the bubble wall 
\be
f^0_{s\pm} (\omega_{s\pm},k_z) = (\e^{\gamma_w(\omega_{s\pm} + v_w k_z)/T}+1)^{-1} = f^0(\omega,k_z) + \gamma_w  {f^0}'(\omega,k_z) (\omega_{s\pm}-\omega) + \O(\partial_z^2),
\label{f0s}
\ee
with $f^0(\omega,k_z)= f^0_{s\pm} (\omega,k_z)$  and $v_w$ the fluid velocity and $\gamma_w=1/\sqrt{1-v^2_w}$ the Lorentz factor. The right-most expression  is to first order in the gradient expansion.  The velocity of the plasma in the wall frame is here approximated by the bubble wall velocity, and  deviations from that induced by the presence of the bubble are neglected. We will return to this point in \cref{sec:bubbles}.
Care was taken in labeling the $k^0$-argument explicitly, but when there is no source of confusion, we will often write $f^0(\omega,k_z) = f^0(k_z)$ as was done in \cref{f0}. In this notation the derivatives are defined via
(for future reference, the notation below is for both bosons and fermions):
\be
f^0(k) = \frac{1}{\e^{k^\mu u_\mu/T} \pm 1}, \quad
f^{0'}(k) \equiv \partial_{(k^\mu u_\mu)} f^0(k) 
, \quad
f^{0''}(k) \equiv \partial_{(k^\mu u_\mu)}^2 f^0 (k)
,
\label{df0}
\ee 
that is, the derivatives are defined with respect to $(k^\mu u_\mu)$.  In the literature also the definition $f^{0'} \to  Tf^{0'}$ and $f^{0''} \to T^2 f^{0''}$ appears, which comes from taking derivatives with respect to $(k^\mu u_\mu/T)$ instead, and which makes $f^{0'}, f^{0''} $ dimensionless. It is always useful to check dimensions and make sure of the conventions.
The primes here should not be confused with $\theta' =\partial_z \theta$ and $m^{2'}=\partial_z m^2$ where the prime instead denotes a spatial derivative.

The source terms can be defined as the contributions that do not vanish in thermal equilibrium, and only depend on the equilibrium distribution $f^0$. Since the CPC and CPV force are of different order in the gradient expansion, the equations for the CPC and CPV perturbations $\delta f^\textsc{cp}_s$  and $ \delta f^\textsc{cpv}_s$ decouple at leading order in the gradient expansion.  Taking the sum and the difference of the Boltzmann equation for particles and antiparticles gives:
\begin{align}
\[  \frac{k_z}{\omega}\partial_z+  F_s^\textsc{cp} \partial_{k_z}\]\delta f^\textsc{cp}_s&=- F_s^\textsc{cp} \partial_{k_z}f_{0}(\omega,\vec k)+\O(\partial_z^2),
\nn\\
\[  \frac{k_z}{\omega}\partial_z+  F_s^\textsc{cp} \partial_{k_z}\]\delta f^\textsc{cpv}_s&=- F_s^\textsc{cpv} \partial_{k_z}f_{0}(\omega,\vec k)
-\frac12   F_s^\textsc{cp} \partial_{k_z} \(\partial_{\omega}f_{0}(\omega,\vec k) (\omega_{s+}-\omega_{s-})
\)
+\O(\partial_z^3) ,
\label{sources1}
\end{align}
where we used \cref{f0s}. The source terms can be identified with the right-hand-side of these equations. The CPV  force gets a contribution from the CPC force term as the energy $\omega_{s\pm}$ has gradient corrections that differ for particles (plus sign) and antiparticles (minus sign).
 Plugging in the expression for the forces \cref{clas_force} and using that $\partial_{k_z} f_0 = \gamma_w v_w f'_0$, this gives
\begin{align}
S_s^\textsc{cp}&=  \gamma_w v_w \frac{{m^2}^{\,\prime}}{2\omega} f'_0, \nn \\
S_s^\textsc{cpv}&= -   s  \gamma_w v_w \( \frac{(m^2 \theta')'}{2 \omega \omega_z} f'_0 - \frac{m^2 \theta' {m^2}^{\,\prime}}{4\omega^3 \omega_{z}}( f'_0 -\gamma_w \omega f''_0)\).
\label{clas_source}
\end{align}

The Boltzmann equations and the semi-classical force are derived for definite spin $s$, which is the spin along the $z$-axis in the frame where the momentum parallel to the wall vanishes $p_\parallel=0$. For the coupling to sphalerons it is convenient to rewrite the source in the helicity basis, which can be defined in the frame where the bubble is at rest.  This was done in Refs.~\cite{Cline:2020jre, Cline:2017qpe} (see also the discussion in \cite{Prokopec:2004ic}) and effectively comes down to replacing 
\be
s \to  s_h = h\gamma_\parallel \frac{k_z}{|k|}\,,
\label{helicity}
\ee
in the semiclassical force (\cref{clas_force}) and source terms (\cref{clas_source}) where  $h =\pm 1$ is the helicity. The difference between the source term in the spin and helicity basis is small \cite{Cline:2017qpe}. For massless fields in the symmetric phase the negative helicity state can be identified with the left-handed chiral state that couples to the sphalerons; however, in a set-up where the masses are non-zero in the symmetric phase the left-handed states are a superposition of both helicities.

\subsubsection{An explicit example}
\label{sec:example_top_source}

\begin{figure}[h!]
    \centering
    \includegraphics[scale=0.6,trim=0 0 0 0]{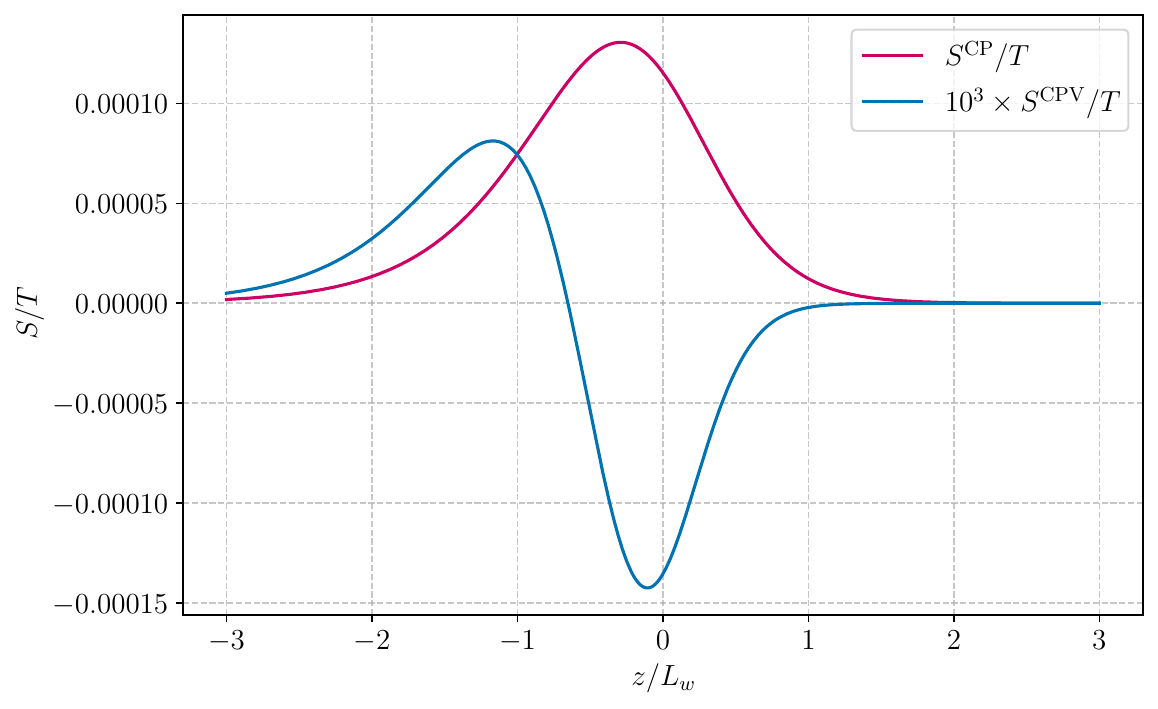}
    \caption{The source terms in units of the temperature as a function of distance in units of the wall width $z/L_w$. Negative $z/L_w$ corresponds to the symmetric phase.}
    \label{fig:source}
\end{figure}

We end this subsection with an explicit example. For the CPV effective top Yukawa interaction \cref{CP_yukawa} we can read off the top mass in the bubble wall background $M_t =m_t \e^{i\theta_t}$ as in \cref{mass_fermion}. For simplicity, we take the Wilson coefficient of the interaction $-y_t c_t \bar Q_L \tilde \vp t_R\(\vp^\dagger \vp\)$ purely imaginary $c_t = i \Lambda^{-2}$. Then
\be
m_t=\frac{y_t}{\sqrt{2}} \phi_b(z) \sqrt{1+ \frac{\phi_b(z)^4}{\Lambda^4}} \approx\frac{y_t}{\sqrt{2}} \phi_b(z) ,\quad
\theta_t = \tan^{-1} \(\frac{\phi_b(z)^2}{\Lambda^2}\).
\ee
For the bubble profile we take the tanh Ansatz \cref{bubble_profile}, which in the wall frame reduces to
\be
\phi_b(z) = \frac{v_n}{2}\(1-\tanh\(\frac{z}{L_w}\)\).
\ee
The tanh Ansatz is commonly employed in the wall velocity and EWBG literature. See \cite{Goncalves:2023svb} for a critical assessment in the context of EWBG. 
The Fermi-Dirac distribution \cref{f0} in the wall frame reads $f^0 = (\e^{\gamma_w (\omega_0 + v_w k_z)/T}+1)^{-1}$ as in \cref{f0s}.
Now we have all the ingredients to calculate the source terms. The result is shown in \cref{fig:source}. Here we took the parameter values  $v_n =1.2 T_n$, $L_w=10/T_n$, $T_n=100\,$GeV, $v_w=0.1$ and $\Lambda=900\,$GeV. The source is different for the different momentum modes, and we choose $k_x =k_y=k_z =T$ in the figure. The CPV source is dominated by the first term in $S^\textsc{cpv}$ in \cref{clas_source}.
The CPV source is a factor $10^3$ smaller than the CPC source.  The suppression of the latter is not just from one extra derivative  (as each derivative is expected to give a factor $1/(L_w T_n)$), but also because CP violation only arises from the dim-6 contribution to the effective mass (as $\theta \sim \frac{\phi_b^2}{\Lambda^2}$).
The number of nodes in the source profile is determined by the number of derivatives. In the absence of rescattering, the perturbations will have the same profile, and the total CP asymmetry integrated over the symmetric phase partially cancels.

\subsection{Flavor source}
\label{sec:flavor}

The flavor source arises from a mismatch between flavor/interaction and mass eigenstates, which in the bubble wall background can give rise to a CPV mixing matrix.  For simplicity, we focus here on a two-flavor bosonic system, and neglect thermal width corrections (thermal masses can be absorbed in the mass matrix). This set-up has been studied in the mass degenerate limit \cite{Cirigliano:2011di, Cirigliano:2009yt}, and away from the resonance in \cite{Konstandin:2004gy}.

Consider the toy model  \cref{M_boson_VIA} with two bosonic `flavors' labeled by $L,R$, with  a CPV interaction with the Higgs field. 
We will work in the mass basis (\cref{L_mass}) in this section; for notational convenience we drop the subscript $m$ on all quantities except for the mass matrix. The kinetic and constraint equation are given in \cref{eq:KBflavorConst,eq:KBflavorKin}. 
At leading order in the gradient expansion, $\e^{-i\diamond}=1$, and $\Sigma = 0$; the collision terms vanish in thermal equilibrium, and the constraint equation reads
\begin{align}
  (k^2-\frac14 \partial_x^2) \Delta^\lambda = \frac12\{M^2_m,\Delta^\lambda\}\,.
\end{align}
Neglecting the $\partial_x^2$ correction\footnote{The $\partial_x^2$ correction can be straightforwardly included in the wall frame, where we can substitute the kinetic equation into the constraint equation to give \cite{Konstandin:2005cd}
\be
\partial_{z} \Delta^\lambda =\frac1{2k_z}[M^2_m,\Delta^\lambda] \quad \Rightarrow \quad
\partial_{z}^2  \Delta^\lambda
=\frac1{2k_z}[M^2_m,\partial_{z}\Delta^\lambda] 
= \frac1{4k_z^2}[M^2_m, [M^2_m,\Delta^\lambda]].
\ee
This gives a correction $m_+^2\to  m_+^2 + m_-^4/k_z^2$ in the dispersion relation for the off-diagonal modes. In the limit of degenerate masses, this corresponds to a small correction, and it was neglected in \cite{Cirigliano:2011di}. }
this gives $(k^2-m_{ij}^2)  \Delta^\lambda_{ij} = 0$ with $m_{ii}^2=m_i^2$ and $m^2_{ij} = m_+^2$ for $i\neq j$ (we will use the same index notation for $\omega_{ij}^2 =k^2+m_{ij}^2$), where we defined $m^2_\pm =\frac12(m_1^2\pm m_2^2)$. The solutions \label{n_in_G} satisfying the sum rule \cref{sumrule_B} are  
\begin{align}
i\Delta_{ij}^> &= 2\pi \sgn(k_0)\delta(k_0^2-\omega_{ij}^2) \frak{g}_{ij}^> = 2\pi \delta(k_0^2-\omega_{ij}^2) 
\[ \theta(k^0) (\delta_{ij} + f_{ij}(\omega_{ij},\vec k,x))+  \theta(-k^0) \bar f_{ij}^T(\omega_{ij},-\vec k,x)\], \nn \\
i\Delta_{ij}^< &= 2\pi \sgn(k_0)\delta(k_0^2-\omega_{ij}^2) \frak{g}_{ij}^<= 2\pi \delta(k_0^2-\omega_{ij}^2) \[ \theta(k^0) f_{ij}(\omega_{ij},\vec k,x)+ \theta(-k^0)   (\delta_{ij} +\bar f_{ij}^T(\omega_{ij},-\vec k,x)) \].
\label{Gnf}
\end{align}
Here $\frak{g}_{ij}^> = \delta_{ij} + \mathfrak{n}_{ij}$ and $\frak{g}_{ij}^< = \mathfrak{n}_{ij}$ generalizing the single species result \cref{g_frak}, and $f_{ij} (k_0,\vec k,x)=\mathfrak{n}_{ij}(|k_0|,\vec k,x)$.
The second expression defines the antiparticle density $\bar f =f^\textsc{CP}$, defined as the CP conjugate of the particle density, as follows from the CP transformation \cref{phi_CP}. This generalizes \cref{project_boson} for several flavors. The transpose in \cref{Gnf} acts on flavor space, that is, $\bar f_{ij}^T = \bar f_{ji}$. Note that our notation differs from \cite{Cirigliano:2011di, Cirigliano:2009yt}, who instead use $\bar f^T = f^\textsc{CP}$.
In thermal equilibrium $f_{ij}^0=\delta_{ij} f_i^{0}$, with $f_i^0(k_z)$ the Bose-Einstein distribution \cref{f0} in the wall frame.

\subsubsection{Kinetic equation}

The next step is to substitute the zeroth order solution of the constraint equation into the first order kinetic equation
\begin{align}
2ik \cdot \partial \Delta^\lambda &=[M_m^2,\Delta^\lambda]- 2i [k\cdot  \Sigma(x), \Delta^\lambda]-\frac{i}{2} \{\partial_x M_m^2, \partial_{k} \Delta^\lambda\} +\C_+[ \Delta^\lambda],
\end{align}
where we have neglected the $[\Pi^\lambda, \Delta^\lambda]$ term, which gives thermal width corrections, and the gradient of the collision term, as is commonly done in the literature \cite{Cirigliano:2011di, Cirigliano:2009yt, Konstandin:2004gy}. Integrating over $\Delta^\lambda$  we can project out the equations for  particle $f_{ij}(\omega_{ij}, k_z,x)$ and antiparticle modes  $\bar f_{ij}(\omega_{ij},k_z,x)$ using \cref{project_boson}. 
Specializing to the bubble wall frame, and suppressing the explicit $(k_z,z)$ dependence of $f$ and $\bar f$ for notational convenience, this gives
\begin{align}
\label{S_flav_0}
(v_1 \partial_z +  F_i  \partial_{k_z}) f_{11}(\omega_1)&= -v_+ (\Sigma_{12}  f_{12}(\omega_+) -\Sigma_{21}  f_{21}(\omega_+))+ \C_{11},  \\
v_+ \partial_z  f_{12}(\omega_+)&=-i\frac{\omega_1^2-\omega_2^2}{2\omega_+}f_{12}(\omega_+)
 -v_+ (\Sigma_{11} -\Sigma_{22}) f_{12}(\omega_+)- \Sigma_{12} (f_{22}(\omega_2) v_2- f_{11}(\omega_1) v_1 ) + \C_{12} ,\nn
\end{align}
and
\begin{align}
\label{S_flav_1}
(v_1 \partial_z +  F_i  \partial_{k_z}) \bar f^T_{11}(\omega_1)&= -v_+ (\Sigma_{12}  \bar f^T_{12}(\omega_+) -\Sigma_{21}  \bar f_{21}^T(\omega_+))+ \C_{11},  \\
v_+ \partial_z  \bar f_{12}^T(\omega_+)&=+i\frac{\omega_1^2-\omega_2^2}{2\omega_+}\bar f^T_{12}(\omega_+)
 -v_+ (\Sigma_{11} -\Sigma_{22}) \bar f^T_{12}(\omega_+)- \Sigma_{12} (\bar f^T_{22}(\omega_2) v_2- \bar f^T_{11}(\omega_1) v_1 ) + \C_{12},\nn
\end{align}
with $\Sigma \equiv \Sigma^z$.
We defined $v_i = \frac{k_z}{\omega_i}$ and  $F_i = -\partial_z \omega_i$ with the same notation for $\omega_i$ as that we used for the masses: $\omega_i =\sqrt{k^2+m_{ii}^2}$ for the diagonal modes and $\omega_+=\sqrt{k^2+m_{+}^2} $ for the off-diagonal modes. Since $m_+^2$ is spacetime independent, there is no force term in the equation for the transverse modes $f_{12}$. The only difference between the $f$ and $\bar f^T$ modes is the sign of the oscillation term in the transverse equation. Transposing the equation for $\bar f^T$, we see that the equation for $f$ and $\bar f= f^\textsc{CP}$ only differ by $\Sigma \to -\Sigma^T$. Hence, CP violation requires  $\Sigma \neq -\Sigma^T$ and resides in the transport equation for the transverse modes. The diagonal modes can obtain CPV densities $f_{ii}-\bar f_{ii} \neq 0$ only via their mixing with the off-diagonal modes, there is no direct source term.

The boundary conditions can be set far away from the wall, where right-movers ($k_z>0)$ are in thermal equilibrium at $z \to -\infty$ and the left-movers  ($k_z<0)$ are in thermal equilibrium at $z \to +\infty$ \cite{Cirigliano:2011di}:
\begin{align}
\lim_{z\to -\infty} f_{ij}(\omega_{ij},k_z>0,z) 
=\delta_{ij} f^0(\omega_{i},k_z),\quad
\lim_{z\to +\infty} f_{ij}(\omega_{ij},k_z<0,z) 
=\delta_{ij} f^0(\omega_{i},k_z),
\end{align}
with $f^0(\omega_{i},k_z)$ the Bose-Einstein distribution in the wall frame given by \cref{f0}.
The flavor source is the $\Sigma$-dependent term in the transverse equations that does not vanish in thermal equilibrium, which only appears in the transverse equation:
\be
S^\textsc{flav}_{12} = - ( f^0(\omega_{1}) v_1- f^0(\omega_{2}) v_2)\Sigma_{12}.
\label{source_flavor}
\ee
It has a CPC and CPV part as defined in \cref{Sigma}.

Suppose that only the left-handed flavor states couple to a baryon-number-violating operator, which is only active in the symmetric phase (mimicking the sphaleron transitions for fermions). The baryon asymmetry is then sourced by the left-handed baryon charge. 
The charge current in the mass basis is
\begin{align}
j^\mu_{ij}(x) &= \int \frac{\dd^4 k}{(2\pi)^4} \langle: \phi^\dagger_{j}(x) \stackrel{\leftrightarrow}{\partial^\mu_x} \phi_{i}(x): \rangle
= \int \frac{\dd^4 k}{(2\pi)^4} k^\mu\( i\Delta^<_{ij}(k,x)+  i\Delta^>_{ij}(k,x)\)\,,
\end{align}
where the colons denote normal ordering.
The charge is then 
\begin{align}
j^0(x) &= 
\int \frac{\dd^3 k}{(2\pi)^3} \( f- \bar f^T\),
\label{current0}
\end{align}
which gives the left-handed baryon charge after rotating to the flavor frame
\be
j^0_{f,LL} = (Uj^0 U^\dagger)_{LL}\,,
\ee
with the subscript $f$ denoting that this is the charge current in the flavor basis, and $U$ the mixing matrix \cref{U}. If $j^0_{f,LL}$ is non-zero in the symmetric phase, it will source the B-violating interactions and create a net baryon asymmetry.

\subsubsection{Discussion of the flavor source term}

Which terms in the transport equation are dominant depends on the parameters of the model.
To get some intuition for this, it is useful to consider 
the various energy scales in the set-up, which are the (inverse) de Broglie wavelength of the plasma particle $k_{\rm dB} \sim T$, the inverse bubble wall length $k_w \sim 1/L_w$, the oscillation scale $k_{\rm osc} \sim (\omega_1-\omega_2)^{-1} \sim  m_-^2/k_{\rm dB}$,
and the collision scale $k_{\rm coll} \sim \alpha T$. 
Here, $\alpha$ denotes a typical interaction strength corresponding to $\mathcal L_{\rm int}$ in \cref{M_boson_VIA}.
From this we can construct the dimensionless parameters
\be
\eps_w =\frac{k_w}{k_{\rm dB} }\sim \frac1{L_w T},\quad
\eps_{\rm osc}  =\frac{k_{\rm osc}}{k_{\rm dB}} \sim \frac{m_-^2}{T^2}, \quad
\eps_{\rm int} =  \frac{k_{\rm int}}{k_{\rm dB} }\sim \alpha.
\label{flavor_eps}
\ee
The gradient expansion is an expansion in $\eps_w \ll 1$, valid in the thick wall regime.

The approach in \cite{Konstandin:2005cd} is to expand around the equilibrium solution $f=f^0+\delta f$ and treat deviations as $\delta f = \O(\eps_w)$, the reasoning being that deviations from thermal equilibrium can only be induced by the passing bubble wall.
With also $\Sigma, F \sim \O(\eps_w)$ we can drop all $\O(\eps_w^2)$ terms to get
\begin{align}
(v_1\partial_z +  F_i  \partial_{k_z})\delta f_{11}&= S^\textsc{cp}_{11}+ \C_{11},\nn \\
v_+ \partial_z\delta f_{12}&=-i\frac{\omega_1^2-\omega_2^2}{2\omega_+}\delta f_{12}
 - ( f^{0}(\omega_1) v_1- f^{0}(\omega_{2}) v_2)\cdot \Sigma_{12}+ \C_{12},
\label{S_flav_approx}
\end{align}
with $S_{ii}^\textsc{cp}= F_i  (\partial_{k_z}f_0(\omega_i))$ the classical CPC force ({\it cf} \cref{clas_force} in the previous subsection, where the equivalent force was found for the fermion system).
Any possible CP asymmetry can only flow into the diagonal modes if the collision term mixes the diagonal and transverse modes.

The above order counting was questioned in Refs.~\cite{Cirigliano:2011di,Cirigliano:2009yt} where it was argued that $\delta f_{12} =\O(\eps_w^0)$. The solution of \cref{S_flav_1} is schematically $\delta f_{12} \propto \int \dd z \, \Sigma$, with $\Sigma = \O(\eps_w)$, but the spatial integration over the source gives a length scale $L_w \sim \O(\eps^{-1}_w)$. This suggests that the full set of eqs. in (\ref{S_flav_0}) should be solved, which allows for a much more efficient transfer of CP asymmetry to the diagonal modes.
The CP asymmetry is maximized for $L_{\rm osc} \lesssim L_w$, or equivalently $\eps_{\rm osc}\lesssim \eps_w$, as in the opposite limit fast oscillations wash out the flavor content. Indeed, with this power counting in mind, Refs.~\cite{Cirigliano:2011di,Cirigliano:2009yt} expand the transport equations around the degenerate mass limit $m_1^2=m_2^2$ for which $\omega_1=\omega_2=\omega_+$, as they are interested in the resonant regime $\eps_{\rm osc}\sim \eps_w$. 
However, we note that in the mass degenerate limit $ ( f^{0}(\omega_{1}) u_1- f^{0}(\omega_{2}) u_2) = \eps_{\rm osc}$, and the source term \cref{source_flavor} is $\O(\eps^2)$. 
Hence, to properly describe the resonance regime, one needs to derive the kinetic equation to 2nd order in the gradient expansion, as there might be other terms contributing at $\O(\eps^2)$ as well.

In a realistic model, flavor dynamics may need to be included for a precise determination of the BAU. The flavor source induces the largest asymmetry in the resonance regime $L_w \sim L_{\rm osc}$, where it may even dominate over the semi-classical source. Several things need to be done to better understand the flavor source, especially in the resonance regime, and make it applicable to realistic EWBG setups. 
As the power counting above showed, a consistent treatment would require to analyze the kinetic equation to second order in the gradient expansion, i.e. one order higher than the current approaches. This would also require extending the treatment beyond the exact mass degenerate limit.  It would be interesting to see how the mechanism can be implemented in realistic EWBG set-ups that go beyond the bosonic toy model discussed here.  In this context, it would be extremely useful to extend the formalism to fermionic systems. A final item on the wish list would be to include thermal corrections.  The thermally generated effective mass terms for fermions connects fermions with the same chirality, which does not happen at tree level, and thus may have quantitatively new effects. The inclusion of thermal widths and masses for fermions is done in the VIA approach. As discussed in the next subsection,  thermal corrections cannot generate a source term at leading order in the gradient expansion. The questions is what happens at higher order in the gradient expansion; can the fermion chiralities act as different flavors and induce a flavor source?


\subsection{The vev-insertion expansion (VIA)}\label{sec:via}

The vev-insertion approximation treats the space-time-dependent mass term as a perturbation, which comes down to a Taylor expansion of the KB equations with the Higgs vev as expansion parameter. Although the VIA-approximation already appeared in early works on EW baryogenesis \cite{Huet:1995mm, Huet:1994jb,Huet:1995sh}, the VIA source term -- which crucially depends on the inclusion of thermal corrections -- was first derived in the closed-time-path formalism in \cite{ Riotto:1998zb, Riotto:1995hh}, and subsequently generalized to include a CP-conserving relaxation rate in \cite{Lee:2004we}.  
Ref.~\cite{Postma:2019scv} computed the next-to-leading order corrections to the source and relaxation rate in the VIA expansion.
The authors of Ref.~\cite{Carena:2000id} tried to avoid the VIA expansion, but their framework is not based on a first-principle CTP derivation, but rather uses a phenomenological approach.

The putative VIA source hinges on a misalignment between the propagation and interaction states induced by thermal corrections. For example, left- and right-handed quarks/leptons receive different thermal corrections, since the weak sector of the SM is chiral; the mass eigenstates then do not have definite chirality, and thus differ from the interaction eigenstates. In this respect, the name `VIA source' is a bit of a misnomer, as the VIA  expansion is not essential; in fact, the vev-dependent masses can be resummed \cite{Kainulainen:2021oqs,Postma:2022dbr}. 
Although these recent works have shown that the VIA source term vanishes at leading order in the gradient expansion, as we will discuss in \cref{sec:viaVanish}, it remains an open question whether thermal effects can induce non-trivial flavor dynamics and generate a non-zero source term at higher order in the gradient expansion.  

Initial criticism of the VIA source questioned the validity of the expansion. For example, it was noted that the source term diverges for degenerate thermal masses and decay widths\footnote{Although whether the source vanishes or blows up depends on the order that the limits are taken, and the source can be regulated by reinstating the $\eps$-prescription.} \cite{Cline:2020jre}, and suggested that it was the result of an unphysical pinch singularity \cite{Kainulainen:2021oqs}. However, the degenerate mass limit is outside the range of validity of the VIA expansion \cite{Postma:2021zux}, and moreover, if the vev-dependent masses are resummed, the mass eigenvalues are explicitly non-degenerate. 

More problematic is that the source term was not derived in a systematic way.
The original calculation \cite{Riotto:1998zb, Riotto:1995hh} treats the vev-dependent masses as a self-energy correction, which are included using two-particle irreducible diagrams. The corresponding Feynman diagram is shown in the left panel of \cref{fig:VIA}. Thermal self-energy corrections are included in the constraint equation and thus in the Wightman functions, but not in the kinetic equation. Although such an approach \emph{might} indicate the order of magnitude of the CPV effect, a more systematic approach is needed to determine the actual coefficient of the source. It has now been conclusively shown that at leading order in the derivative expansion the VIA source vanishes, and thus that the coefficient is zero.
Ref. \cite{Kainulainen:2021oqs}  pointed out the non-local and non-one-particle irreducible (non-1PI) nature of the self-energy diagrams, and showed that the VIA source vanishes at all orders in the derivative expansion if the interactions with the thermal bath are mediated by a vector-like theory such as QCD. The results were then generalized to chiral theories in \cite{Postma:2022dbr, Postma:2021zux} -- which is important as the original literature found a VIA source that is chiral -- and it was shown that the vanishing of the source term still holds, at least at leading order in the derivative expansion.

In hindsight, it is not surprising that the leading-order source term vanishes, 
otherwise even a constant background could give a non-zero source.
The tricky part was realizing that the original VIA source is a zeroth order effect in the gradient expansion \cite{Kainulainen:2021oqs, Postma:2021zux}. This got obscured, as the original VIA source term was derived in the position space rather than Wigner space basis. 
To identify the CPV part of the source, a derivative expansion was done, but crucially this expansion is not a gradient expansion in the slowly evolving background.

\subsubsection{Vanishing of the leading-order VIA source}\label{sec:viaVanish}

\begin{figure}
     \centering
     \begin{subfigure}[t]{0.4\textwidth}
         \centering
         \includegraphics[width=0.65\textwidth]{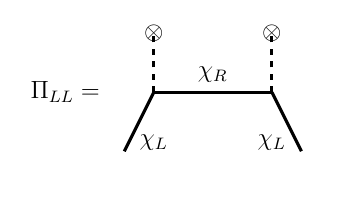}
         \caption{Self-energy $\Pi^\lambda_{LL}$ used in the collision term in the kinetic equation in the original VIA literature \cite{Riotto:1998zb, Riotto:1995hh}. The diagram is not one-particle-irreducible. Thermal loop corrections to the self-energy are neglected in the kinetic equation. }
     \end{subfigure}
     \hspace{1.5cm}
     \begin{subfigure}[t]{0.48\textwidth}
         \centering
         \hspace*{-0.8cm}    
         \includegraphics[width=1.15\textwidth]{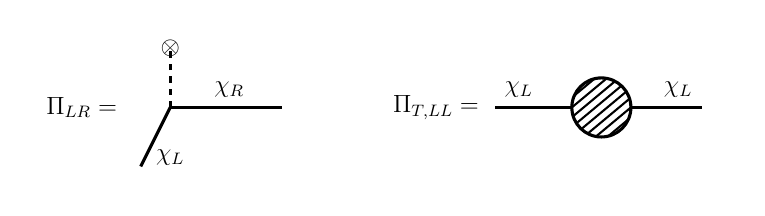}
         \caption{Self-energy $\Pi_{LR}$ used in the kinetic equation in the recent VIA analysis \cite{Postma:2022dbr}; more precisely, $\Pi^h_{LR}$ is absorbed in the mass matrix. The diagram is one-particle-irreducible. In addition loop level thermal corrections $\Pi^h_{T,LL}$ (thermal masses) and $\Pi^\lambda_{T,LL}$ (thermal widths) are included.}
     \end{subfigure}
        \caption{Self-energy diagrams used in the original derivation (plot (a) on the left), and in the recent reevaluation that showed that the leading order VIA source vanishes (plot (b) on the right).}
        \label{fig:VIA}
\end{figure}

In this section we will show that the  VIA source vanishes at leading order in the gradient expansion (but to all orders in the VIA expansion), following the treatment in \cite{Postma:2022dbr, Postma:2021zux}.  The vev-dependent mass contributions are included in the mass matrix -- the mass corrections can be perturbatively expanded in a series of 1PI diagrams -- and the self-energies contain the thermal corrections.  For simplicity we will focus on the two-flavor bosonic system, but after projection onto chirality eigenstates, the equations and analysis for a fermion set-up are qualitatively similar and can be found in \cite{ Postma:2022dbr}.

Consider the same toy model with two bosonic flavors as used for the discussion of the flavor source in the previous subsection. The Lagrangian is given in \cref{M_boson_VIA}. Following the literature we solve the KB equations  \cref{constraint_B,kinetic_B} in the flavor basis; for notational convenience we drop the explicit $f$ subscript. The off-diagonal entries in the mass matrix \cref{M_boson_VIA} depend on the background Higgs vev $v$, and they violate CP in the bubble wall background if complex-valued. The vev-insertion approximation consists of an expansion in the small (compared to the temperature) off-diagonal masses $m^2_{LR} = (m^2_{RL})^*=\O(\eps)$. Thermal mass corrections are absorbed in the mass matrix; to first approximation they correct the diagonal $m^2_{LL}$ and $m_{RR}^2$ masses.  

The assumption in the original VIA literature is that the interactions with the thermal bath are fast, and to a good approximation the bath can be regarded as a collection of particles in flavor eigenstates. The constraint equation is not solved explicitly, but it can be checked that the Wightman functions that are used are solutions of the constraint equations with thermal (loop level)  self-energies  that are flavor diagonal and of the form: 
\be \Pi_{\alpha\beta}^\lambda =\mathfrak{g}^\lambda_{\alpha} \gamma_\alpha \delta_{\alpha\beta},
\qquad \gamma_\alpha = -4ik_0 \Gamma_\alpha,
\label{gamma_VIA}
\ee
with $\alpha,\beta =L,R$ labeling the boson flavors, $\Gamma_\alpha$ the thermal width, and $\mathfrak{g}_{\alpha}^< =\mathfrak{n}_\alpha(k,x),\quad \mathfrak{g}_{\alpha}^> =1+\mathfrak{n}_\alpha(k,x)$. 
As already mentioned above, these thermal contributions to the self-energy are not included in the kinetic equation. To see the vanishing of the VIA source, they should be included consistently in both, as was done in \cite{Postma:2022dbr}, which we follow in this review, and illustrated in the right panel of \cref{fig:VIA}.
The vanishing of the 0th order VIA source term does not depend on the specific structure of the self-energies, and holds true for other choices as well.

The KB equations at leading order in the derivative expansion  (setting $\e^{-i\diamond} =1$) can be solved exactly for the off-diagonal mass matrix, and there is no need to actually perform a vev-insertion expansion. The Green's functions and self-energies are $2\times2$ matrices in flavor space. The solution of the constraint  \cref{constraint_B} with  self-energy \cref{gamma_VIA} is
\begin{align}
  \Delta^\lambda_{LL}&= \frac{1}{\D_+ \D_-} \frac{\gamma_R \gamma_L}{\rho_R}
                  \(\mathfrak{g}^\lambda_L +\mathfrak{g}^\lambda_R \frac{\rho_R}{\gamma_L}
                  |m_{LR}|^4\),\nn \\
  \Delta^\lambda_{LR}&=  \frac{ m_{LR}^2  }{\D_+ \D_-} \(\gamma_R \mathfrak{g}_R^\lambda
                  (k^2-m_{LL}^2) +\gamma_L \mathfrak{g}^\lambda_L (k^2-m_{RR}^2)
                  +\frac12 \gamma_R\gamma_L(\mathfrak{g}^\lambda_R-\mathfrak{g}^\lambda_L)\),
\nn \\
\Delta^t_{LR} &=\frac{ m_{LR}^2  }{\cal{D}_+\cal{D}_-} \Big((k^2-m_{LL}^2) \gamma_R (\mathfrak{n}_R+\frac12)
           +(k^2-m_{RR}^2) \gamma_L (\mathfrak{n}_L+\frac12) - |m_{LR}|^{4} \nn \\
  &\hspace{2cm} + \frac14\gamma_L \gamma_R
(2\mathfrak{n}_R-2\mathfrak{n}_L+1) + (k^2-m_{RR}^2)(k^2 -m_{LL}^2) \Big),
\label{Gresummed_VIA}
\end{align}
 $\Delta^{\bar t}_{LR} = -(\Delta^t)^*_{RL}$, and the other flavor structures ($\Delta_{RR}, \Delta_{RL}$) follow from permutation of the indices.  Here 
\be
\D_\pm = (k^2 -m_{LL}^2 \pm \gamma_L/2) (k^2 -m_{RR}^2 \pm \gamma_R/2) -|m_{LR}|^{4}.
\ee
The poles in the denominator $\D_+, \D_-$ of the resummed propagators are shifted by the off-diagonal mass squared $|m_{LR}|^2$, and there is no divergence in the limit of equal diagonal (thermal) masses $m^2_{LL} =m^2_{RR}$ and vanishing decay widths $\gamma_\alpha \to 0$.

The number current can be isolated by integrating over the Wightman functions \cite{Riotto:1998zb} 
\be
 \frac 1 2\partial_\mu \int \frac{\dd^4 k }{(2\pi)^4} ik^\mu \(\Delta^>(k,x)+\Delta^<(k,x) \)
 =- i\langle\phi^\dagger(x)\overset{\leftrightarrow}{\partial^\mu} \phi(x) \rangle =
- \partial_\mu \langle j^\mu(x)\rangle.
 \ee
 Note that the integrand is just the left-hand-side of the kinetic \cref{kinetic_B}.  The current is thus non-conserved if the integration  gives a non-zero result, and we identify this with the source term. The integration over 4-momenta only picks up the leading-order terms in the gradient expansion, effectively setting $\e^{-i \diamond} =1$, as all other terms vanish upon integration.
Explicitly, the source is $S=-\int\frac{\dd^4 k }{(2\pi)^4} \bar S$ with
\begin{align}
\bar S &=  [M^2, \Delta^>+\Delta^<]+ [\Pi^>+\Pi^<,
                                           \Delta^h] +
                                       \{\Pi^> , \Delta^<\}-\{\Pi^< ,
                                       \Delta^>\}.
     \label{barS}
\end{align}                                        

The KMS relation \cref{KMS} is not satisfied
in the flavor basis if there is mixing, as can be seen from the explicit solution, and the collision term in the kinetic equation can then be non-zero.
Let's calculate the flavor diagonal source $\bar S_{LL}$, substituting the solutions \cref{Gresummed_VIA} into the source \cref{barS}. The commutator with $\Delta^h$ vanishes, while the two other terms yield
\begin{align}
[\delta M^2,  (\Delta^>+\Delta^<)]_{LL} &=2 | m_{LR}|^4\frac{ \gamma_L\gamma_R (\mathfrak{n}_L-\mathfrak{n}_R)}{\D_+ \D_-} ,\nn \\
  \{\Pi^> , \Delta^<\}_{LL}-\{\Pi^< ,
\Delta^>\}_{LL}&=- 2 | m_{LR}|^4\frac{ \gamma_L\gamma_R (\mathfrak{n}_L-\mathfrak{n}_R)}{\D_+
            \D_-} .
            \label{barS_terms}
\end{align}
Adding them up, it follows that the source term cancels exactly. It can be checked explicitly that the cancellation also holds for the flavor off-diagonal part.  The end result is thus that the full VIA source term, resummed over all vev insertions, vanishes at leading order in the derivative expansion:
\begin{align}
\bar S =0.
\end{align}

We can compare this with the derivation used in the original VIA papers  \cite{Riotto:1995hh,Lee:2004we}.  The problem with that approach does not stem from a pathology in the vev-insertion expansion itself (as long as it is limited to its domain of validity), but instead from using non-1PI diagrams which obscures a proper counting of all contributions, and from not consistently including the thermal corrections in the self-energy. The self-energy is expanded to 2nd order in VIA, as indicated by the subscript, and taken as
\be \Pi^\lambda_{\alpha\alpha,(2)} = -m^2_{\alpha\beta} \Delta^\lambda_{\beta\beta, (0)}m^2_{\beta\alpha},
\ee
where the Wightman functions are the 0th order VIA expressions, which are diagonal in flavor space.
The diagonal source
originates from the collision term in $\bar S$ as all other terms vanish. To directly compare with the position space results in the literature \cite{Riotto:1998zb, Riotto:1995hh}, an inverse Wigner transform can be done
\begin{align}
\bar S_{LL}^{(2)}\big|_{\rm usual} &  = 
\{\Pi_{LL,(2)}^> , \Delta_{LL,(0)}^<\}-\{\Pi_{(LL),(2)}^< ,
                                                      \Delta_{LL,(0)}^>\} \nn \\ &=-|m_{LR}|^4
                                     \(\{ \Delta_{RR}^>,\Delta_{LL},(0)^<\}  -\{
                                     \Delta_{RR}^<,\Delta_{LL,(0)}^>\}\)
   =- 2 | m_{LR}|^4\frac{ \gamma_L\gamma_R (\mathfrak{n}_L-\mathfrak{n}_R)}{\D_+
            \D_-} . \label{eqn:colsource}
\end{align}
This is the same expression for the collision term as \cref{barS_terms} before, except that now there is no additional term to cancel against.  The original, non-systematic approach gives the same parametric dependence but it cannot give the correct coefficient, and it thus fails to see the vanishing of the source.

\subsection{Summary of \cref{sec:CP}}

This section has given a detailed review of the derivation of the source terms in the transport equations. We have decided to include many technical details, as results are scattered over the literature, often using different notation as well, making it hard (and time-consuming) to piece it all together. Because of this, the discussion has been very technical. For those mainly interested in the conclusions, we summarize the important results.

\begin{itemize}
\item The VIA source term vanishes at leading order in the gradient expansion. Whether thermal corrections can lead to non-trivial flavor dynamics at higher order has not been fully worked out yet. The VIA source terms should  not be used for the calculation of the CP-asymmetry, and conclusions drawn in the literature on the viability of baryogenesis models based on the VIA source, including work \cite{deVries:2017ncy, deVries:2018tgs} by the authors, should be reconsidered. 
\item The best understood CPV source is the semi-classical/WKB source \cref{clas_source}, which is a consequence of the changing  mass parameters across the bubble wall. This is an effect that is always present for fermions, and can arise for a single fermion with a complex mass at 2nd order in the gradient expansion.  In contrast, 
for a single (complex) boson, no CPV force is generated at this order.
\item Flavor dynamics and flavor oscillations can add to the source term, and effects are enhanced in the resonance regime. The flavor source has only been calculated in toy models, for fermions only away from the resonance, and the resonance regime is not fully understood yet (it may require including higher-order gradient corrections for an accurate description). 
\end{itemize}

It would be interesting to derive the source term including all thermal and flavor effects for physically well-motivated systems. Lacking these results, the safest bet is to calculate the CP-asymmetry with the semi-classical source. For a hierarchical (thermal) mass spectrum flavor effects are suppressed and this should give a good approximation, while for a more degenerate spectrum the classical source term gives a lower bound on the asymmetry (barring accidental cancellations) as its effect is always present. The discussion in this chapter should make it clear that the theoretical understanding of the CP-violating source is not yet satisfactory and requires more study. 

In the next section we discuss how to solve the Boltzmann equations with the classical CPC and  semiclassical CPV source terms. 
b
			
    \newpage
    \section{Boltzmann equations for the CP asymmetry and bubble wall velocity}\label{sec:bubbles}

In the previous section we have seen how to derive the evolution equations for the Wightman functions in the non-equilibrium CTP approach. Integrating the kinetic equations over energy, a set of quantum Boltzmann equations for the distribution functions for  particles $f$ and antiparticles $\bar f$ is obtained. In this section we review the various approaches to solving the Boltzmann equations, which in general are a set of complicated partial integro-differential equations. We will focus on the fermionic set-up with  CP-conserving (CPC) classical and CP-violating (CPV) semi-classical source terms, as discussed in \cref{sec:semiclassical}, and neglect any flavor-mixing effects. 

The Boltzmann equation for the particle distributions in the presence of a bubble wall moving in the $z$-direction is
\begin{equation}
	\partial_t f_i + v_z \partial_{z} f_i + F_i\partial_{k_z}f_i =  -C_i[f_j]\,,\label{eq:Boltzmann}
\end{equation}
with $v_z=k_z/\omega$ and $\omega=\sqrt{k^2+m^2}$ the energy, and $i,j$ species labels.
We focus on a single species with negative helicity, and for notational convenience we have dropped the  helicity indices.
The force can be split into the CPC  and CPV  contributions $F_i = F_i^\textsc{cp}+F_i^\textsc{cpv}$.  The force term  was derived from the CTP formalism in the previous section, see \cref{clas_force,helicity}, where we specialized to the wall frame (and the time-derivative in the Boltzmann equation is absent).  The classical force is the same for particles and antiparticles, whereas the semiclassical force flips sign for the conjugate antiparticle density $\bar f$.  We have reinstated the collision term $C_i$  on the right-hand-side, where we have chosen the convention for the sign consistent with e.g. Refs~\cite{Laurent:2022jrs, Dorsch:2024jjl}.
In principle, the collision term $C_i$ can be derived from the KB equations as well \cite{Prokopec:2004ic}, but in practice the interaction rates are determined from kinetic theory. The source terms drive the system out of equilibrium as the bubble passes by, while the plasma interactions, encoded by the collision term, relax it back to thermal equilibrium. 

As the collision term contains phase space integrations over distribution functions, see \cref{C_integral} below for the explicit expression, this makes the Boltzmann equation an integro-differential equation that is difficult to solve even numerically. To deal with this, the distribution function is split into a (local) equilibrium piece plus fluctuations, as we did in the previous section \cref{f0s}, and then the collision term is linearized in the perturbations $\delta f$. As the CPV source is higher order in the derivative expansion and much weaker than the CPC classical source, the equations for the CPC and CPV perturbations decouple at leading order in the gradient expansion, and they can be solved independently. The Boltzmann equations in the wall frame become
\be
\[v_z \partial_{z}+  F_i^\textsc{cp} \partial_{k_z}\]\delta f_i^X=S_i^X -C_i^X[f_j],
\label{eq_fluc}
\ee
with $X=\;$CP, CPV for the respective perturbations, and 
\be
F_i^\textsc{cp}= -\frac{\partial_z m_i^2}{2\omega}.
 \label{eq:forcevw}
\ee
The source terms are given in \cref{clas_source,helicity} in the wall frame. As the CPC and CPV perturbations are decoupled by taking  the $f \pm \bar f$-combinations of the Boltzmann equation for particles and anti-particles, the collision term is $C_i^X[f_j] = C_i[f_j] \pm C_i[\bar f_j]$, with plus (minus) for the CPC (CPV) combination.

The CPC perturbation enters the calculation of the bubble wall velocity. The CPV perturbation describes the chiral asymmetry, i.e. a net number of left-handed quarks/leptons minus left-handed anti-quarks/leptons,
relevant for baryogenesis, which biasses the sphaleron transitions, see \cref{sec:sphalerons}. As the two calculations decouple, the bubble wall velocity can be determined first, and imposed as an external parameter on the baryogenesis calculation. It is an important input, as it strongly affects the predicted value of the baryon asymmetry (see e.g. \cite{Bodeker:2004ws, Fromme:2006cm,Cline:2020jre, Cline:2021dkf, Cline:2017qpe,Bruggisser:2017lhc,  DeVries:2018aul, Dorsch:2021ubz}).

This section reviews the calculation of the bubble wall velocity and of  the CP-asymmetry. 
Although the linearization in $\delta f$ greatly simplifies the collision terms, the Boltzmann equations are still a set of PDEs that are difficult to solve. 
The next step is then to take a specific Ansatz for the perturbations, and work with moments of the Boltzmann equations. An alternative approach is to expand the deviation from equilibrium on an orthogonal basis of functions. 
As reviewed in this section, various Ans\"atze for the parameter dependence of $\delta f$ exist.
We will comment on how to deal with supersonic bubble wall velocities, a long-standing issue that was recently resolved.

The system can be analyzed in the wall frame, in the plasma frame, or in a covariant (frame-independent) formalism. We will use the latter two approaches in this section, following the choices made in the literature under review (to minimize sign errors), but note that the switch between these last two is particularly simple -- see the discussion in \hyperlink{box2}{Box 2} below.

\subsection{Bubble wall velocity}\label{sec:Bubblevw}

The bubble wall velocity is determined by an interplay between the outward pressure from the vacuum energy release and the interactions with the particles in the plasma, which can provide friction and hydrodynamic backreaction. The wall velocity is typically determined by assuming that the system reaches a steady state, where the outward and inward pressure exactly balance.  
In the  foundational works~\cite{Moore:1995ua, Moore:1995si} a procedure for the computation of the wall velocity was developed. We will first outline the computation of $v_w$ in the approach of~\cite{Moore:1995ua, Moore:1995si}, and then comment on more recent improvements. 

For consistency of notation we denote the energy by $\omega$, but note that in the literature the use of $E$ is prevalent. In this subsection on the bubble wall velocity  $\delta f = \delta f^\textsc{CP}$ is the CPC fluctuation.

\subsubsection{Set-up for the calculation}

For definiteness, consider a phase transition in which only the Higgs field obtains a vev; the equations can be straightforwardly generalized to a multifield set-up (see e.g. \cite{Laurent:2022jrs}).
The equation of motion for the classical Higgs field $\phi$ in the presence of a thermal bath is 
\begin{equation}
	\Box \phi + V'(\phi) + \sum_i \frac{dm_i^2}{d\phi} \int \frac{d^3k}{(2\pi)^3 2\omega} f_i(k,x) = 0,
\end{equation}
where the sum is over all plasma particles. Decompose the distribution functions into
an equilibrium and a non-equilibrium part as $f(k,x) = f^0(k,x)+\delta f(k,x)$, the equation can be rewritten in terms of the fluctuations as
\begin{equation}
	{\rm EOM} \equiv \Box \phi + V_{T}'(\phi) + \sum_i \frac{dm_i^2}{d\phi} \int \frac{d^3k}{(2\pi)^3 2\omega} \delta f_i(k,x) = 0,
    \label{EOM}
\end{equation}
with $V_T$ the effective potential including thermal corrections in \cref{eq:VT}. 
We identify the last term as friction from out-of-equilibrium particles, and the most significant contributions come from the heaviest particles which have the largest coupling to the Higgs field. 

The distribution functions evolve according to the Boltzmann equation in \cref{eq:Boltzmann} with classical force term \cref{eq:forcevw},
causing deviations from equilibrium.\footnote{In \cite{Moore:1996bn, Moore:2000wx} the gauge bosons instead follow classical/overdamped evolution, but this approach is not commonly followed in the wall velocity literature.} 
If particle $i$ has a CP-violating mass, $m_i$ denotes the absolute value of its mass.
In the previous section, we have seen how to derive the Boltzmann equations and this leading-order (in the gradient expansion) CPC classical source term from first principles, {\it cf} \cref{Boltz_fermion} and Refs.~\cite{Blaizot:2001nr, Konstandin:2014zta}. 
The CPV force is higher order in the derivative expansion and therefore neglected. The collision term in the Boltzmann  equations relaxes the distribution function back to equilibrium. It encodes particle scatterings, and can be written for particle $i$ as \cite{Arnold:2000dr}
\begin{align}
	C_i[f] = \sum_{jkl} \frac{1}{2} \int \frac{d^3k_2 d^3 k_3 d^3k_4}{(2\pi)^9 2 \omega_2 2 \omega_{3} 2 \omega_{4}} & |\mathcal M_{ij\rightarrow kl}(k_1,k_2; k_3,k_4)|^2 (2\pi)^4 \delta^4(k_1 + k_2 - k_3 - k_4)  \nonumber \\
		& \times \left( f_1 f_2 (1 \pm f_3)(1 \pm f_4) - f_3 f_4(1 \pm f_1) (1 \pm f_2) \right),
\label{C_integral}
\end{align}
where the upper sign is for fermions, and the lower sign for bosons. $M_{ij \rightarrow kl}$ denotes the matrix element for particles $i,j$ scattering into $k,l$.
The momenta $k_{1,2,3,4}$ correspond to particles $i,j,k,l$ respectively. The collision terms are typically computed at leading log accuracy \cite{Moore:1995si, Arnold:2000dr, Kozaczuk:2015owa, Dorsch:2021nje}.\footnote{See \cite{Arnold:2003zc, Jeon:1994if} for a computation of transport coefficients with full leading order collision terms. To our knowledge, no computation of the wall velocity or baryon asymmetry with full leading-order collision terms has yet been performed.} 
As remarked above, the non-linear dependence on $f$ in the collision term makes  \cref{eq:Boltzmann} very challenging to solve in general, and therefore an approximation scheme needs to be employed. A common approach is to linearize the collision terms in the fluctuations $\delta f$. In the early works \cite{Moore:1995ua, Moore:1995si} the so-called fluid Ansatz was made for $\delta f$, which we discuss in \cref{sec:fluid_ansatz} below.

The perturbations are sourced by the mass change due to the wall that passes by. 
Using the kink Ansatz for the wall profile \cref{bubble_profile},
the wall velocity and wall width $L_w$ are then determined by minimizing two moments of the scalar equation of motion in \cref{EOM}
\begin{equation}
	\int {\rm EOM} \,\phi' \,dz =0, \qquad 	\int {\rm EOM} \,\frac{z}{L}\phi'\, dz =0\,.
\end{equation}
These conditions encode, respectively, that the total pressure on the wall should be zero and that the wall does not get stretched. In Ref. \cite{Moore:1995si}, $v_w$ is determined in the SM with a light Higgs mass with and without the kink Ansatz \cref{bubble_profile} finding differences in $v_w$ below $10\%$.

\subsubsection{Fluid Ansatz}
\label{sec:fluid_ansatz}

We will work with the covariant notation introduced in \cite{Dorsch:2024jjl, Dorsch:2021ubz}. 
The original results in \cite{Moore:1995ua,Moore:1995si} were derived in the plasma frame; this approach is recovered by choosing the fluid velocity vector appropriately. The starting point is the relativistic Boltzmann equation (for simplicity we drop the species index) in frame-independent form
\be
k^\mu \partial_\mu \delta f =  -C[\delta f] +  S[f^0].
\label{BE_rel}
\ee
Compared to \cref{eq_fluc} a factor of energy has been absorbed in the definition of the collision and source term on the right-hand-side.
When taking expressions for the rates and the collision term from the literature, it is important to check the normalization that is used. See in this context also the discussion below \cref{Znormalization}.
The $F^\textsc{CP}\partial_{k_z} \delta f$-term is dropped, which is higher order in the perturbations.  In the fluid Ansatz the perturbations $\delta f$ are expanded in powers of momenta. 
Truncating this expansion at first order, as is done in
\cite{Moore:1995ua,Moore:1995si}, the fluid Ansatz reads 
\begin{equation}
	f = \frac{1}{e^{(k^u u_\mu +\delta )/T} \pm 1},
        \qquad \delta = -\left(\mu+ k^\mu u_\mu \delta T/T + k^\mu \delta u_\mu  \right),
\label{fluid_ansatz}
\end{equation}
with $u_\mu$ the fluid velocity. In the absence of perturbations, $f =f^0$ is the Bose-Einstein distribution for bosons or Fermi-Dirac distribution for fermions in frame-invariant notation \cref{f0}.
The three independent perturbations $(\mu,\delta T,\delta u_\mu)$  are the chemical potential, temperature fluctuation, and velocity perturbation respectively, corresponding to the Lagrange multipliers for particle number, energy, and momentum.  Since $u_\mu \delta u^\mu=0$, the velocity perturbation corresponds to only one degree of freedom which can be parameterized as  $\delta u^\mu= \bar u^\mu \delta v$, with $\bar u$ a space-like unit vector perpendicular to the fluid velocity, see \hyperlink{box2}{Box 2}. 
An advantage of this approach is that the perturbations have a clear physical interpretation. The perturbations can be split into $\delta =\delta_{l} +\delta_{h}$, where the background piece $\delta_{l}$ are the fluctuations of the light fields, and $\delta_h$ the fluctuations of heavy fields relative to the light ones. The light fields only couple weakly to the Higgs field, and their source term can be neglected, which implies $\mu_{l}=0$. 
However, the non-equilibrium background fluctuations $\delta_l$ are still generated via collisions with the heavy fields.

\begin{tcolorbox}[float*=t,hypertarget=box2,before skip=0.5cm,after skip=1cm,colback=blue!5!white,colframe=blue!75!black,toptitle=0.1cm,bottomtitle=0.1cm,title= Box 2: Frames]

We work in the approximation, valid for large bubbles, that the bubble wall can be approximated by a planar wall. We choose the convention, following most of the literature, that the bubble wall is moving along the negative $z$-direction with velocity $v_w$, that is $\phi_b=\phi_b(v_w t +z)$. The physics is of course convention independent, which can be made explicit by working in a frame-invariant way \cite{Dorsch:2024jjl, Dorsch:2021ubz}.
For this, introduce the plasma velocity $u^\mu$, the space-like vector perpendicular to the fluid velocity (in the plasma frame pointing in the direction of the moving bubble wall) $\bar u^\mu$ via $\bar u^\mu u_\mu =0$ with $\bar u^\mu \bar u_\mu =-1$. For a steady-state planar wall the background then becomes a function of $\xi$, $\phi_b =\phi_b(\xi)$ with $\xi \equiv x^\mu w_\mu$ and  $w^\mu = \gamma_w(v_w u^\mu- \bar u^\mu)$.  The signs are defined such that the wall is moving in the negative $\xi$-direction, that is, the symmetric phase corresponds to $\xi < 0$.  

We can define two frames that are particularly useful for calculations and that are both used in the literature.  In the \emph{wall frame} the bubble is at rest, whereas in the \emph{plasma frame} the plasma infinitely far from the bubble is at rest. For small bubble wall velocities, it is a good approximation to take the plasma at rest (also near the bubble) in the plasma frame, and the two frames are related via a Lorentz boost with velocity $v_w$.  In this approximation, the relevant quantities in the plasma frame and wall frame  are then 
\begin{align}
&{\rm plasma} \;\; {\rm frame}: &  u^\mu &=(1,0), & \bar u^\mu &=(0,1), & w^\mu &= \gamma_w(v_w,-1), & \xi &=\gamma_w(v_wt+z), \nn \\
&{\rm wall} \;\; {\rm frame}: &  u^\mu &=\gamma_w(1,v_w), & \bar u^\mu &=\-\gamma_w(v_w,1), & w^\mu &= (0,-1), & \xi &=z, 
\label{frame_vectors}
\end{align}
with $\gamma_w =(1-v_w^2)^{-1/2}$ the Lorentz factor for the transformation between the two frames. Here we used the notation $x^\mu=(t,0,0,z) = (t,z)$, as the planar wall dynamics is effectively $(1+1)$-dimensional.  To switch from the covariant formalism to the plasma frame we can thus simply set $\xi=z$.

The plasma velocity is affected by the presence of the bubble, and for relativistic velocities it is crucial to take this into account, as discussed in \cref{sec:singularity}. The plasma velocity in the plasma frame then is $u^\mu =\gamma_{\rm pl}(1,v_{\rm pl}(z))$ with $v_{\rm pl}(\pm \infty )=0$ and $\gamma_{\rm pl} =1/\sqrt{1-v_{\rm pl}^2}$.

\end{tcolorbox}

For a steady-state planar wall the background only depends on $\xi$, (see \hyperlink{box2}{Box 2} for definitions), with $\xi <0$ corresponding to the symmetric phase in front of the bubble.
Using $\partial_\mu=w_\mu \partial_\xi$, the linearized Boltzmann equation becomes 
\be
(-f^{0'}) k^\mu w_\mu \partial_\xi \left(\mu+ k^\mu u_\mu \delta T/T + k^\mu \delta u_\mu  \right)
=   -C[\delta f] +  S[f^0],
\label{BE_fluid}
\ee
with $f^{0'}$ defined in \cref{df0}.
To get three independent equations for the three perturbations, the perturbed Boltzmann equation is integrated over $\int \dd^3 k \, (w/\omega)$ with respective weight factors $w=(1,u_\nu k^\nu ,\bar u_\nu k^\nu )$. The integration over an odd number of factors $(k^\mu \bar u_\mu)$ vanishes.
The coefficients are defined through
\be
\int \frac{\dd^3k}{\omega}(-f^{0'}) (k^\mu u_\mu)^m (k^\nu \bar u_\nu)^n = 4\pi T^{m+n+1} \frac{\bar c_{[m,n]}}{n+1},
\ee
with $\bar c_{[m,n]}=0$ for $n$ odd. Note that the integration measure, the weights, as well as the Boltzmann function is frame-invariant, and thus so are the coefficients $c_{m+n+1}$ -- they are most easily calculated in the plasma frame.  
For massless particles the coefficients can be found analytically and expressed in terms of 
\be
c_a = T^{-a} \int \dd k\, k^a (-f^{0'}) = \frac1{2\pi^2} a!\, \zeta_a (1-2^{1-a}), \quad (a\geq 2),\label{eq:c}
\ee
where we used that $(\sqrt{3}k^\mu \bar u_\mu)^2 =3k_z^2 =k^2$ and $\zeta_a$ denotes the Riemann zeta function. Explicitly $\bar c_{[1,0]}= c_2$ and $\bar c_{[2,0]} = 3 \bar c_{[0,2]} = c_3$.
For example, the integration of \cref{BE_fluid} with weight $w=1$ gives 
\be
\int \frac{\dd^3 k}{\omega} (-f^{0'}) k^\mu \gamma_w(v_w u_\mu-\bar u_\mu) \partial_\xi \left(\mu+  k^\nu u_\nu \delta T/T + k^\nu \delta u_\nu  \right)
=\gamma_w T^3 (4\pi) \( c_2 v_w \partial_\xi \frac{\mu}{T} + c_3 v_w\partial_\xi \frac{\delta T}{T} - \frac{c_3}{3}\partial_\xi \delta v\).
\ee
The three moment equations can be written in compact form in terms of the perturbation vector $q=(\mu/T,\delta T/T,-\delta v)^T$ 
\be
A \cdot \partial_\xi q +  \Gamma \cdot q = \bar S\,,
\label{eq_A3}
\ee
with
\be
A =\gamma_w\( \begin{array}{ccc}
v_w c_2 & v_w c_3 & \frac13 c_3 \\
v_w c_3 & v_w c_4 & \frac13 c_4\\
\frac13 c_3 & \frac13 c_4 & \frac13 v_w c_4
\end{array} \right).
\ee
The matrix of rates $\Gamma$ and source vector $\bar S$ are given in terms of moments of the (linearized) collision term and source term: $\Gamma_{ij} q_j=(4\pi T^{3})^{-1}\int \dd^3 k\, (w^i/E) C $ and $\bar S^i= (4\pi T^{3})^{-1}\int \dd^3 k\, (w^i/E) S$. Likewise, moment equations for the perturbations of the light fields can be derived \cite{Dorsch:2024jjl}, which allows to solve $\delta_l$ separately from $\delta_h$. The relevant rates for top quarks and $W$-bosons are listed in Ref.~\cite{Moore:1995si}.

\subsubsection{Singularity at the sound speed}
\label{sec:singularity}

In the fluid approximation as described in the previous subsection, the background $f^0 =f^0(k)$ is taken constant in space-time. 
This approximation works well for small bubble wall velocities but breaks down for larger values. The equations derived in Ref.~\cite{Moore:1995si} feature a singularity when the wall velocity equals the speed of sound $c_s = 1/\sqrt 3$.  The local thermal equilibrium temperature and velocity profile, which can be found from energy-momentum conservation and boundary conditions from hydrodynamics, vary across the bubble wall profile. In the fluid approximation this effect is not regarded as background but rather absorbed in the fluctuations of the light fields $\delta_l$. So far, this is a matter of choice, but the problem is caused  by the linearization of the Boltzmann equations in $\delta =\delta_l+ \delta_h$  \cite{Laurent:2022jrs, Dorsch:2021ubz, Dorsch:2021nje}, which assumes that the background fluctuations $\delta_l$ are small. The background velocity profile can be calculated for the fluid Ansatz \cref{fluid_ansatz}, and on rather general grounds it can be shown to be proportional to $\delta v_l \propto (1-3v_w^2)^{-1}$ \cite{Dorsch:2024jjl}, which is not a small perturbation for wall velocities approaching the speed of sound.

The singularity at the sound speed can thus be traced to taking the background $f^0$ constant and to linearizing the collision terms in $\delta f$, and it is not resolved by e.g. expanding $\delta$ to higher order in momenta and including more perturbations.
Two different approaches have been developed to obtain non-singular equations for $\delta f$, based on an expansion around a spacetime-dependent background. 
Ref.~\cite{Laurent:2022jrs} abandons the fluid Ansatz of \cref{fluid_ansatz}, and writes the distribution function as 
\begin{equation}
    f(k,x) = f^0(k,x) + \delta f(k,x), \qquad f^{0}(k,x) =\frac{1}{\exp{\left[k_\mu u^\mu(x)/T(x) \right]}\pm 1}\,.
    \label{cheb}
\end{equation} 
Here $u^\mu(x)$ and $T(x)$ now denote the spacetime-dependent local thermal equilibrium plasma velocity and temperature respectively, determined from the equation for energy-momentum conservation. In the wall frame $u^\mu(z) = \gamma_{\rm pl}(z) (1 ,0 ,0 ,v_{\rm pl}(z))$, and the spatial dependence is only on the $z$-coordinate (and $\xi =z$ to translate to the covariant notation).
The fluctuations $\delta f(k,x)$ are kept general, and are expanded in an orthogonal basis of Chebyshev polynomials. 
The advantage of this approach is that the solution does not depend on an Ansatz for the shape of $\delta f$, and the convergence of the solution can be controlled by the number of basis polynomials. The disadvantage is that the physical interpretation of the fluid Ansatz is lost, and moreover the method is computationally more expensive than the one based on the fluid Ansatz. The approach of~\cite{Laurent:2022jrs} has been automated in the package {\tt WallGo} \cite{Ekstedt:2024fyq}, including the computation of the matrix elements and collision integrals. 
This makes the computation of $v_w$ accessible for any model, and this source of uncertainty can thus be alleviated in the computation of the BAU.

Motivated by the approach described just above, Ref. \cite{Dorsch:2024jjl} updated the fluid Ansatz in a similar way, by expanding around the local equilibrium background rather than the constant background used in the original approach:
\begin{equation}
    f_i(k,x) =\frac{1}{\exp{\left[k_\mu u^\mu(x)/T(x) + \delta \right]}\pm 1}, \qquad 
     \delta = -\left(\mu+k^\mu u_\mu \delta T/T + k^\mu \delta u_\mu  \cdots \right),
    \label{fluid_ansatz2}
\end{equation}
where the background velocity and temperature profile are now also obtained from the equations for energy-momentum conservation. Compared to the original fluid Ansatz in \cref{fluid_ansatz}, in the new Ansatz the effect of the light species, $\delta_l$, is absorbed in the evolving background $f^0(k,x)$. Consequently, the collision term is only linearized in the small perturbations of the heavy fields $\delta =\delta_h$. The dots denote perturbations that are higher order in the momentum expansion which, as argued above, are not crucial for the resolution of the singularity at the sound speed.
The improved fluid Ansatz retains the clean physical interpretation of the perturbations, and is computationally more efficient than the expansion in Chebyshev polynomials.  Currently, knowledge of the convergence of the solution in the number of basis functions in \cref{fluid_ansatz2} is lacking.
Results from \cite{Dorsch:2021nje}, obtained using the original fluid Ansatz, \emph{with} linearized background temperature and plasma velocity, suggest that higher order corrections in the momentum expansion in $\delta$ give only small corrections to the friction.

Finally, we mention the parameterization $f_{\pm} (\omega_{s\pm},\vec k,x) = f_{0s\pm}(\omega_{s\pm},\vec k)+ \delta f_s(k,x)$ with $f_{0s\pm}(\omega_{s\pm})$ in \cref{f0s} \cite{Fromme:2006wx,Kainulainen:2021oqs, Cline:2020jre}. The inclusion of the CPV fluctuation, and the moment equations, are discussed in \cref{sec:BE_baryogenesis}. Here we just note that although the plasma velocity is approximated by the wall velocity (in the wall frame), the energy $\omega_{s\pm}$ includes the derivative correction to the dispersion relation, and in this sense the expansion is not around a constant background. As the momentum-dependence of $\delta f_s$ is left unspecified, additional assumptions are needed to close the system of moment equations. The collision terms are linearized in the fluctuations.
Nevertheless, there is no singularity at the speed of sound if non-relativistic corrections to the Boltzmann equations are included \cite{Cline:2020jre}. It is not clear how the background $f_{0s\pm}(\omega_{s\pm},\vec k)$ compares to the local thermal equilibrium backgrounds used in the `Chebyshev' and improved fluid Ansatz approaches \cref{cheb,fluid_ansatz2}.

\subsubsection{Approximations of $v_w$}
\label{sec:appprox_vw}
The computation of $v_w$ is technically and numerically involved.
Even though some of the technical difficulties have been lifted by the automation of the computation in {\tt WallGo}, 
 it is useful to have some simpler (and numerically inexpensive) estimates. 

The computation of $v_w$ greatly simplifies both in the limit of vanishing collision as well as large collision rates in \cref{eq:Boltzmann}. The regime in which particle collisions become negligible is known as the ballistic limit \cite{Moore:1995si, Liu:1992tn, BarrosoMancha:2020fay, Wang:2024wcs, Ai:2024btx}. In this limit interactions are sufficiently weak to allow particles to traverse the bubble wall without scattering, and the Boltzmann equation \eqref{eq:Boltzmann} can be solved exactly.
Assuming that particles are in thermal equilibrium with the plasma far from the wall, the resulting expressions for the friction exerted by transmitted and reflected particles can directly be derived. Crucially, the boundary conditions must account for the spacetime dependence of the temperature and fluid velocity. When such effects are included, the ballistic approximation yields an upper limit on the frictional force and, correspondingly, a lower limit on the wall velocity \cite{Ai:2024btx}.

The opposite is the limit of a very large collision rate. Now the distribution functions remain close to equilibrium, and the Boltzmann equation becomes trivial, as $\delta f \approx  0$. This limit is called `local thermal equilibrium' (LTE). 
In this simplification, only the scalar field equation of motion \cref{EOM} and energy-momentum conservation for the fluid temperature and velocity 
\begin{align}
	T^{30} &= w \gamma^2 v = \tilde c_1, \nonumber \\
	T^{33} & = \frac 1 2 (\partial_z \phi)^2 - V_T(\phi, T) + w \gamma^2 v = \tilde c_2,
    \label{T33}
\end{align}
have to be solved, which are given here in the wall frame. Here, $w$ denotes the enthalpy, and $\tilde c_1$ and $\tilde c_2$ are constants that can be obtained by solving the hydrodynamics equations for a bubble with a constant velocity (the tilde is meant to distinguish the constants from the ones introduced in \cref{eq:c}).
The force that decelerates the wall is a purely hydrodynamic backreaction effect, which originates from the heating of the fluid \cite{Konstandin:2010dm, BarrosoMancha:2020fay, Balaji:2020yrx, Ai:2021kak}.
As LTE corresponds to conservation of entropy, it provides an additional hydrodynamic matching relation, which allows a determination of $v_w$ purely from hydrodynamics \cite{Ai:2021kak, Ai:2023see}.
The LTE result can be understood as an upper bound on $v_w$, as the inclusion of out-of-equilibrium effects will enhance the friction. 
Ref.~\cite{Eriksson:2025owh} demonstrated that this is an unsaturated upper bound, as fluctuations of the scalar field itself will inevitably provide an entropy-generating contribution.
For a given model, the ballistic and LTE approximations nevertheless provide a range of possible wall velocities. To find a more precise estimate requires a solution of the Boltzmann equation.

\subsubsection{Runaway walls}\label{sec:run}
A question that has attracted a lot of attention in the literature, is under which conditions the bubble wall reaches a steady state solution. It was pointed out in Ref.~\cite{Bodeker:2009qy} that the mass gain from particles entering the bubble might not provide sufficient friction to stop the wall from accelerating, possibly resulting in bubble walls which keep accelerating to speeds very close to the speed of light, until they collide with each other. It was demonstrated in Ref.~\cite{Ai:2024shx} that in certain cases the hydrodynamic friction effect can slow down walls that would be considered runaway walls according to the criterion of \cite{Bodeker:2009qy}.
To determine whether the bubble walls really keep accelerating, or whether they reach a terminal (but possibly large) velocity, it is necessary to identify next-to-leading-order (in the couplings) contributions to the pressure, which can become relevant for very fast walls \cite{Bodeker:2017cim, Hoche:2020ysm, Azatov:2020ufh, Gouttenoire:2021kjv, Ai:2023suz, Long:2024sqg}. 
For the purpose of electroweak baryogenesis, this question is less relevant  as the baryon asymmetry vanishes in the limit $v_w\rightarrow 1$ \cite{Cline:2020jre}, as discussed in \cref{sec:supersonic}. This suggests that a computation of $v_w$ including hydrodynamic backreaction and leading-order friction should be sufficient, although a study of the impact of the various approximations made in the computation (such as the leading log approximation of the collision terms, the linearization in the perturbations, or the Ansatz for the bubble wall profile) has not been performed.

\subsubsection{Some examples of the wall velocity and wall width in models with a FOPT}
\begin{figure}[h!]
    \centering
    \includegraphics[scale=0.6,trim=0 0 0 0]{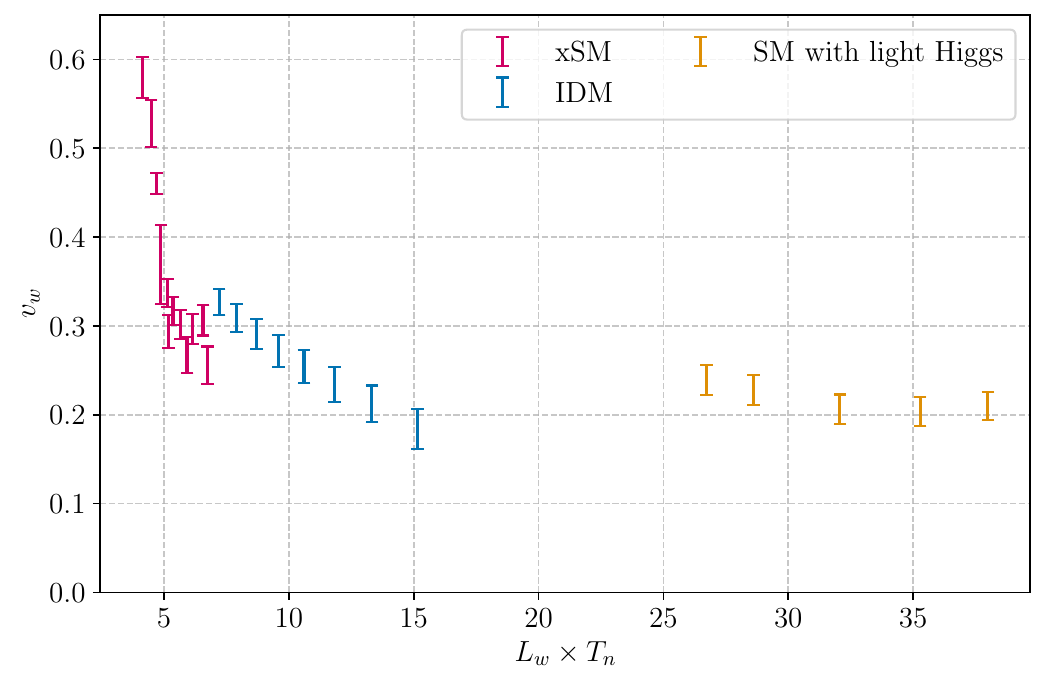}
    \caption{Examples of $v_w$ and $L_w \times T_n$ for the Standard Model coupled to a singlet (xSM), Inert Doublet Model (IDM) and the Standard Model with a (unphysically) light Higgs; determined with {\tt WallGo} \cite{Ekstedt:2024fyq}. See the main text for a description of the benchmark points.
    }
    \label{fig:vwLW}
\end{figure}    
\Cref{fig:vwLW} shows the value of $v_w$ and $L_w$ for some benchmark points in the following three models:
\begin{itemize}
    \item{The Standard Model extended by a $Z_2$-symmetric singlet; for a parameter choice where the phase transition is two-step. $v_w$ is computed for the second step of the PT and $L_w$ corresponds to the wall width of the Higgs profile. The BM points shown here are a small subset of the ones obtained in \cite{Laurent:2022jrs}, and correspond to a singlet mass of $m_s = 85$ GeV and varying portal coupling. }
    \item{The Inert Doublet Model (IDM), implemented as in \cite{Blinov:2015sna}. The benchmark points shown here are obtained from BM1 of \cite{Blinov:2015sna} and by varying the mass of the new heavy scalars $m_A, m_{H^\pm}$.}
    \item{The Standard Model with a small Higgs mass, such that the EWPT becomes first order. The points shown in the graph have a Higgs mass between $0$ and $81 \,{\rm GeV}$. These points were presented in \cite{Ekstedt:2024fyq} to compare against \cite{Moore:1995si}, but of course do not correspond to a realistic BSM scenario.
    }
\end{itemize}
In all cases, $v_w$ was computed with {\tt WallGo} \cite{Ekstedt:2024fyq}, using the linearization of \cref{cheb}. The out-of-equilibrium friction effects from the top quarks and the $W$- and $Z$-gauge bosons are taken into account. The number of basis polynomials in the momentum expansion was $N=11$. The error bars denote the truncation error in the solution to the Boltzmann equation. 

In all benchmark points shown, the friction is sufficient to stop the wall from running away. The wall velocities range from $v_w=0.2$ to 0.55, where the largest values occur in the xSM, where the phase transition is strongest. 
The values reported here should still be taken with some caution. For the IDM, for example, the friction effect of the heavy scalar has not been taken into account, and in all three models -- as common in computations of the wall velocity -- the $W$- and $Z$-bosons are treated as identical. Nevertheless, the benchmark points indicate the typical range of bubble wall velocities.


\subsection{Boltzmann equations and baryogenesis}
\label{sec:BE_baryogenesis}

In this subsection we review how to solve the Boltzmann equation in \cref{eq:Boltzmann}, with the CPV force $F^\textsc{CPV}$ derived in \cref{sec:CP} and given in \cref{clas_force}, to find the CP asymmetry that feeds into the baryon yield. For the CP asymmetry we are interested in helicity eigenstates rather than spin eigenstates and we use the replacement \cref{helicity}; for massless fermions in the symmetric phase the negative helicity state ($s_h=-1$) corresponds to the left-handed chiral state. 

The distribution function can be split into an equilibrium piece plus fluctuations, see \cref{df0}. We can use e.g. the fluid Ansatz for $\delta f$ in \cref{fluid_ansatz}, or the generalization \cref{fluid_ansatz2}, to accurately calculate the CPV part of the distribution functions. 
However, only the sphaleron coupling to the chemical potential is known, see \cref{sec:sphalerons}, and it is unclear how the higher moments of the perturbation affect the final BAU. As in practice the chemical potential is all that is used to obtain the baryon yield, it is tempting to include it as the sole perturbation. However, this does not give a consistent solution. As argued in Ref. \cite{Cline:2000nw}, 
with only the $\mu$ perturbation the distribution function describes a species in kinetic, but not in chemical equilibrium. As the force acts in the direction perpendicular to the wall, this breaks the isotropy of the system, and the total perturbation will be anisotropic in the momenta.  Hence, in the prevalent `generalized velocity perturbation' approach, as described in \cref{sec:gen_velocity} below, on top of the chemical potential a generic `velocity perturbation' is introduced.

\subsubsection{Generalised velocity perturbation}
\label{sec:gen_velocity}

The most commonly used parameterization of the distribution function for the calculation of the CP-asymmetry (as we focus on the CPV perturbation we drop $\delta f^\textsc{cp}$)  \cite{Fromme:2006wx,Cline:2020jre,Kainulainen:2021oqs} is 
\begin{align}
f_{s\pm} (\omega_{s\pm},\vec k,z) &= f^0_{s\pm}(\omega_{s\pm},\vec k) \pm \delta f^\textsc{cpv}_s( k,z) \nn \\
&= (\e^{\beta(\gamma_w(\omega_{s\pm} +v_w k_z)\mp \mu( k,z))}+ 1)^{-1} \pm \delta f( k,z),
\label{gen_vel}
\end{align}
where we used the Fermi-Dirac distribution for fermions. The energy is $\omega_{s\pm} =  \omega \mp s\frac{m^2\theta'}{2\omega \omega_{z}}$, as in \cref{energy_semi}, with $\omega =\sqrt{k^2 +m^2}$ and the plus (minus) for particles (antiparticles); that is, the energy of the background field includes the first order gradient corrections. 
In the right-hand-side the fluctuation is denoted as $\delta f^\textsc{cpv}_s$. On the 2nd line we rewrote the fluctuation $\delta f^\textsc{cpv}_s( k,z)$ in terms of $(\mu,\delta f)$. The condition $\int\dd^3k \, \delta f=0$ then defines $\mu$ as the chemical potential that parametrizes perturbations in the local number density. In the gradient expansion $f^0_{s\pm} = f^0(\omega) + \gamma_w \beta f^{0'}(\omega_{s\pm}-\omega) + \O(\partial_z^2)$  as in \cref{f0s}, with $f^0 = f_{s\pm} (\omega,\vec k)$ the distribution for $k^0=\omega$.

In this subsection we follow the analysis of  Ref. \cite{Cline:2020jre} that generalizes earlier treatments \cite{Cline:2000nw,Fromme:2006wx} to relativistic wall velocities. The starting point is the Boltzmann equation for the fluctuations \cref{eq_fluc} with CPV source given (in the wall frame) by \cref{clas_force}.  For now, we focus on a single species with negative helicity, and for notational convenience drop both the species and helicity index. Specializing to the wall frame, and rewriting $\delta f^\textsc{cpv}$ in terms of the perturbations $(\mu,\delta f)$ then gives
\be
L[\mu,\delta f]= -\frac{k_z}{\omega_0}( f^{0'} \partial_z \mu -\partial_z \delta f)
-F^{\rm CP} (\gamma_w v_w  f^{0''} \mu- \partial_{k_z}\delta f ) = S^\textsc{CPV}- C^\textsc{CPV}\,,
\label{Boltz_lin_df}
\ee
with $L$ the Liouville operator, and $f^{0'},f^{0''}$  defined in \cref{df0}.
The CPV source term on the right-hand side is \cref{clas_source}, and the collision term relevant for the CPV equation is further specified below.

We can take moments of the Boltzmann equation to solve for the perturbations. We define the averages
\begin{align}
\langle X \rangle = \frac1{N_1} \int \dd^3 k\, X\,,\qquad [ X ] = \frac1{N_0} \int \dd^3 k\, X f_0\,.\
\label{average}
\end{align}
The normalization factors $N_1 =\int \dd^3 k \,f^{0'}|_{m=0}$ is chosen to simplify the form of the collision term. The average $[..]$ is used to define the truncation scheme, and the normalization  $N_0 =\int \dd^3 k \,f^0$  then brings this scheme into a simple form, as we discuss below. The average defined in \cref{average} is not  Lorentz invariant and the definition of the perturbations are frame dependent. The zeroth and first moment equations are taken as the averages $\langle ... \rangle$ and $\langle (k_z/\omega) ... \rangle$ of the linearized Boltzmann equation in \cref{Boltz_lin_df}. As the energy-momentum dependence of $\delta f$ is not specified, this gives equations for the `velocity perturbations' $u_1 =\langle (k_z/\omega) \delta f\rangle$ and $u_2 = \langle (k_z/\omega)^2 \delta f\rangle$, which are independent variables.

To get a closed set of equations, the assumption is made that one can replace
\be
\langle X \delta f\rangle = \frac1{N_1} \int \dd^3 k\, X \delta f \,\,\to\,\, \langle \frac{k_z}{\omega} \delta f \rangle \, [\frac{\omega}{k_z} X]=
u_1 \( \frac{1}{N_0}  \int \dd^3 k\, \frac{\omega}{k_z} X f^0\),
\label{assumption_av}
\ee
for any $X$ that does not depend on $\delta f$. With this substitution the velocity averages factorize. The specific form of the rule is motivated by the simple and linear relation between the velocity perturbations:  $u_2 =R u_1 =[k_z/\omega] u_1$; this relation defines the coefficient $R$ which can be calculated exactly; $R=-v_w$.
 In addition, it allows to replace $\langle \delta f/(2\omega)^2 \rangle \to \bar R u_1$ in the Boltzmann equation, where the integral expression for $\bar R = [(2k_z\omega)^{-1}]$ can be evaluated numerically. 
The assumption \cref{assumption_av} is ad hoc, and it is not clear how good of an approximation it is; this could be tested for specific choices of $\delta f$ that have an explicit momentum dependence. Recently, the analysis was extended to higher moments of the Boltzmann equation in \cite{Kainulainen:2024qpm}, where it was shown that the CP asymmetry found from higher moment calculations can significantly differ from the 2-moment equations described in this subsection, and an alternative truncation scheme was introduced. It would be useful to compare the truncation scheme against other approaches. In the fluid Ansatz and Chebyshev approach the perturbations are expanded in a momentum expansion respectively in a basis of Chebyshev polynomials, as discussed in \cref{sec:singularity} in the context of the calculation of the bubble wall velocity.

The result of the truncation \cref{assumption_av} is a closed set of equations for the chemical potential and velocity perturbation $u_1$. This can be written in matrix form, in terms of the perturbation vector $q=(\mu,u_1)^T$ :
\be
A \partial_z q + (\partial_z m^2) B q = \bar S - \bar C\,,
\label{result_df}
\ee
with 
\begin{align}
A =\begin{pmatrix}
-D_1 & 1 \\ -D_2 & R
\end{pmatrix}\,, \quad
B= \begin{pmatrix}
v_w\gamma_w Q_1& 0 \\ v_w\gamma_w Q_2 & \bar R
\end{pmatrix}\,, \quad
\bar S= \begin{pmatrix}
S_1 \\ S_2
\end{pmatrix} \,.
\end{align}
The coefficients  are defined through the averages
\be 
D_l = \Big\langle \(\frac{k_z}{\omega}\)^l f^{0'}\Big\rangle, \quad
Q_l = \Big\langle \(\frac{k_z}{\omega}\)^{l-1} \frac{f^{0''}}{2\omega} \Big\rangle, \quad
S_l=\Big\langle \(\frac{k_z}{\omega}\)^{l-1} S^\textsc{CPV}\Big\rangle.
\label{Dl}
\ee
As  the averaged value $\langle X \rangle$ has a mass dimension that is one higher than the mass dimension of $X$, the mass dimension of the various quantities are
\be
{\rm dim}(q) =1, \quad
{\rm dim}(A) =0,\quad
{\rm dim}(B) =-2,\quad
{\rm dim}(\bar S) ={\rm dim}(\bar C) = 2.
\ee
The collision vector is defined analogously to the source term. The components can be expressed in terms of the interaction rates $C_{i1} = (N_0/N_1) \sum_{i,j} \bar \Gamma_i s_{ij} \mu_j/T$ and $ C_{i2}= \bar \Gamma_{\rm tot} u_1 -v_w  \C_{i1} $ with $\bar \Gamma_{\rm tot} =\sum_i \bar \Gamma_i$ the total decay rate.\footnote{We have defined the rates with a minus sign compared to \cite{Cline:2020jre}, to obtain a Boltzmann equation $v_w \rho' - D_{\rm eff} \rho''= ...$ with a minus sign in front of diffusion coefficient $D_{\rm eff} $, and the prime denoting $z$-derivative. This agrees with naive application of Fick's law $\partial_\mu j^\mu = \partial_t n + \nabla \cdot \vec j = v_w n'- D n''$ if we identify $D_{\rm eff} = D$. The explicit solution \cref{ns_semi} only depends on the combination $D \Gamma$, and does not depend on the sign choice.}
Here $i$ denotes that this is the collision term that is added to the equations for particle $i$ with $s_{ij} =1 (-1)$ if species $i$ is in the initial (final) state of the interaction; the subscript $1 (2)$ gives the top (bottom) component of $\bar C$. Explicit expressions for the rates can be found in Refs.~\cite{Cline:2020jre,Cline:2000nw,Fromme:2006wx}.

\subsubsection{Equation for the chemical potential}

\Cref{result_df} can be solved numerically, or using an approximate analytical approach. Here, we first simplify the equations further, and write them as a single equation (per species) for the chemical potential, or equivalently, number density. As the chemical potential is needed as input for the sphaleron equation, this is exactly what we are after. Hence, the goal is to eliminate the velocity perturbation, and this can be done with the additional assumption that the $(\partial_z m^2) B q$-term can be neglected.
This approximation is equivalent to neglecting the $(\partial_z m^2) \partial_{k_z} \delta f$-term in the Boltzmann equation (\cref{eq:Boltzmann}). This same approximation is commonly made in the fluid Ansatz; as $\delta f \propto S^\textsc{CPV} \propto (\partial_z m^2) $, this term is higher order in the derivative expansion. Defining ${\rm eq}_1$ and ${\rm eq}_2$ as the top and bottom row of the matrix \cref{result_df}, taking the combination $-v_w {\rm eq}_1/D_1 + (v_w^2 {\rm eq}_1' +v_w {\rm eq}_2')/(D_1 \Gamma_{\rm tot}) $ gives
\be
v_w \mu'-D_{\rm eff} \mu''  = -C_{\rm eff} + S_{\rm eff}, 
\label{eq_eff}
\ee
with
\be
D_{\rm eff}= \frac{D_2-v_w^2 D_0}{\bar \Gamma_{\rm tot}  D_0}, \quad
S_{\rm eff} = \frac{S_1}{D_0} +\frac{v_w S_1' + S_2'}{ D_0 \bar \Gamma_{\rm tot}}, \quad
C_{\rm eff} =\frac{C_1}{- D_0}.
\label{Deff}
\ee
Here we used that $R=-v_w$, $D_1=-v_w D_0$, and the explicit form for $C_2$ given above.
The numerical results for the baryon asymmetry in Ref.~\cite{Cline:2020jre} indicate that \cref{eq_eff} is a good approximation of the full system \cref{result_df} for thick walls with $L_w T \gtrsim 10$, which is the region of validity of the gradient expansion. 

For small velocities, we can drop the $v_w^2$-term in the diffusion term and $D_{\rm eff}= {D_2}/({\bar \Gamma_{\rm tot} D_0})$. $D_{\rm eff}$ is typically identified with the diffusion constants $D_i$  -- for quarks $D_q =6/T$ and for the Higgs $D_h=20/T$ \cite{Joyce:1994zn,Cline:2000nw} -- which implicitly defines $\bar \Gamma_{\rm tot}$. Using that  $S_1 \ll S_2$ as $S_1$ is the average of a total derivative up to small corrections, the source term can be approximated by
$S_{\rm eff} ={ S_2' D_{\rm eff}}/{D_2}$.   This is the form that appears in many papers in the literature that derive the source for WKB-excitations, see e.g. \cite{Cline:2000nw}.
The interaction term is 
\be
C^{\rm eff}_{i1} =\kappa_i  \sum_{i,j} \Gamma_i s_{ij} \mu_j, \quad \kappa_i =\(-\frac{N_0}{N_1 D_0 T}\)_i.
\ee
For massless fermions $\kappa_i \approx 1.1$, and this factor is often not included or set to one in the literature.

The chemical potential can be directly related to the asymmetry in the number densities, using that the number densities for antiparticles $\bar f_{s_h}(k,\mu) = f_{s_h}(k,-\mu)$. We denote the number density of particles minus antiparticles by $\rho_{is_h}$ (in the literature often $n_{is_h}$ is used).
For sphaleron transitions we are only interested in left-handed number densities, and we drop the helicity index.  With these definitions, we then get for $\rho_i \equiv n_i- \bar n_i $
\begin{align}
\rho_i &=g_i\ \int_0^\infty {d^3k\over (2\pi)^3} 
\left[f(k,\mu)-f(k, -\mu)\right]
=    \mu_i \frac{T^2}{6}  \times \frac{6  g_i}{\pi^2}
\int_{m/T} \dd x \, x \sqrt{x^2-\frac{m^2}{T^2}} \frac{\e^x}{(1\pm \e^x)^2}
+ \O(\mu_i^3)
\nn \\
&\equiv \frac{\mu_i T^2}{6} k_i (m_i/T) + \O(\mu_i^3).
\label{eq:muvsn}
\end{align}
In the massless limit $k_i(0)$ counts the degrees of freedom $g_i$: for a Weyl fermion or a real scalar $k_i=g_i=1$.
This gives the equation
\be
v_w \rho_i' -D_i^{\rm eff} \rho_i'' + \sum_{i,j}  \kappa_i \Gamma_i s_{ij} \frac{k_i}{k_j} \rho_j=  \frac{k_i T^2}{6} S_i^{\rm eff} .
\label{n_eq_eff}
\ee

\subsubsection{The set of transport equations}

So far we have mainly focused on the Boltzmann equation for a single species, but the interactions described by the collision term will spread the CP asymmetry over the other species in the plasma. It is thus necessary to solve for the full set of Boltzmann equations. 

The source varies over a length scale set by the bubble wall width $L_w$. Interactions with an interaction length $L_{\rm int} \sim v_w/ \Gamma$ much smaller than this, can be treated as in thermal equilibrium. An important example are the gauge interactions in the SM. For much larger interaction length scales, on the other hand, the interactions can be neglected -- such as the Yukawa interactions of the 1st and 2nd generation quarks and leptons. Also the electroweak sphaleron interactions are slow, which allows to calculate the BAU in a two-step process: first determine the CP asymmetry from the Boltzmann equations (the subject of this section), and then the baryon asymmetry from the transport equation for the sphaleron rate (the subject of the next section).

Which set of plasma species and which interactions to include is model dependent. To be specific in this section, we focus on a CPV top quark source generated by the dim-6 operator \cref{CP_yukawa}, and we only consider SM particles and interactions. We follow closely the analysis in \cite{DeVries:2018aul}.
Even though the model considered in \cite{DeVries:2018aul} can not explain the observed BAU with current EDM constraints, it still provides an illustrative example of a set of transport equations for the CP-asymmetry.
The net number density  of third-generation quarks is denoted by $t= \rho_{t_R}, \, b = \rho_{b_R},\, q=\rho_{t_L} + \rho_{b_L}$,  the third-generation leptons  by 
$\tau = \rho_{\tau_R},\, l=\rho_{\nu_L} + \rho_{\tau_L}$. 
The Higgs number density is defined as $h= \rho_{H^0} + \rho_{H^+}$, 
where we have parameterized the Higgs doublet as $\vp = (H^+, H_0)^T$.
As the gauge interactions and Higgs self-interactions are fast, the chemical potentials of the up and down components of SU(2)$_L$ doublets and the components of the Higgs doublet are equal. The light leptons effectively decouple as they only interact through small Yukawa couplings. It is a good approximation to neglect all quark Yukawa interactions except for the top quark. The light quarks still interact through fast strong sphaleron interactions and their densities are related through\footnote{$q_{1}$ and $q_2$ denote, respectively, the first- and second-generation left-handed doublet, and $u,d,s,c,b$ the right-handed quarks.} 
\be
q_1=q_2 = -2 u = -2d = -2s =-2c =-2b\,.
\label{ss}
\ee
As such, we can eliminate all but one of the light quark number densities. Finally, since the left- and right-handed quarks have approximately equal diffusion constants (dominated by strong interactions), baryon number is locally conserved on the time scale of the transport equations and
\be
	t+ b + q + c + s + q_2 + u + d + q_1 =	t+ u + q  =0\,,\label{localbc}
\ee
which can be used to eliminate the transport equation for the up quark. As as result, the number densities of all quarks directly follow from $q$ and $t$. 

In contrast, lepton number is only conserved globally \cite{Chung:2009cb}. Right-handed leptons diffuse more easily than left-handed leptons, and therefore $D_l \neq D_\tau$. It was shown in \cite{DeVries:2018aul} that better accuracy is obtained if $l$ and $\tau$ are kept as independent degrees of freedom, but for the current set-up it is a reasonably good approximation to neglect this difference and assume local lepton number conservation as well.

The full set of transport equations becomes (we have absorbed $k_i \kappa_i$ in $ \Gamma_i$ in \cref{n_eq_eff}), 
\begin{align}
\hat \partial q &= 
+ \Gamma_M^{(t)}\mu_M^{(t)}
+ \Gamma_M^{(b)}\mu_M^{(b)}
+ \Gamma_Y^{(t)} \mu_Y^{(t)}
+\Gamma_Y^{(b)} \mu_Y^{(b)}
 -2 \, \Gamma_{\rm ss} \, \mu_{\rm ss}
- S_t \, ,
\nn\\
\hat \partial t  &=
-\Gamma_M^{(t)}\mu_M^{(t)} -  \Gamma_Y^{(t)} \mu_Y^{(t)}
+ \Gamma_{\rm ss} \, \mu_{\rm ss}
+S_t
\,,
\nn \\
\hat   \partial l &= 
+ \Gamma_M^{(\tau)}\mu_M^{(\tau)}
+\Gamma_Y^{(\tau)} \mu_Y^{(\tau)}
\, ,      
\nn \\
\hat \partial \tau  &=
-\Gamma_M^{(\tau)}\mu_M^{(\tau)}- \Gamma_Y^{(\tau)} \mu_Y^{(\tau)}
\,,
                         \nn \\ 
 \hat \partial h &= 
                      +\Gamma_Y^{(t)} \mu_Y^{(t)}-\Gamma_Y^{(b)}  \mu_Y^{(b)}\, +\Gamma_Y^{(c)} \mu_Y^{(c)}-\Gamma_Y^{(\tau)}  \mu_Y^{(\tau)}\,,
 \label{transport}
\end{align}
with $\hat \partial q = v_w q'-D_q q'' $ and likewise for the other fields. Comparing with \cref{n_eq_eff} the top source is $S_t =({k_t T^2}/{6}) S_t^{\rm eff}$, which is opposite for left- and right-chiral fields; strictly speaking, it is opposite for the two helicity states, which for the massless quarks in the symmetric phase can be identified with the chirality states. Here, $S_t^{\rm eff}$ is given in terms of $\bar S_i = \langle ({k_z}/{\omega})^{l-1} S^\textsc{CPV}\rangle$ in \cref{Deff}. The semi-classical source $S^\textsc{CPV}$ appearing in this average was derived in \cref{clas_source}, which for the CPV top Yukawa interaction can be calculated with the inputs of \cref{sec:example_top_source}. 

The interactions included are  the chirality-flip rate $\Gamma_M^{(i)}$ for species $i$,  the Yukawa rate $\Gamma_Y^{(i)}$, and the strong sphaleron rate $\Gamma_{ss}$. The complete expressions for the interaction rates, source terms, masses, and diffusion constants entering the equations can be found in Ref.~\cite{deVries:2017ncy,Cline:2000nw,Cline:2020jre} and references therein. However, in Ref.~\cite{deVries:2017ncy} $\Gamma_M$ was calculated in the VIA approximation, which is incorrect; rather, the rates estimated in \cite{Huet:1995sh} should be used. The Yukawa rates can be found in  Ref.~\cite{Joyce:1994zn}.
The rescaled chemical potentials (up to a factor $6/T^2$) for these interactions are
\begin{alignat}{2}
\mu_M^{(t)} &= \left( \frac{t}{k_t} - \frac{q}{k_q} \right), 
&\quad \mu_Y^{(t)} &= \left( \frac{t}{k_t} - \frac{q}{k_q} - \frac{h}{k_h} \right), \nonumber \\
\mu_M^{(\tau)} &= \left( \frac{\tau}{k_\tau} - \frac{l}{k_l} \right), 
&\quad \mu_Y^{(\tau)} &= \left( \frac{\tau}{k_\tau} - \frac{l}{k_l} + \frac{h}{k_h} \right),
\end{alignat}
and 
\be
\mu_{\rm ss}= \sum_{i=1}^3 \( \frac{2q_i}{k_{q_i}} -\frac{u_i}{k_{u_i}} -\frac{d_i}{k_{d_i}}\).                                                 \ee
The $k_i(m_i/T)$-functions relating the chemical potentials to the number densities are defined via $\rho_i=T^2\mu_i k_i/6 + {\mathcal O}(\mu_i^3)$ as in \cref{eq:muvsn}.

One might be tempted to neglect the $\tau$ Yukawa coupling as well, and decouple the quark and lepton sector.  However, this is not a good approximation, and may to lead to an $\O(10)$ error in the CP asymmetry \cite{DeVries:2018aul}. The reason is that the strong sphaleron transitions are very effective in washing out the CP asymmetry in the quark sector \cite{Giudice:1993bb, Tulin:2011wi,Huet:1995sh}. Setting $\mu_{ss} \simeq 0$ it can be shown that together with local baryon number conservation, the chiral asymmetry vanishes for massless quarks (i.e. when setting $k_i =k_i(0)$).  The sphalerons are only in equilibrium away from the bubble walls, and not all of the quark asymmetry is washed out. Part of the asymmetry that is generated by the top CPV source term can also be
 transferred to the lepton sector via the lepton Yukawa interactions, where it is safe from strong sphaleron washout. It turns out that this latter contribution may be non-negligible. 

The  density of left-handed fermions that sources the electroweak sphaleron transitions is  $n_L = \sum_i (q_i + l_i)\,,$. To find this asymmetry, the set of Boltzmann equations can be solved by an approximate semi-analytic method \cite{Huet:1995sh,White:2015bva,deVries:2017ncy}, which was shown to give an excellent fit to the full equations.
To illustrate the method while keeping the notation light, we will give the results for a single species. That is, we review the solution of the equation
\be
v_w  \rho' - D \rho''+   \Gamma \rho= S,
\label{BE_solution}
\ee
with $z < 0$ corresponding to the symmetric phase.
The equation \cref{eq:baryonNrDE} for the baryon density $\rho_b$ discussed in \cref{sec:sphalerons} is of the same form. The idea is to solve the equation in both the broken and symmetric phase, approximating the rates as constants (distinguished with subscripts $b,s$) but allowing for the $z$-dependence of the source. The solutions are then matched at $z=0$, and together with the requirement that the perturbations vanish far away from the bubble wall, this fixes all coefficients.

For the bubble wall profile \cref{bubble_profile} the semi-classical source is mainly effective in the broken phase and for simplicity we neglect it in the symmetric phase.\footnote{\label{baryon_sol}To adapt to the baryon number density equation \cref{eq:baryonNrDE}, where the source is only active in the symmetric phase, we can interchange the symmetric and broken phase solution (taking into account that $z\to -z$). As the sphaleron rate vanishes in the broken phase, the root $\gamma_1=0$ and the solution is constant in the broken phase.}  The first step is to solve the homogeneous equations with the Ansatz $\rho = \e^{\alpha z}$, which gives a positive and negative root $(v_w \pm \sqrt{4  D    \Gamma+ v_w^2})/({2  D})$, which we denote by $\gamma_i$ and $\alpha_i$ (with $i=1 \,(2)$ the positive (negative) roots) in the symmetric and broken phase  respectively.
The solutions in the symmetric and broken phase, indicated by the subscript $s$ and $b$, that are finite at $z = \pm \infty$ are 
\be
\rho_s|_{z <0} = c \e^{\gamma_1 z}, \qquad
\rho_b|_{z> 0} = \sum_{i} x_i \e^{\alpha_iz} \( \int_0^{z}\e^{-\alpha_i y} S \dd y - \beta_i\)\,.
\ee
The constants $x_{1,2}$ and $\beta_1$ can be obtained demanding that $\rho_b|_{z> 0}$ is a solution of the full inhomogeneous equation \cref{BE_solution}, and by demanding that  the solution is finite. This gives
\be
x_2= -x_1 = \frac{1}{ D(\alpha_1 -\alpha_2)}\,, \qquad
\beta_1 = \int_0^{\infty} \e^{-\alpha_1 y} S \dd y\,.
\label{x1x2}
\ee
Matching the broken and symmetric phase solution and their first derivative at $z=0$ fixes the remaining coefficients
\be
c =-\frac{\beta_1}{D(\alpha_2-\gamma_1)} \,,\qquad
\beta_2 = \beta_1 \frac{(\alpha_1-\gamma_1)}{(\alpha_2-\gamma_1)} \,.
\ee
 Ultimately, we need the chiral asymmetry in the symmetric phase as this will be transferred into a baryon asymmetry by sphaleron transitions. 
Putting everything back together, the symmetric phase chiral asymmetry is given by
\begin{align}
 \rho_s|_{z<0} &= -\frac{\beta_1\e^{\gamma_1 z}}{D(\alpha_2-\gamma_1)} 
=\frac{2 \e^{\gamma_1z}
}{ (\sqrt{4 D \Gamma_b + v_w^2} + \sqrt{4 D \Gamma_s + v_w^2})}\int_0^{\infty} \e^{- \alpha_1 y/L_b} S \dd y\,.
\label{ns_semi}
\end{align}

\subsubsection{Supersonic bubbles}
\label{sec:supersonic}

The bubble wall velocity is an important input parameter for the baryon yield. It is expected that
the asymmetry approaches zero both for very small velocities, as the system remains close to thermal equilibrium, and for large velocities, as there is no time for the CP asymmetry produced inside the wall to diffuse into the symmetric phase where electroweak sphalerons are active. In recent years there has been renewed interest in the latter limit: how does the asymmetry develop for bubble wall velocities around or above the speed of sound?
This is motivated by the possibility of observing gravitational waves (GW) produced during the first-order phase transition at future GW experiments such as the upcoming LISA observatory. Strong phase transitions with a large latent heat, which generically produce faster expanding bubbles, have more energy to transfer into gravitational waves, and thus can give a larger GW signal.

It was long believed that efficient electroweak baryogenesis is incompatible with supersonic bubble velocities, i.e. velocities exceeding the speed of sound in the plasma $c_s\simeq 1/\sqrt{3}$. Early works  based on the fluid Ansatz discussed in \cref{sec:fluid_ansatz} indeed confirm this expectation. It was found that the asymmetry quenches at the speed of sound. For the CP asymmetry the effect of the background $\delta_l$ drops out, and unlike the calculation of the bubble wall velocity in \cref{sec:fluid_ansatz}, there is no singularity at the speed of sound. The vanishing of the solution, instead, can be traced back to the form of the Liouville operator in the fluid Ansatz. 

Approximating the source with a delta-function located at the center of the bubble, the set of equations in \cref{eq_A3} can be solved analytically  \cite{Moore:1995ua,Moore:1995si}. This approach was generalized in the previous subsection (for a single species) to take the explicit $z$-dependence of the source into account. In both cases, the first step is to solve the set of homogeneous Boltzmann equations. Defining $\chi_i$ and $\lambda_i$ as the three eigenvectors and eigenvalues of $-A^{-1}\Gamma$, that is $\Gamma \chi_i = \lambda_i A \chi_i$, the solution of the homogeneous equation (\cref{eq_A3} with the source term set to zero) is a superposition of factors $\e^{-\lambda_i z}$; here $z<0$ corresponds to the symmetric phase.
It was found that for subsonic bubble wall velocities, two of the eigenvalues are positive and one is negative, and there is a well-behaved solution both inside and outside the wall. As the product of eigenvalues is proportional to 
\be
\det(A) \propto \frac{c_4 v_w}{3}(c_2c_4-c_3^2) \left(\frac13 -v_w^2\right)\,,
\label{flip_lambda}
\ee
there is a sign flip of the eigenvalues for supersonic speeds $v_w^2> 1/3$,  all eigenvalues become positive, and there is no finite solution outside the bubble.

The fluid Ansatz in \cref{fluid_ansatz2} thus leads to the conclusion that the baryon asymmetry vanishes for bubble velocities larger than the sound speed.  The same conclusion was reached in early works using the generalized velocity 
perturbation approach of \cref{gen_vel} \cite{Fromme:2006wx}; however, these results are based on the small velocity limit for (some of) the coefficients  in the transport \cref{result_df} and cannot be trusted at larger velocities \cite{Cline:2020jre}. 
The lore that the bubble wall velocity cannot exceed the sound velocity for baryogenesis was recently questioned, as  a  sudden quench is not expected upon physical grounds. Indeed, Ref.~\cite{Cline:2020jre} argued that diffusion is a microscopic process, and particles do not collectively move at the speed of sound. Some fraction of the particles still travel fast enough to make it in front of the wall before being swept up by the expanding bubble. Consequently, one would expect the baryon asymmetry to continuously decrease as a function of $v_w$ without a sudden quench for supersonic velocities. Instead of the fluid approximation, Ref.~\cite{Cline:2020jre} used the generalized velocity perturbation of the previous subsection. The set of equations  (\ref{result_df}) was solved numerically for a system of top quarks with a complex mass -- providing the CP asymmetry -- and a tanh-Ansatz for the bubble wall. No discontinuity was found at $v_w =1/\sqrt{3}$, the speed of sound. Baryogenesis is possible with relativistic bubble wall velocities, although the efficiency decreases for supersonic bubbles.  This can be understood as the $A$-matrix in the Boltzmann equations for the perturbations \cref{result_df} only becomes singular, that is ${\rm det}(A)=0$, for wall velocities $v_w \to 1$ that equal the speed of light.

The calculation in Ref.~\cite{Cline:2020jre} depends on the ad hoc assumption in \cref{assumption_av}, and although the results are consistent with physical expectations, it prompted  Ref.~\cite{Dorsch:2021ubz} to revisit the  problem. In the fluid approximation with three perturbations, only normalizable solutions exist inside the bubble for subsonic 
velocities. Therefore, Ref.~\cite{Dorsch:2021ubz} extends the moment expansion to higher order.
Working in a general frame, the generalized form of the perturbation in the fluid expansion is
\be
\delta f=f'_{0}(p^\mu u_\mu) ( \omega^{(0)} + p^\mu \omega_\mu^{(1)} + p^\mu p^\nu\omega_{\mu\nu}^{(2)} +...)\,.
\label{df_generalized}
\ee
Due to the symmetries for the system with a planar wall, at each order $n$ in the momentum expansion there are $n+1$ independent perturbations in $\omega^{(n)}$. For the moment expansion, the weight factors are chosen as $(p^\mu u_\mu)^a(p^\nu \bar u_\nu)^b$ with $a,b>0$ and $(a+b) \leq n$. 
The resulting set of equations in \cref{eq_A3} for $n=2$ is now for a 6-dimensional fluctuation vector $q$ with a corresponding $ 6\times 6$ $A$-matrix.  

The numerical solution of the 6-dimensional system is  continuous at the speed of sound. In fact, the additional perturbations push the sign flip of the eigenvalue, where the solution becomes zero, to $v_w \to 1$. In addition, the moment expansion converges, although this should be checked for different parameter choices and set-ups. There seems, a priori, no reason why the expansion in momentum in \cref{df_generalized} should work, as typically $p^\mu \sim T$ is not small. That the first three perturbations in this expansion are not sufficient at large velocities might also indicate that the momentum expansion is not the best approach. Going to higher moments comes at the cost of more collision integrals that must be calculated, although many have to be calculated only once and can then be recycled for other models and set-ups.

Another approach that allows for a more general parameterization of the perturbations is to expand $\delta f$ in Chebyshev polynomials, see the discussion around \cref{cheb}. This has been done for the CPC equation that determines the  bubble wall velocity \cite{Laurent:2022jrs}, where it was shown to have a good convergence, but not yet for the CPV equations. Although the generalized fluid Ansatz and the expansion in Chebyshev polynomials can help determine the CP asymmetry to higher precision, it is not clear how the higher moments of the perturbations couple to the sphaleron rate (see the next section). The resulting improved precision therefore does not immediately transfer to a more precise determination of the final baryon yield. These approaches are nevertheless  useful to determine the accuracy of the commonly used generalized velocity perturbation of \cref{sec:gen_velocity}.

\subsection{Summary of \cref{sec:bubbles}}

The CPC and CPV force terms enter in the Boltzmann equations at different order in the gradient expansion, which implies that the bubble wall velocity can be determined first, and imposed as an external parameter for the calculation of the CP asymmetry. Taking the details of the model and the FOPT as input -- the effective potential and the nucleation temperature, as discussed in \cref{sec:FOPT} -- the {\tt WallGo} package \cite{Ekstedt:2024fyq} calculates the bubble wall velocity $v_w$. Alternatively, one can expand the deviation from equilibrium in momenta and take moments of the Boltzmann equations, as described in \cref{sec:fluid_ansatz}.
To avoid a singularity at the sound speed, it is important to avoid linearization in the background velocity and temperature profiles.
Analytic approximations for $v_w$ exist in the limit of vanishing collisions and in the limit of very large collisions, as reviewed in \cref{sec:appprox_vw}.

The CP asymmetry in left-handed fermions that feeds into the equation for the baryon yield, \cref{eq:baryonNrDE} in the next section, is commonly computed using  the generalized velocity perturbation of \cref{sec:gen_velocity}. For thick bubble walls for which the gradient expansion holds, the Boltzmann equations reduce to a single equation (for each species) for the chemical potential \cref{eq_eff}, or equivalently, for the number density \cref{BE_solution}. The set of Boltzmann equations can be solved with very good accuracy using a semi-analytic approximation \cref{ns_semi}; the generalization to multiple species can be found in Refs.~\cite{Huet:1995sh,White:2015bva,deVries:2017ncy}.

\begin{figure}[t]\label{fig:Ybimprovement}
     \centering
      \includegraphics[width=0.40\textwidth]{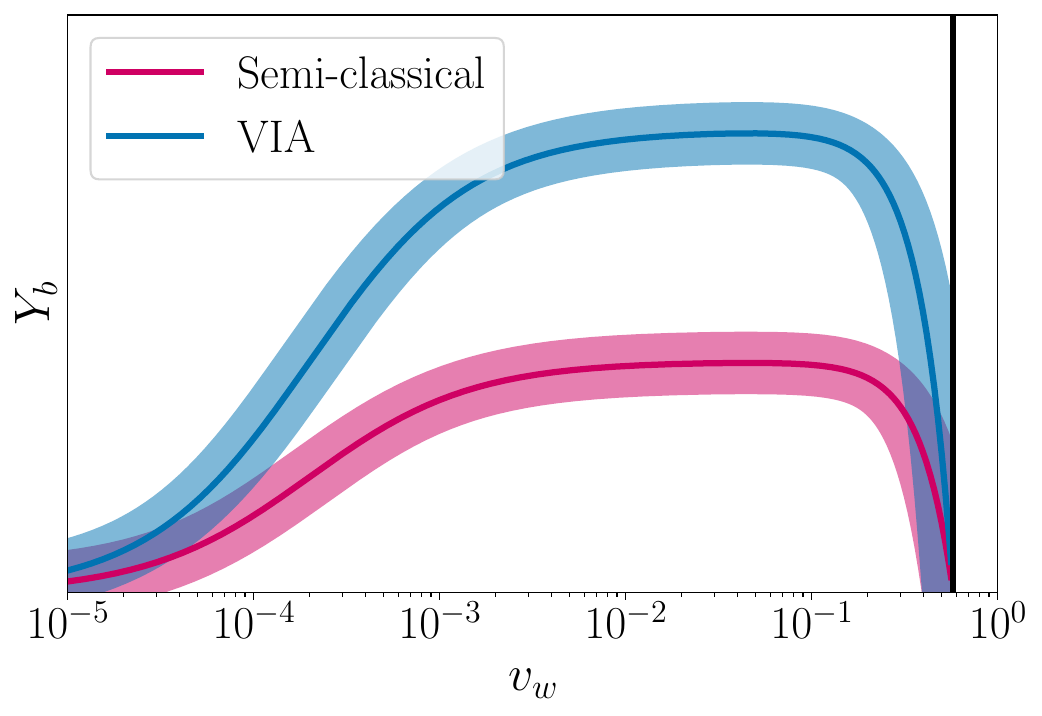}
      \hspace{5mm}
      \includegraphics[width=0.40\textwidth]{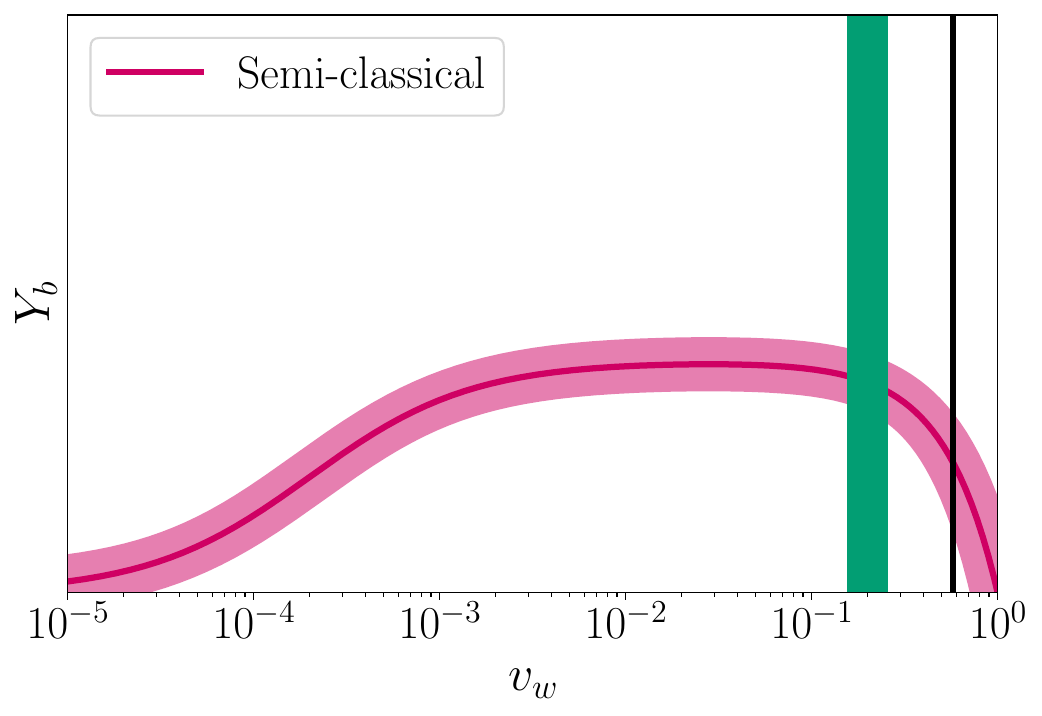}
  
        \caption{A sketch of the improvement of the baryon yield calculation made in recent years. The left plot shows schematically the yield $Y_b$ versus the bubble wall velocity $v_w$ based on predictions from a few years ago, while the right plot shows the current understanding (see main text for more explanations). The speed of sound is indicated by the black line. }
        \label{fig:VIA}
\end{figure}

There are other approaches to solving the CPV Boltzmann equation, such as the generalized fluid Ansatz. 
However, a more precise determination of the distribution function does not necessarily translate to a more precise determination of the baryon yield (yet), as it is unclear how the full distribution function couples to the sphaleron rate. It would still be very useful to compare the various approaches to obtaining the distribution function, as this can shed light on the validity of the (ad-hoc) approximation made in the generalized velocity perturbation approach of \cref{sec:gen_velocity}, and it can help determine the error in this part of the calculation of the baryon asymmetry.

We have illustrated the progress made in recent years in computing the baryon asymmetry in a typical EWBG scenario in Fig.~\ref{fig:VIA}. The left panel depicts the status from a few years ago and shows a sketch of $Y_b$ computed with the semi-classical and VIA source arising from the same mechanism of CP violation. Although the uncertainty of the calculation is not known, we expect rather large errors bands, as shown in the plot.  The obtained value of $Y_b$ differs significantly between the sources. In addition, the bubble wall velocity was typically treated as a free input parameter and not computed within the specific EWBG scenario. The asymmetry was expected to vanish for supersonic values of $v_w$. In the right panel we sketch the current status. The VIA source should not be considered and the semi-classical source remains (although, depending on the scenario, a flavor source could be present). In addition, the bubble wall velocity is now explicitly computed (in this sketch it is taken to be around $0.2$, as indicated by the green band) 
and thus the value of $Y_b$ is predicted to much higher accuracy. In addition, although not the case in the example, scenarios with supersonic wall velocities could still produce nonzero values for $Y_b$.  A better understanding of the phase transition dynamics, and the determination of the nucleation temperature, have reduced the uncertainty of the calculation reducing the error band.

	\newpage
	\section{Converting the CP-asymmetry to a baryon asymmetry}\label{sec:sphalerons}
        
Conversion of the CP-violating number density computed in \cref{sec:bubbles}, induced by the source terms discussed in \cref{sec:CP}, happens via the so-called \emph{electroweak sphaleron} process.
At the classical level, baryon number and lepton number are both conserved in the Standard Model. However, at the quantum level, they are violated by the chiral anomaly \cite{tHooft:1976rip}. This non-perturbative quantum process gives rise to the baryon and lepton number changing currents 
\begin{equation}
    \partial_\mu J^\mu_{B,L} = \frac{n_f}{32 \pi^2}g^2 F^a_{\mu\nu} \tilde F^{a\mu\nu},
\end{equation}
where $n_f$ is the number of fermionic families, $g$ is the SU(2) gauge coupling, and $\tilde F^{a\mu\nu} = \epsilon^{\mu\nu\rho\sigma}F^a_{\rho\sigma}/2$, with $F^{a\mu\nu}$ the SU(2) gauge field strength (the hypercharge field is irrelevant as there are no non-trivial vacua for Abelian gauge fields).
Since the current is the same for baryons and leptons, $B-L$ is conserved, while $B+L$ is not.

The change in baryon number between times $t_i$ and $t_f$ is given by
\begin{equation}
    B(t_f) - B(t_i) = n_f \int_{t_i}^{t_f} dt \int d^3 \mathbf{x} \,\frac{g^2}{32\pi^2}  F^a_{\mu\nu} \tilde F^{a\mu\nu} = n_f [n_{\rm CS}(t_f) - n_{\rm CS}(t_i)]\,.
\end{equation}
The integral on the right hand side takes an integer value \cite{Chern}, called the Chern-Simons, or winding number. In the vacuum, the Higgs field is at the minimum of its potential, and the SU(2) field strength vanishes. Such a configuration corresponds to an integer value of $n_{\rm CS}$. Two vacua with different values of $n_{\rm CS}$ are topologically distinct and can only be reached through a large gauge transformation -- the fields need to move through a configuration with a (large) non-zero energy, see, for example, Fig.~1 of \cite{Amoroso:2020zrz} for an illustration.

The static field configuration corresponding to a maximum of the energy, i.e. on top of the barrier between adjacent vacua, is called the \emph{sphaleron} solution. In the SM, the energy of this configuration is estimated as \cite{Klinkhamer:1984di}
\begin{equation}
    E_{\rm sph} \sim 2.8 \frac{4\pi v}{g}\,,\label{eq:Esph}
\end{equation}
with $v$ the classical Higgs field value in the broken phase. Note that $E_{\rm sph}$ is gauge dependent \cite{Patel:2011th}, but this dependence  was argued to be small in Ref.~\cite{Garny:2012cg}. We will comment on this further below. The sphaleron energy can receive corrections in BSM models with extended Higgs sectors (see e.g. Ref.~\cite{Spannowsky:2016ile}). 
At zero temperature, transitions between vacua with different $n_{\rm CS}$ would proceed via instanton tunneling \cite{Belavin:1975fg}  with probability
\begin{equation}
    P_{\rm tunnel} \sim e^{-4\pi/\alpha_W} \sim 10^{-162}\,,
\end{equation}
where $\alpha_W = g^2/(4\pi)$ is the weak coupling constant. 
This rate is minuscule and therefore baryon-number-violating processes do not occur at zero temperature and the deuteron and ${}^3$He nucleus remain stable (the proton would remain stable even for larger instanton tunneling rates as electroweak instantons violate baryon number by three units).

The crucial realization in Ref.~\cite{Kuzmin:1985mm} that the sphaleron rate is less suppressed at high temperatures and that baryon-number violation might thus occur during the electroweak phase transition led to the invention of electroweak baryogenesis. 
In the broken phase, the transition rate is dominated by thermal fluctuations rather than quantum tunneling. 
In the symmetric phase, the rate associated to change of the Chern-Simons number is unsuppressed. 
Although the sphaleron does not exist in the symmetric phase, the rate is still referred to as the hot sphaleron rate.

In the following subsections, we will discuss the equations governing the generation of the baryon number density, and provide an expression for the BAU in terms of the CP-asymmetry obtained in section~\ref{sec:bubbles}.
We will also discuss the values of the sphaleron rates in the symmetric and broken phase that enter these equations.

\subsection{From a CP-asymmetry to the baryon asymmetry}
Typically, the electroweak sphaleron is slow compared to the CP-violating physics, and the generation of the baryon asymmetry can be treated as a two-step process. We will make this assumption here, but see Ref.~\cite{DeVries:2018aul} for a computation where the baryon asymmetry is solved simultaneously with the transport equations.\footnote{
If there is a CPV coupling of $W\tilde W$ to the Higgs doublet \cite{Dine:1990fj}, as in \cref{WWcoupling}, this effectively gives a CPV contribution to $\mu^0_L$ in the bubble wall background. In that case, the full set of Boltzmann equations needs to be solved in a single step, although as far as we are aware, in the literature only the baryon equation is solved (and the plasma interactions are neglected).}
Let us thus assume that a CP-asymmetry has been formed in front of the bubble wall. 
In the generalized velocity perturbation framework, summarized in \cref{sec:gen_velocity},
the CP-asymmetry is described by $\mu_L$, the chemical potential of light-handed particles, including all quarks and leptons that have been significantly produced in the transport equations. 

The evolution of the baryon asymmetry is then described by \cite{Cline:2000nw, Cline:1997vk, Cline:2020jre}\footnote{If one were to compute the baryon asymmetry with the (extended) fluid Ansatz, as in \cite{Dorsch:2021ubz}, 
one has to recalculate how the sphaleron couples to the different perturbations in the fluid.}
\begin{equation}\label{eq:baryonNrDE}
    v_w \gamma_w\frac{d \rho_b}{dz}  =-\frac{n_f}{2} \Gamma_{\rm sph} \left( \mu^0_{L} + \mu_L \right) =-
    \frac{n_f}{4} \tilde \Gamma_{\rm sph}  \rho^0_L +{\mathcal R} \rho_b,
\end{equation}
where $\mathcal R=(15/4)(\tilde \Gamma_{\rm sph}/2)$ is the SM relaxation coefficient for the decay of baryon number through weak sphalerons, as discussed below. The  last equality assumes small chemical potentials, \emph{cf}. \cref{eq:muvsn}, and we have neglected finite mass corrections and set $k_i=k_i(0)$.
We use the notation $\tilde \Gamma_{\rm sph}=(6/T^3) \Gamma_{\rm sph}$,   $\Gamma_{\rm sph}$ denotes the sphaleron rate per unit volume, and $n_f =3$ is the number of fermionic families. 
In some references, e.g. Refs.~\cite{Huet:1995sh,Lee:2004we}, an additional diffusion term $D_b$ is included on the LHS. The resulting expression for the BAU shows that this term can be neglected if $D_b \tilde \Gamma_{\rm sph} \ll v_w^2$ which is a good approximation and we drop the term here as well.
As discussed in \cite{Cline:2021dkf}, there are disagreements in the literature about the normalization of \eqref{eq:baryonNrDE}, with e.g. \cite{Huet:1995sh} differing from \eqref{eq:baryonNrDE} by a factor 4 on the RHS, and \cite{Lee:2004we, DeVries:2018aul} differing by a factor 2.
We use the normalization of \eqref{eq:baryonNrDE}, as $\Gamma_{\rm sph}$ then corresponds to the sphaleron diffusion rate obtained in lattice simulations, as demonstrated explicitly in \cite{Moore:1996qs}.  Both forms of the baryon asymmetry evolution equation -- in terms of the chemical potential and in terms of the baryon number density -- are found in the literature.

The terms with $\mu_L^0$ and $\rho^0_L$ describe the generation of the baryon number from a CP-asymmetry, as discussed in the previous section, which is put in as an initial condition (thus serves as a source term) in the equation for the baryon number density. The chemical potential $\mu_L$ sums over all the left-handed quark and leptons.
As the gauge interactions are fast, the chemical potentials of all components of the SU(2) doublets and SU(3) doublets are the same, and 
\be
\mu_L =\sum_i (\mu_{q_{L_i}} + \mu_{l_i}) = \frac{6}{T^2} \frac12  \sum_i ( \rho_{q_{L_i}} +\rho_{l_i}) \equiv  \frac{6}{T^2} \frac12 \rho_L\,.
\label{muL}
\ee
Here $\mu_{q_{L_i}}$ and  $\mu_{l_i}$ are the chemical potential and number density of the left-handed (color triplet) quark  and lepton doublets, and $\rho_{q_{L_i}}$ and  $\rho_{l_i}$ the corresponding number densities (the factor $1/2$ arises from the 2 degrees of freedom in a doublet). The sum $i$ is over the $n_f=3$ families. 
The last equation defines $\rho_L$, which is the number density of left-handed doublets.
To produce a baryon dominated universe with  $\rho_b>0$ requires a negative initial CP asymmetry $\rho^0_L <0$ to bias the sphaleron rates for baryon production; the sign of $\rho^0_L$ is determined by the sign of the CP violating interaction.

The term with $\mu_L$ in the first (with $\mathcal \rho_b$ in the second) expression of \cref{eq:baryonNrDE} is the relaxation term, which takes into account the inverse sphaleron transitions that wash out the asymmetry. To express it in terms of the number density of baryons, sometimes written as $\mu =A{\rho_b}/{T^2}$,
it should be taken into account that the left- and right-handed quark number densities equilibrate $\rho_{q_L}=\rho_{q_R}$ on the timescale of sphaleron interactions. The Yukawa interactions are not fast enough compared to the sphaleron rate to also equilibrate the right-handed leptons in the symmetric phase (although for the $\tau$ lepton this is borderline). We can then relate the chemical potential (and number density) of left-handed color-triplet quarks and leptons, that couple to the sphalerons, to the baryon number density:
\be
\rho_b=  \sum_i (\rho_{q_{L_i}}+ \rho_{q_{R_i}}) = 2 \sum_i \rho_{q_{L_i}},
\quad \& \quad
\rho_b=
\rho_{l_L} =\sum_i \rho_{l_{L_i}} \quad \Rightarrow \quad
\mu_L = \frac{6}{T^2} \frac12  \sum_i (\rho_{q_{Li}} +\rho_{l_i}) =
 \frac{6}{T^2} \frac54 \rho_b,
\ee
where in the last expression we used \cref{muL}. Here $i$ again is the family index.
For the SM field content this gives $A =15/2$ and ${\mathcal R} = A (n_f/6)(\tilde \Gamma_{\rm sph}/2)=15/4(\tilde \Gamma_{\rm sph}/2) $. The extension to the broken phase, and the inclusion of additional light (s)quarks and/or (s)leptons can be found in \cite{Cline:2000nw}. 

The baryon asymmetry normalized by the entropy density $s =\frac{2\pi^2}{45} g_{* s} T^3$ is obtained integrating \cref{eq:baryonNrDE} over $z$. Approximating the sphaleron rate to be only active in the symmetric phase (see \cref{washout} for a discussion), the asymmetry is constant in the broken phase and given by \cite{Lee:2004we,Cline:2011mm, Cline:2020jre}
\be
Y_b =\frac{\rho_b}{s} = - \frac{n_f \tilde \Gamma_{\rm sph,s}}{4s v_w \gamma_w} \int_{-\infty}^0 \rho_L(z) \e^{z {\mathcal R}/(v_w \gamma_w)} \dd z,\label{eq:YbFinal}
\ee
with $\tilde \Gamma_{\rm sph,s}$ the sphaleron rate in the symmetric phase, see \cref{eq:SphsBest} below. 
Finally, we have reached our goal of finding an expression for the baryon asymmetry $Y_b$. It has been quite a derivation!

\subsection{Symmetric phase sphaleron rate}
In the symmetric phase, the sphaleron rate is given by the diffusion rate
\begin{equation}
    \Gamma_{\rm{sph, s}} = \lim_{V,t \to \infty} \frac{\langle[n_{\rm{CS}}(t) - n_{\rm{CS}}(0)]^2 \rangle }{Vt}.
\end{equation}
One of the main challenges for an accurate and efficient computation of this rate, is the construction of an appropriate effective theory for the \emph{classical} equations of motion \cite{Grigoriev:1988bd, Moore:1997cr, Ambjorn:1997jz}. In early studies of the sphaleron rate, equations of motion corresponding to the dimensionally reduced SU(2)+Higgs sector of \cref{eq:SM3dEFT} of the SM were simulated on the lattice \cite{Tang:1996qx, Ambjorn:1997jz}.
These classical equations of motion were assumed to be appropriate due to the high occupation numbers of the IR modes.
However, the non-classical `hard' modes (with momenta of the scale $\sim T$) strongly affect the problem via UV-divergent Landau damping \cite{Bodeker:1995pp, Arnold:1996dy}.
In \cite{Bodeker:1998hm}, these divergences were cured by constructing an effective theory for the modes at the scale $g^2 T$, and it was demonstrated -- to leading log accuracy in the gauge coupling $g$ -- that the time-evolution can be described by a Langevin equation for the overdamped gauge modes, depending on the action \cref{eq:SM3dEFT} and the SU(2) color conductivity.
This results in the following perturbative estimate
\begin{equation}
    \Gamma_{\rm sph, s} \sim g^2 \log{\left(\frac 1 g\right)} (g^2 T)^4 .
\end{equation}
The effective theory was extended to include the Higgs field in \cite{Moore:2000mx}.
Numerous lattice computations of the sphaleron rate have been performed \cite{Moore:2000mx, DOnofrio:2012phz, DOnofrio:2014rug} (as some of these were performed before the Higgs was detected, the Higgs has the wrong mass in some of these computations). 
We report here the result from \cite{Annala:2023jvr}, which was determined for a Higgs mass in agreement with the LHC measurement 
\begin{equation}\label{eq:SphsBest}
    \Gamma_{\rm sph, s} = (13.9 \pm 0.1)\alpha_W^5 T^4 \quad \Rightarrow \quad
    \tilde  \Gamma_{\rm sph, s}= 6\times(13.9 \pm 0.1)\alpha_W^5 T,
\end{equation}
where factors of the logarithm of the gauge couplings are absorbed in the numerical prefactor.

\subsection{Broken phase sphaleron rate}
\label{sec:broken_sphaleron}

In the broken phase, the sphaleron transition is suppressed compared to the symmetric phase, but much less so than at zero temperature. The reason is that the fields can make a thermal jump over the barrier, with probability
\begin{equation}
    P_{\rm thermal} \sim e^{-E_{\rm sph}/T},
\end{equation}
which becomes dominant over the tunneling rate for $T \gtrsim M_W/(2\pi)$. 

Several perturbative and non-perturbative computations of the broken phase sphaleron rate $\Gamma_{\rm sph,b} $ have been done. 
The broken phase sphaleron rate $\Gamma_{\rm sph, b}$ was determined at one-loop in perturbation theory in \cite{Arnold:1987mh, Khlebnikov:1988sr, Carson:1989rf}. 
Here, we write the result found in \cite{Arnold:1987mh}, written in the form given in \cite{Moore:1998ge} (with the normalization corresponding to the one used in \cref{eq:baryonNrDE}); note also that the result in \cite{Khlebnikov:1988sr} does not include the factor $\omega_-$,
\begin{equation}
	\Gamma_{\rm sph, b}= 4 T^4 \frac{\omega_-}{g v} \left(\frac{\alpha_W}{4\pi} \right)^4 \left(\frac{4\pi v}{gT} \right)^7 \mathcal{N}_{\rm tr}(NV)_{\rm rot}  e^{-E_{\rm sph}/T} \kappa, \label{eq:sphaleronBrokPert}
\end{equation}
with $v$ the Higgs vev in the broken phase. $\omega_-$ denotes the unstable frequency of the sphaleron solution and it is estimated as $\omega_- \sim g v f(\lambda /g^2)$ \cite{Carson:1990jm}, where $f(\lambda/ g^2) \sim 0.7 $ is a mildly varying function of $\lambda/g^2$, and the value 0.7 corresponds to $\lambda/g^2 = 0.04$.
The factors $\mathcal N_{\rm tr} \sim 26 $ and $(NV)_{\rm rot} \sim 5.3 \times 10^3$ are normalization and volume factors (computed for $\lambda = g^2$ in \cite{Arnold:1987mh}).
$\kappa$ denotes the one-loop fluctuation determinant evaluated on the sphaleron background, computed in \cite{Khlebnikov:1988sr, Carson:1990jm}.
$\kappa$ is a function of $\lambda /g^2$, and equals $\kappa \sim 0.3$ for $g^2 \sim \lambda$, but becomes significantly smaller for larger or smaller values of $\lambda/g^2$.

Note that, just like the sphaleron energy, the perturbative broken phase sphaleron rate of \cref{eq:sphaleronBrokPert} is gauge-dependent. 
In \cite{Burnier:2005hp,Li:2025kyo} a gauge-independent derivation of the sphaleron rate in the 3-dimensional effective theory of \cref{eq:SM3dEFT} was presented. Gauge independence can be obtained by performing a consistent perturbative expansion in the EFT. Ref.~\cite{Li:2025kyo} presents the leading order result of the sphaleron rate; a further improvement can be obtained by including the fluctuation determinant.
The rate computed in \cite{Li:2025kyo} is appropriate for theories with new physics that is heavy enough to be integrated out in the EFT for the sphalerons.
In theories with \emph{light} new physics, 
the sphaleron rate typically receives corrections \cite{Ahriche:2007jp, Funakubo:2009eg, Fuyuto:2014yia}, 
but a gauge-independent computation is currently lacking.

A non-perturbative (and gauge-independent) determination of the sphaleron rate in the broken phase was performed in \cite{Moore:1998ge, Moore:1998swa} for 3 different values of $\lambda / g^2$.
Also in the broken phase, the sphaleron rate follows from the Langevin-like equations derived in \cite{Bodeker:1998hm, Moore:2000mx}.
The lattice result is compared to the perturbative result of \eqref{eq:sphaleronBrokPert} \cite{Arnold:1987mh}, with $v$ obtained in the two-loop effective potential, and under the assumption $\kappa = 1$. 
The non-perturbative result is approximately a factor $e^{-3.6}$ smaller than the perturbative result. Part of the difference is explained by the inclusion of the dynamical pre-factor, which quantifies the fraction of crossings that lead to a settling in the new vacuum 
which is not included in the perturbative result. The assumption $\kappa \sim 1$ is an overestimate of the fluctuation determinant, which gets corrected in the non-perturbative result. 
The agreement between lattice and the one-loop result is also $x$-dependent; \cite{DOnofrio:2014rug} observes a much better agreement for the physical Higgs mass. 
The results of the recent lattice simulations \cite{Annala:2025aci} of the broken phase sphaleron rate can be found in \cite{annala_2025_15599387}, and
from there one can obtain the following fit (in terms of the parameters $y$ and \emph{for fixed} $x$, as defined in \cref{eq:xy})
\begin{equation}\label{eq:fitSph}
    \ln \hat \Gamma \sim (377 \pm 9)y - (42.1 \pm 0.1), \qquad x = 0.025,
\end{equation}
with
\begin{equation}
    \Gamma_{\rm sph, b} = \frac{(g_3^2)^5}{\sigma_{\rm el}} \hat \Gamma, \qquad \sigma_{\rm el} = \frac{m_D}{3\gamma}, \qquad \gamma = \frac{2g^2 T}{4\pi} \left[\ln{\frac{m_D}{\gamma}}  + 3.041 + \mathcal O\left(\frac{1}{\ln(1/g)}\right) \right],
\end{equation}
with $m_D$ the Debye mass and $g_3$ as used in \cref{eq:SM3dEFT}.
Fitting functions for three other values of $x$ are provided in \cite{Annala:2025aci}. However, the case with $x \gtrsim 0.025$ is of less relevance for EWBG, as for these values of $x$ the generated baryon asymmetry gets washed out, as we will see in the following subsection.
To obtain an accurate estimate of the broken-phase sphaleron rate for $x \neq 0.025$, we recommend to read off the value from \cite{annala_2025_15599387}.

\subsection{Baryon number washout}\label{washout}
Before going through the effort of deriving and solving the transport equations in a BSM model, it is useful to estimate beforehand whether the phase transition is actually strong enough for EWBG.
A potential obstacle to successful EWBG is the washout of the generated baryon number inside of the bubbles. 
This can happen when the phase transition is too weak, and the sphaleron rate inside  the bubbles is not sufficiently suppressed \cite{Arnold:1987zg}.
When using \cref{eq:YbFinal} to compute $Y_b$, this check is mandatory, as the solution \cref{eq:YbFinal} assumes that the sphaleron rate cuts off immediately in the broken phase. 

A rough estimate for the non-washout criterion (see e.g.  \cite{Quiros:1999jp, Konstandin:2013caa}) can be obtained by assuming that the sphaleron rate does not vary significantly over the time scale of the transition and solving eq.~\eqref{eq:baryonNrDE} (with $\mu^0_L =0$) in the broken phase as
\begin{equation}
    \rho_b (z)\sim  \rho_b(0)
    \e^{ - \frac{ {\mathcal R}}{2 v_w}z}
    =\rho_b(0)
    \e^{ - \frac{ n_fA\Gamma_{\rm sph, b}}{2 v_w T^3}z},
\end{equation}
where $\rho_b(0)$ is the baryon number formed in the symmetric phase.
Plugging in the approximate expression for $\Gamma_{\rm sph,b} \sim T^4 e^{-E_{\rm sph}/T}$ and the expression of the sphaleron energy \cref{eq:Esph} at the time of nucleation setting $v=v_n$, and demanding that the exponent is larger than e.g. $-10$, one obtains the well-known criterion
\begin{equation}\label{eq:washout}
    \frac{v_n}{T} \gtrsim 1.
\end{equation}
There are several caveats to this criterion. 
First, allowing for a washout factor of $e^{-10}$ requires a very large $\rho_b(0)$, corresponding to very strong CPV sources, which are possibly incompatible with experimental constraints. It was pointed out in \cite{Patel:2011th} that demanding instead that the washout factor is at most $1/e$ results in order $10\%$ corrections to the right-hand-side of \cref{eq:washout}. Moreover, as noted above, the expression for the electroweak sphaleron energy \cref{eq:Esph} is gauge-dependent, making the washout criterion \cref{eq:washout} gauge-dependent as well. 
Further gauge-dependence could enter when the nucleation temperature is not computed in a gauge-independent way, \emph{cf.} \cref{sec:VTaccuracy}.

The sphaleron rate is computed to leading order in a gauge-independent approach in \cite{Li:2025kyo}. 
The washout criterion is determined from the requirement that the sphaleron rate should freeze out \emph{above} the percolation temperature, which can also be derived in a gauge-independent manner. The result of the perturbative computation of \cite{Li:2025kyo} is remarkably close to the lattice result of \cite{Annala:2025aci}, which finds the following bound on $x$ in \cref{eq:xy}
\begin{equation}
    x(T_c) \lesssim 0.025.
\end{equation}
This criterion can be straightforwardly applied to any BSM theory which maps into the effective theory of \cref{eq:SM3dEFT}.
For theories that do not map into \cref{eq:SM3dEFT}, a perturbative estimate of the washout criterion could be obtained by a computation analogous to the one of \cite{Li:2025kyo} for the appropriate effective theory.


\subsection{Summary of \cref{sec:sphalerons}}
After computing the CP-asymmetry from the Boltzmann equation in \cref{sec:bubbles}, we can finally compute the baryon asymmetry.
This is done by solving \cref{eq:baryonNrDE}, with the solution given in \cref{eq:YbFinal}.
The result $Y_b$ depends on the symmetric phase sphaleron rate, the CP-asymmetry in the symmetric phase and the wall velocity.
It is crucial to note that \cref{eq:YbFinal} is valid only if the sphaleron rate in the broken phase is sufficiently suppressed. For models that map onto the Standard Model-like effective theory described by \cref{eq:SM3dEFT}, this condition corresponds to 
$x(T_c)\lesssim 0.025$. In other models, this bound must be reevaluated by computing the sphaleron rate in a gauge-independent manner and comparing it to the Hubble rate.

The symmetric and broken phase sphaleron rates can be obtained most accurately on the lattice. Simulations have been performed for the Standard Model-like EFT of \cref{eq:SM3dEFT}.
The best value for the symmetric-phase rate is given by \cref{eq:SphsBest}. 
The broken phase sphaleron rate can be read off from the data set \cite{annala_2025_15599387}.
For EWBG models that do not map onto \cref{eq:SM3dEFT} -- e.g. two-step phase transitions -- the sphaleron rate receives corrections.
As long as lattice results are unavailable for such models, one has to resort to a perturbative computation of the sphaleron rate. Care should be taken to avoid gauge-dependence.

Finally, we remark that we have assumed that the CP-asymmetry has been computed in the generalized velocity perturbation (\emph{cf} \cref{sec:gen_velocity}) in this section.
If the asymmetry gets computed in the extended fluid Ansatz, like in \cite{Dorsch:2021ubz}, the interactions of the sphaleron with the fluid and \cref{eq:baryonNrDE} have to be recomputed.

    \newpage
	\section{Experimental probes and constraints}\label{sec:experiment}

An attractive feature of EWBG is its testability. The crucial BSM ingredients to fulfill the Sakharov conditions (with the potential exception of baryon-number violation) can be readily probed in existing and future experiments. In this section, we discuss the most promising ways to test EWBG. We focus on the identification of new sources of CP violation through electric dipole moments, on collider measurements of the modified scalar sector that induces the electroweak FOPT, and the detection of gravitational waves that are induced at the same phase transition.

\subsection{Constraints from electric dipole moment measurements}
\label{sec:EDM}

Within the context of electroweak baryogenesis, the second Sakharov condition can be interpreted as requiring CP violation from beyond-the-Standard-Model (BSM) physics. That is, even with a strong first-order phase transition the amount and structure of CP violation from the CKM mechanism and the topological QCD theta term are not sufficient to generate sufficient baryon asymmetry. The search for electric dipole moment (EDMs) provides generally the most powerful probes of BSM CP violation and therefore plays an important role in the falsification and hopefully detection of EWBG models. In fact, some developments in EWBG model building have been inspired to avoid the stringent constraints from EDM searches, for example two-step phase transitions. In this section, we will briefly review the field of EDMs and how to connect BSM sources of CP violation to experimental measurements. This is not trivial, because EDM experiments typically involve complex systems such as molecules, atoms, and hadrons. Excellent dedicated reviews on EDMs can be found in Refs.~\cite{Pospelov:2005pr, Engel:2013lsa}.

In non-relativistic electrodynamics the interaction of a fermion with spin $\vec S$ (e.g. $\vec S =\frac12 \vecS \sigma $ for a spin-1/2 system) with an electric field $\vec E$ and magnetic field $\vec B$ is described by the Hamiltonian
\be\label{EDM0}
H = - \mu (\vec S \cdot \vec B) - d (\vec S \cdot \vec E), 
\ee
with $\mu$ and $d$ the magnetic and electric dipole moment respectively.  Turning on a magnetic (electric) field puts a torque on the system leading to spin precession with angular velocity $\omega = 2\mu |\vec B| \sin \theta\,$  ($\omega = 2 d |\vec E| \sin \theta$) where $\theta$ is the angle between the spin polarization and the external fields. This forms the basic idea of many EDM experiments. 
The essential idea is to measure the spin precession in presence of both an electric and magnetic field and then flip the direction of the electric field and see if there is a shift in the precession frequency. This approach of course only works for an electrically neutral system as otherwise the particle would accelerate out of the setup.

The vectors $\vec S, \vec B$, and $\vec E$ transform, respectively, under a time-reversal $T: \, t \to -t$ and parity $P:\, \vec x \to - \vec x$
 as
\begin{align}
&T: & \vec S& \to -\vec S,&
\vec B &\to -\vec B, &
                     \vec E &\to \vec E, \nn \\
                              &P: & \vec S& \to \vec S,&
\vec B &\to \vec B, &
\vec E &\to -\vec E\,.
\end{align}
The transformations of $\vec E, \vec B$ can be deduced from the Lorentz force $\vec F = m \vec a = q(\vec E + \vec v \times \vec B)$, under $T$: $\vec a \to \vec a $ and $\vec v \to -\vec v $, while under $P$: $\vec a \to -\vec a $ and $\vec v \to -\vec v $. Spin is a pseudovector and behaves as angular momentum $\vec L = \vec r \times \vec p$. From these properties we can induce that 
magnetic dipole moments (MDMs) are invariant under $T$ and $P$, but EDMs violate both $T$ and $P$.  The CPT theorem tells that $T$ violation is equivalent to CP violation and thus measurements of EDMs are typically interpreted as probes of CP violation.
The EDM appearing in \cref{EDM0} is typically not measured directly for elementary particles. An exception is the muon EDM that has been probed in the muon g-2 experiment at BNL \cite{Muong-2:2008ebm}. 

It will be useful to see how the EDM appears in a Lorentz invariant form. For example, for an elementary or composite (like the neutron) spin $1/2$ fermion, the relativistic EDM operator can be written as 
\begin{align}\label{EDMdim5}
\mathcal L = - \frac{d}{2} \bar \Psi\sigma^{\mu\nu}i\gamma_5 \Psi\,F_{\mu\nu}\,,
\end{align}
where $d$ is the EDM with mass dimension $-1$. Limits on $d$ for various particles are typically provided in units of $e\,\mathrm{cm}$. 

In the context of EWBG, the goal is to compute $d$ in terms of CP-violating sources that drive the generation of the chiral asymmetry in front of the bubble wall. For elementary SM fermions (a quark or lepton) we can draw some general lessons from EFT arguments. The EDM operator in \cref{EDMdim5} flips the chirality of the fermion (it couples $\Psi_L$ to $\Psi_R$) and therefore is not invariant under SU(2)$_L$ gauge symmetry. A gauge-invariant extension of the EDM operator necessarily involves the Higgs field \cite{DeRujula:1990db}. For example, gauge-invariant operators relevant for the EDMs of charged leptons can be defined as
\begin{align}\label{EDMdim6}
\mathcal L = - \frac{C_{eB}}{\sqrt{2}} \bar L \sigma^{\mu\nu} e_R\,\varphi\,B_{\mu\nu} - \frac{C_{eW}}{\sqrt{2}} \bar L \sigma^{\mu\nu} \tau^a e_R\,\varphi\,W^a_{\mu\nu}\,,
\end{align}
in terms of the first-generation lepton doublet $L=(\nu_L\,e_L)^T$, the Higgs field $\varphi$, and, respectively, the U(1)$_Y$ and SU(2)$_L$, field strengths $B_{\mu\nu}$ and $W_{\mu\nu}^a$. After electroweak symmetry breaking we can read off the electron EDM as
\begin{align}
d_e = v \left( \cos \theta_W\,\mathrm{Im}[C_{eB}]-\sin \theta_W\,\mathrm{Im}[C_{eW}]\right)\,,
\end{align}
in terms of the Higgs vev, $v$, and the Weinberg angle $\theta_W$. Very similar operators can be written down for the EDMs of quarks (and second- and third-generation charged leptons). The task is then to compute the Wilson coefficients ($C_{eB}$ and $C_{eW}$) in terms of the underlying EWBG model. 

The main lesson from this exercise is that EDMs of elementary fermions are actually dimension-six operators due to the presence of the Higgs field in \cref{EDMdim6}. As such, in BSM models EDMs typically scale as $1/\Lambda^2$ where $\Lambda$ is the scale of BSM physics. Furthermore, EDMs require a chirality flip and it is often assumed that EDMs scale with the fermion mass,  i.e.  $d_e \sim m_e/\Lambda^2$. While this is often true in explicit BSM models, there are examples where so-called chiral enhancements occur. A good example is provided by scalar leptoquark models where $d_e \sim m_t/\Lambda^2$ is possible albeit with a loop suppression \cite{Dekens:2018bci}. 

The fact that elementary EDMs are dimension-six operators can also be seen from a different perspective. If we treat the SM as a low-energy EFT of a more fundamental theory at a higher energy, we can extend the SM with higher-dimensional operators containing just SM fields and obeying the SM Lorentz and gauge symmetries. The resulting EFT, called the SMEFT (see also \cref{sec:SMEFT_FOPT}), contains an infinite number of operators which can be ordered by their mass dimension.
The first operator has dimension five and is related to neutrino Majorana masses (the famous Weinberg operator) and is not relevant for EDMs. At the level of dimension-six operators there are, however, many operators that violate CP and can contribute to EDMs. Depending on the UV-model under consideration, the dominant contribution to EDMs of elementary or composite systems can come from different dimension-six operators.

 All dim-6 SMEFT operators have been constructed \cite{Grzadkowski:2010es}. We do not wish to provide an exhaustive list of operators (see e.g. Refs~\cite{Dekens:2013zca,Kley:2021yhn,Kumar:2024yuu} for such an analysis) but it is useful to discuss various classes of operators. 
\begin{itemize}
\item{\textbf{Dipoles.}} These operators contain the operators in \cref{EDMdim6} for all SM fermions. They can be flavor-conserving and flavor-violating but the latter only contribute to EDMs with additional electroweak loops. In addition, there appear so-called chromo-electric dipole moments (CEDMs) of quarks involving the gluonic field strength
\begin{align}
\mathcal L_{\mathrm{qCEDM}} = - \frac{C_{dG}}{\sqrt{2}} \bar Q \sigma^{\mu\nu} t^a d_R\,\varphi\,G^a_{\mu\nu} - \frac{C_{uG}}{\sqrt{2}} \bar Q \sigma^{\mu\nu} t^a u_R\,\tilde \varphi\,G^a_{\mu\nu}\,.
\end{align}
The dipole operators are generated in many BSM models at the one-loop level. Famous examples are supersymmetric scenarios where quark and lepton EDMs and quark CEDMs are induced by one-loop diagrams with external fermions and SM gauge bosons and virtual sfermions, charginos, neutralinos, and gluinos \cite{Ellis:1982tk,Buchmuller:1982ye}. 

\item{\textbf{Yukawa.}} These operators multiply the structure of the SM Yukawa interactions by the gauge-singlet combination $\varphi^\dagger \varphi$ giving rise to:
\begin{equation}\label{eq:dim6Y}
\mathcal L_Y = - \bar Q_LY_d\varphi d_R\,(\varphi^\dagger \varphi) - \bar Q_LY_u\tilde \varphi u_R\,(\varphi^\dagger \varphi) - \bar L_L Y_e\varphi e_R\,(\varphi^\dagger \varphi)\,,
\end{equation}
where the Yukawa couplings $Y_{d,u,e}$ are $3\times3$ matrices in generation space. Within the SM itself, going to the mass basis implies that all fermion-Higgs interactions are flavor-diagonal and CP-conserving but this is no longer the case in the presence of the dim-6 Yukawa structures. For EDM purposes the most important effect is that the flavor-diagonal fermion-Higgs couplings are no longer CP conserving (if $Y_{d,u,e}$ have an imaginary part) and contribute to EDMs of light quarks and leptons through two-loop Bar-Zee diagrams (see \cref{EDMexamples}). In the context of EWBG, the dim-6 top-Yukawa has been a commonly used source of CP violation \cite{Huber:2006ri,Balazs:2016yvi,deVries:2017ncy,Fuchs:2020uoc}, as we did in our vanilla model in the introduction ({\it cf} \cref{CP_yukawa}), but also the $\tau$-lepton variant and lighter quarks and leptons \cite{DeVries:2018aul, Fuchs:2020uoc} have been considered.

\item{\textbf{Gauge or Higgs-gauge interactions}.} There are 6 independent CPV dim-6 operators involving no fermions. 4 of these involve 2 Higgs fields and 2 field strengths, and 2 contain just 3 field strengths \cite{Buchmuller:1985jz}: 
\begin{align}
\label{HiggsGauge}
\mathcal L &=
 -g^2 C_{\varphi  \tilde W} \, \varphi^\dagger \varphi  \, \tilde W^i_{\mu \nu} W^{\mu \nu}_i 
 -g'^2 C_{\varphi \tilde B} \, \varphi^\dagger \varphi     \,  \tilde B_{\mu \nu} B^{\mu \nu}
 - g_s^2 C_{\varphi  \tilde G}\, \varphi^\dagger \varphi    G^a_{\mu \nu} \tilde G^{\mu \nu}_a
\notag
\\
& - g g' C_{\varphi \tilde{W} B} \, \varphi^\dagger \tau^i  \varphi     \,  \tilde W^i_{\mu \nu}  B^ {\mu \nu} +  \frac{C_{\tilde G}}{3} \, f_{abc} \tilde G_{\mu\nu}^a G^{\nu\rho}_b G^{c\,\mu}_\rho  + 
\frac{C_{\tilde W}}{3}  \ \eps_{ijk} \tilde W_{\mu\nu}^i W^{\nu\rho}_j W^{k\,\mu}_\rho \,. 
\end{align}
Some of these operators have been considered as a source for EWB in Ref.~\cite{Dine:1990fj}. 
After EWSB, the first four operators lead to CPV interactions between at least one Higgs bosons and various combinations of photon, $Z$, and $W^{\pm}$ bosons. In addition, the third operator shifts the value of the QCD $\bar \theta$ term. The last 2 operators generate CPV cubic, quartic, and quintic gauge interactions. The operator with coefficient $C_{\tilde G}$ is often called the Weinberg operator (not to be confused with the dimension-five operator that generates neutrino masses). It is for example induced in two Higgs doublet \cite{Weinberg:1989dx} or leptoquark models \cite{Abe:2017sam}. We will see below that EDM constraints on operators in Eq.~\eqref{HiggsGauge} are very stringent.

\item{\textbf{Four-fermion operators.}} There exist a multitude of CPV dim-6 operators containing four fermion fields. For EDM purposes the most important are four-quark operators involving (light) quark fields and semi-leptonic operators involving 2 quark and 2 electron fields (see Ref.~\cite{Ardu:2025rqy} for the realization of these operators in SMEFT). 
The former contribute to EDMs of nucleons and diamagnetic atoms, while the latter contribute to paramagnetic systems. 
Very schematically, at low energies, they can be represented as
\begin{equation}
\mathcal L_{4\Psi} = C_S\,\bar \Psi\Psi\,\bar \Psi' i\gamma^5 \Psi'+C_T\,\bar \Psi \sigma^{\mu\nu}\Psi\,\bar \Psi' i \sigma_{\mu\nu}\gamma^5 \Psi'\,,
\end{equation}
where $\Psi$ and $\Psi'$ denote various quark and leptons (see e.g. \cite{deVries:2012ab} for a more thorough discussion). In the SM itself, after integrating out heavy quarks and gauge bosons, four-quark and semi-leptonic four-fermion operators provide the dominant contribution to EDMs \cite{Khriplovich:1981ca,Ema:2022yra}.
In left-right symmetric \cite{Dekens:2014jka} and leptoquark models \cite{Dekens:2018bci}, respectively, CPV four-quark and semi-leptonic operators induce large EDMs. 
\end{itemize}

Depending on the explicit BSM scenario under consideration different CPV dim-6 operators can be induced and the resulting EDM phenomenology can be quite different. Before discussing this in more detail, we first discuss the various classes of EDM experiments that provide the most stringent constraints. 

\subsubsection{Classes of EDM experiments}

Almost all EDM experiments do not actually measure EDMs of elementary particles directly. The experiments that set the most stringent limits involve polar molecules, diamagnetic atoms, and neutrons. From the experimental results, molecular, atomic, nuclear, hadronic, and particle physics theory is then required to set limits on the fundamental sources of CP violation, see for example Ref.~\cite{Degenkolb:2024eve} for a recent global analysis. For this review, we will divide experiments into several distinct classes based on the sources of CP violation they are most sensitive too. 
\begin{itemize}
\item \textbf{Neutron EDM}. This was historically the first system used for an EDM experiment in the experiment by Ramsey \cite{Smith:1957ht} using a neutron beam. Interestingly, this experiment was originally done to test parity symmetry in the strong interaction. Nowadays, the best limit on the neutron EDM is obtained using ultracold neutrons at the PSI experiment \cite{Abel:2020pzs}
\begin{equation}\label{eq:EDMLimitNeutron}
d_n < 1.8 \cdot 10^{-26}\,e\,\mathrm{cm}.
\end{equation}
Unfortunately, mainly due to problems related to the neutron source, the experimental improvement has slowed down in recent years with only a factor 2 improvement in a decade. Ongoing experiments such as those at ILL, PSI, and Oak Ridge aim to improve the limit of \cref{eq:EDMLimitNeutron} by a factor 10 \cite{Athanasakis-Kaklamanakis:2025xcg}. The neutron EDM can be used to set limits on CP-violating operators involving quarks and gluons such as the QCD $\bar \theta$ term, quark EDMs, quark chromo-EDMs, the Weinberg three-gluon operator, and four-quark operators. 
\item \textbf{Diamagnetic systems.} Diamagnetic atoms are neutral systems where the total angular momentum of the electrons combines to zero. This suggests that these systems are sensitive to the EDM of the atomic nucleus.
However, the Schiff shielding theorem \cite{Schiff:1963zz} says that the EDMs of charged constituents that combine to a neutral system (such as an atom) are effectively screened from external electric fields, implying that the nuclear EDM does not contribute to the atomic EDM. The Schiff shielding is based on non-relativistic particles (a very good approximation for the nucleus) and point-like constituents. It is the latter assumption that is violated and the Schiff screening theorem is (somewhat) avoided by the finite size of the nucleus. More precisely, an atomic EDM is induced by the difference between the electric dipole radius and the charge radius which is often called the nuclear Schiff moment. The screening theorem implies that the EDM of the atom is suppressed by the EDM of the nucleus by  roughly a factor $Z^2 (r_{\mathrm{nucleus}}/r_{\mathrm{atom}})$. This is a severe $10^{-10}$ penalty for hydrogen but a milder $10^{-4}$ suppression for a heavy system such as ${}^{199}$Hg \cite{Griffith:2009zz,Graner:2016ses}, which currently is the most stringent diamagnetic system, yielding the bound
\begin{equation}\label{Hgbound}
d_{^{199}\mathrm{Hg}} < 6.3 \cdot 10^{-30}\,e\,\mathrm{cm}\,,
\end{equation}
obtained nearly 10 years ago.

Certain diamagnetic atoms are expected to have an enhanced sensitivity to nuclear CP violation due to octopole deformation of the nucleus \cite{Spevak:1995zem}. One such candidate is ${}^{225}$Ra for which an experimental bound has been set of $d{^{225}\mathrm{Ra}} <1.2 \cdot  10^{-23}$ e cm \cite{Bishof:2016uqx}. While still many orders of magnitude weaker than the limit on $d_{^{199}\mathrm{Hg}}$, this is partially compensated by the octopole enhancement by roughly two orders of magnitude. The Schiff moments of atomic nuclei can also be measured in molecular systems where big enhancements are possible (see the next item). For example, the CenTREX collaboration aims to measure the Schiff moment of ${}^{205}$Tl using the TlF polar molecule \cite{Grasdijk_2021}. Other experiments focus on radioactive molecules \cite{Arrowsmith-Kron:2023hcr}. 

A key difficulty to interpret diamagnetic EDMs in terms of underlying sources of CP violation, is the calculation of the nuclear Schiff moment. For most hadronic CPV operators such as the $\bar \theta$ term or  quark chromo-EDMs, the Schiff moment is dominated by CPV nuclear forces. The associated nuclear many-body calculation is notoriously difficult leading to large uncertainties \cite{deJesus:2005nb,Engel:2013lsa, Yanase:2020agg}. In some cases, different calculations do not even agree on the sign such that the theoretical range of the Schiff moment includes zero. This makes it hard to use the impressive experimental limit to test models of EWBG. 

\item \textbf{Paramagnetic atoms and molecules.} Paramagnetic systems have an unpaired electron and are therefore sensitive to the electron EDM. The Schiff shielding theorem is overcome by the fact that in heavy systems electrons are relativistic \cite{Ginges:2003qt}. In large systems, the large internal electric fields can enhance the EDM of the system with respect to the EDM of the electron. This was already true for the atomic measurements with ${}^{205}$Tl \cite{Regan:2002ta} but is much more effective in polar molecules. Within polar molecules, relatively small external electric fields can lead to giant internal fields enhancing the EDM signal (in practice, an electron energy level splitting). Since the first EDM measurement performed on the polar molecule YbF \cite{Hudson:2011zz}, limits have improved by two orders of magnitude \cite{ACME:2018yjb,Roussy:2022cmp} and more progress is expected \cite{NL-eEDM:2018lno,Alarcon:2022ero,Athanasakis-Kaklamanakis:2025xcg}. The current best limit on the electron EDM comes from the HfF$^+$ JILA experiment \cite{Roussy:2022cmp} 
\begin{equation}
d_e < 4.1 \cdot 10^{-30}\,e\,\mathrm{cm}\,.
\end{equation}
We stress that this is not a direct measurement of the electron EDM, but an inferred limit based on a theoretical calculation of a molecular matrix element connecting the measured level splitting to the electron EDM, see e.g. Refs.~\cite{Meyer:2008gc,NL-eEDM:2021wtn}. 

While paramagnetic EDM experiments are typically interpreted as a limit on the electron EDM, the measured level splitting is actually a combination of various CP-violating contributions. Next to the electron EDM, there are contributions from CPV electron-nucleus interactions which can be important. In the SM itself, the phase of the quark-mixing matrix induces EDMs of a polar molecule mainly through this effective electron-nucleus interaction \cite{Ema:2022yra} and the EDM of the electron is negligible \cite{Hoogeveen:1990cb}. In certain BSM scenarios, for example scalar leptoquarks, the CPV electron-nucleus interactions are induced by semi-leptonic electron-quark dimension-six operators and can be much larger than the electron EDM contributions \cite{Dekens:2018bci,Fuyuto:2018scm,Ardu:2025rqy}.

Finally, it is good to mention that paramagnetic experiments have improved much more than other systems in recent years. While paramagnetic systems are mainly sensitive to (semi-)leptonic CP violation, also purely hadronic CP-violating sources can, through multi-loop diagrams, contribute to paramagnetic EDMs. At present such indirect limits are not yet competitive with limits derived form the neutron EDM and diamagnetic EDMs, but if paramagnetic experiments improve by two orders of magnitude this will change \cite{Flambaum:2019ejc, Mulder:2025esr}. 

\item \textbf{Storage ring experiments.} Traditional EDM experiments aim to measure the change in the spin precession of the system in the presence of an external electric field. This is not possible with charged particles which would just escape the experimental environment. Therefore most experiments use neutral systems and this brings all the resulting complications of the Schiff shielding theorem. It was realized that EDMs of charged particles can be measured in a storage ring \cite{Farley:2003wt}. The best direct limit on the muon EDM has been set by the g-2 collaboration in such a setup \cite{Muong-2:2008ebm} and this is a rare case of a direct EDM measurement of an elementary particle.\footnote{However, a stronger indirect limit arises from the contribution of the muon EDM to atomic and molecular EDMs \cite{Ema:2021jds}.} A dedicated storage ring muon EDM experiment has been proposed \cite{Hu:2024wuc, Adelmann:2025nev}.

Storage ring experiments could perform direct measurements of the EDMs of the proton and light nuclei. Light nuclei are attractive because the nuclear theory is well under control and the interpretation of the measurement is far cleaner than that of large diamagnetic atoms. In addition, the EDM of a deuteron can be significantly larger than the EDM of a neutron or proton because of CP-violating nuclear forces. It has been argued that measuring the EDMs of several light nuclei could separate the QCD $\bar \theta$ term from BSM operators \cite{Dekens:2014jka}. 
This could be very important in testing scenarios of EWBG. We refer to Refs.~\cite{Alarcon:2022ero,Athanasakis-Kaklamanakis:2025xcg,Dutsov:2025kbd} for discussion of the experimental prospects.
\end{itemize}

\subsubsection{EDMs in the Standard Model}

One of the reasons why EDMs are powerful ways to search for BSM physics in general and mechanisms of EWBG in particular, is that they are `background-free' observables. That is, the only confirmed source of CP violation in the form of a complex phase in the CKM matrix induces EDMs that are too small to be measured with current and foreseen experiments. However, it is still interesting to discuss how small CKM-induced EDMs are.\footnote{We do not call these Standard Model EDMs to differentiate from EDMs induced by the QCD $\bar \theta$ term.} This is a long-standing issue with, remarkably, still new developments. 

Because the CKM phase appears in the flavor-changing charged quark weak current, it is easy to see that EDMs of elementary quarks vanish at one loop electroweak order because all such diagrams are proportional to $\sim V_{ij} V^*_{ij}$ and are thus real. Alternatively this can be understood from the fact that any CPV observables must be proportional to the Jarlskog determinant \cite{Jarlskog:1985cw} and thus requiring at least 4 electroweak vertices. In principle (chromo-)EDMs of quarks could be expected to appear at two-loop level but the sum of all these diagrams vanishes \cite{Czarnecki:1997bu}. At three-loop level the first quark EDMs are induced but they are very small \cite{Czarnecki:1997bu} leading to a neutron EDM of $\mathcal O(10^{-34}\,e\,\mathrm{cm})$, a severe 8 orders of magnitude below present limits. A more efficient path is possible. Integrating out heavy gauge bosons and quarks leads to a strangeness-changing four-quark operator proportional to the CKM phase. By using chiral perturbation theory it is then possible to compute long-distance (long distance here means $1/m_K$ where $m_K$ is the kaon mass) meson-loop contributions to the neutron EDM at a level of $\mathcal O(10^{-31,-32}\,e\,\mathrm{cm})$ \cite{Khriplovich:1981ca,Seng:2014lea}, still too small to be observed. 

The electron EDM is even smaller and appears at 4-loop level \cite{Hoogeveen:1990cb, Pospelov:1991zt} with a tiny value $\mathcal O(10^{-41}\,e\,\mathrm{cm})$, roughly 11 orders of magnitude below the current limit. Similarly to the neutron EDM, long-distance meson loops can enhance this to $\mathcal O(10^{-40}\,e\,\mathrm{cm})$ \cite{Yamaguchi:2020eub}.
However, this contribution, while often quoted, is essentially irrelevant as the electron EDM is measured inside atoms or molecules and much larger contributions to the atomic/molecular EDMs are induced through CKM-induced CPV electron-nucleus interactions. Only very recently an efficient two-loop path leading to a CPV kaon exchange between electrons and nucleons was identified. This leads to polar molecule EDMs that are roughly 5 orders of magnitude below the present limits \cite{Ema:2022yra}. Interesting, but still too small to be seen in the near future. 

While not confirmed experimentally it is likely that there is another source of CP violation in the neutrino-mixing matrix. Since the mechanism of neutrino masses is not understood it is unclear what this means for EDMs. In a type-1 seesaw mechanism there can be additional sources of CP violation in the sterile neutrino sector which, depending on the sterile neutrino mass, can lead to somewhat enhanced EDMs. However, they are again too small to be detected \cite{Archambault:2004td, Abada:2024hpb}. 

Finally, we discuss the QCD $\bar \theta$ term for which EDMs are very important. In fact, the non-observation of EDMs directly leads to the strong CP problem and forms the main motivation for axions, one of the hottest topics in present-day particle physics (based on the number of appearing papers). We will not review all the physics of the QCD $\bar \theta$ term as excellent sources exist, see e.g. Ref.~\cite{Hook:2018dlk}, but instead discuss the connection to EDMs. Focusing on the lightest two quarks, the relevant part of the QCD Lagrangian is given by
\begin{equation}
\mathcal L_{m,\tb} = -\left(e^{i \rho}\,\bar q_L M q_R + e^{-i \rho}\,\bar q_R M q_L \right)- \frac{\theta g_s^2}{64\pi^2}\epsilon^{\mu\nu\alpha\beta}G_{\mu\nu}^a G_{\alpha\beta}^a\,\,\,, 
\end{equation}
in terms of  a mass matrix $M = \bar m (1+\varepsilon \tau^3)$ where $\bar m=(m_u+m_d)/2$ is the average quark mass and $ \varepsilon=(m_u-m_d)/(m_u+m_d)$ the normalized mass splitting, the strong coupling constant $g_s$, the overall phase $\rho$, and the vacuum angle $\theta$. By performing an anomalous U(1)$_A$ transformation to eliminate the gluonic $\theta$ term in favor of a complex mass term proportional to $\tb = \theta+2\rho$ (this is the physical combination that is measured in experiments) we can write
\begin{equation}\label{theta}
\mathcal L_{m,\tb} = -\bar m \left( \bar q q+ \varepsilon\,\bar q \tau^3 q - \frac{1-\varepsilon^2}{2}\tb\,\bar q i\gamma^5 q\right)\,\,\,,
\end{equation}
where we already assumed $\tb \ll 1$. From this we can directly estimate the neutron EDM using naive dimensional analysis. The neutron EDM should be proportional to 
\begin{equation}
    d_n \sim \frac{e \bar m\,(1-\varepsilon^2)}{2 M^2_{\rm{had}}}\tb\,,
    \end{equation}
    where $M_{\rm{had}}$ is some hadronic scale inserted to get the right dimension for the neutron EDM and we inserted an electromagnetic coupling $e$. While the exact value of $M_{\rm{had}}$ is not clear, if we set it to the neutron mass and use $\bar m \sim 5$ MeV and $\varepsilon \sim -1/3$, we obtain
    \begin{equation}
d_n \sim 10^{-16}\,e\,\mathrm{cm}\,,
    \end{equation}
    and thus the current nEDM limit \cite{Abel:2020pzs} sets $\bar \theta \lesssim 10^{-10}$. While this estimate is naive, the same calculation has been done with QCD sum rules \cite{Pospelov:1999ha}, chiral perturbation theory \cite{Crewther:1979pi,Ottnad:2009jw,Mereghetti:2010kp}, and lattice QCD \cite{Dragos:2019oxn,Liang:2023jfj} essentially coming to the same conclusion: $\bar \theta$ is tiny.\footnote{The impressive progress made in paramagnetic molecular EDM experiments can be used to set an independent constraint $\tb <10^{-8}$ \cite{Flambaum:2019ejc,Mulder:2025esr}. Another round of improvements might make these experiments more sensitive to $\bar \theta$ than the neutron EDM.} 

The strong CP problem is the problem (a better word might be puzzle instead of problem) why $\bar \theta$ is so small. Interestingly, within the SM the smallness of $\bar \theta$ is technically natural because radiative corrections to $\bar \theta$ are small. The first radiative corrections induced by the CKM phase appear at high loop order and are at the level of $\bar \theta \sim 10^{-16}$-$10^{-18}$ \cite{Ellis:1978hq, Khriplovich:1985jr,Gerard:2012ud}. This is no longer true in generic extensions of the SM and this implies that generic scenarios of EWBG \textit{worsen} the strong CP problem. The reason is that $\bar \theta$ is a marginal coupling and threshold corrections to it do not necessarily decouple if new degrees of freedom are made heavy \cite{deVries:2018mgf}. Simply put: new sources of CP violation in BSM models in general lead to large threshold corrections to $\bar \theta$. This problem is typically neglected in the EWBG literature as it can be avoided by considering an infrared relaxation of $\bar \theta \rightarrow 0$ through an axion mechanism. Of course, this is a rather big assumption and should be kept in mind when studying explicit EWBG scenarios. Interestingly, measurements of several nonzero EDMs could pinpoint whether an axion mechanism is active or not \cite{deVries:2021sxz} without measuring an actual axion. In the rest of this review we will ignore the $\bar \theta$ term and assume that the strong CP problem is solved in some unspecified manner, be it fine tuning or axions.

\subsubsection{Examples of EDM constraints.}\label{EDMexamples}

To illustrate how to use EDMs to constrain EWBG we discuss a few examples. We stress that these are far from complete and that every explicit EWBG model requires its own analysis. That being said, many parts of the analysis are universal and some general lessons can be learned from the given examples. 

\begin{figure}[t!]
    \centering
    \includegraphics[width=0.8\textwidth]{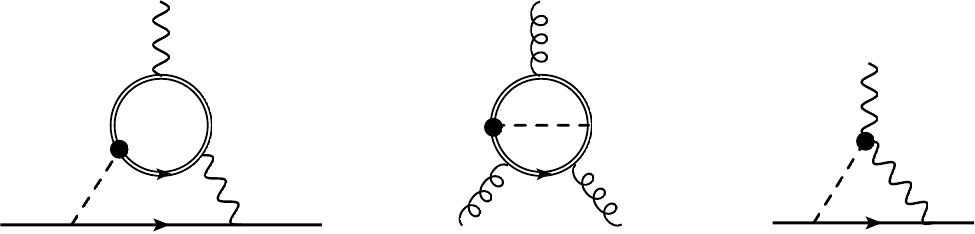}
    \caption{Various diagrams contributing to CPV effective operators. The single solid line denotes an electron, the double sold line a top-quark, the dashed lines a Higgs, the wavy lines a photon or a virtual $Z$, and the curly line a gluon. The black circle denotes a CPV vertex. 
    Left: a two-loop Barr-Zee diagrams connecting a CPV top-Higgs Yukawa to the electron EDM. Middle: A two-loop diagram connecting the same Yukawa to the three-gluon Weinberg operator. Right: a one-loop diagram connecting a CPV Higgs-gauge-gauge coupling to the electron EDM.}
    \label{fig:EDM1}
\end{figure}

We begin with a rather minimal example where CP violation originates in a single dimension-six Yukawa interaction of the form in \cref{eq:dim6Y}.
 The imaginary part of $Y_i$ gives rise to a CP-violating fermion-Higgs interaction. We first consider couplings to top quarks. The strongest limits come from two-loop Barr-Zee diagrams \cite{Barr:1990vd} shown in Fig.~\ref{fig:EDM1}. The left diagram induces the EDM of the electron (or other charged fermions) 
\be
\frac{d_e}{e} = -  m_e\frac{8 n_c Q_t^2\alpha_{\rm em}}{(4\pi)^3} g(x_t) \times {\rm Im} (Y_t)\,, \qquad
g(x_t) = \frac{x_t}{2} \int_0^1 \dd x \frac{1}{x(1-x) -x_t} \ln\( \frac{x(1-x)}{x_t} \)\,,
\ee
where $n_c=3$, the number of colors, $Q_t =2/3$ the top quark charge, 
 $x_t = m_t^2/m_h^2$, and numerically the two-loop function is $g(x_t) \approx 1.4$. 
As explained above, the electron EDM is not measured directly but extracted from the molecular measurement on HfF$^+$. In particular, the eEDM induced energy splitting is given by
\begin{equation}
\omega_{\mathrm{HfF}} = (34.9 \pm 1.4)\,(\mathrm{mrad}/s)\,\frac{d_e}{10^{-27}\,e\,\mathrm{cm}}+\dots\,,
\end{equation}
where the coefficient is computed with molecular many-body theory \cite{Skripnikov:2017,Fleig:2018bsf} that is relatively well understood. 
The dots denote terms from, for example, CPV semi-leptonic electron-nucleon interactions which are not relevant in this particular scenario. The experimental limit on the frequency \cite{Roussy:2022cmp} then leads to the eEDM bound $d_e < 4.1 \cdot 10^{-30}\,e\,\mathrm{cm}$\,. This gives a limit $|v^2 {\rm Im} (Y_t)| < 4\cdot 10^{-4}$ which essentially kills this source for EWBG \cite{Bodeker:2004ws, Huber:2006ri, deVries:2017ncy}. Note that the eEDM limits were roughly 250 times weaker at the time of earlier reviews such as Ref.~\cite{Morrissey:2012db} illustrating the experimental progress. 

A possible, although not pretty, way to avoid the stringent electron EDM bound is to assume there is another contribution to the electron EDM that cancels the contribution from the CPV top Yukawa. Such scenarios have for instance been considered in Refs.~\cite{Fuyuto:2019svr,Kanemura:2020ibp,Enomoto:2021dkl}. 
Alternatively, it can be argued that the SM electron Yukawa coupling has not been measured and could still be zero \cite{Brod:2013cka}, although this requires an alternative electron mass mechanism. In such, admittedly contrived, cases we have to consider the limits from other EDM experiments. In particular, the middle diagram of \cref{fig:EDM1} induces a contribution to the CP-violating gluonic Weinberg operator
\be
\mathcal L_W = \frac{C_{\tilde G}}{3} \, f_{abc} \tilde G_{\mu\nu}^a G^{\nu\rho}_b G^{c\,\mu}_\rho\,, 
\ee
where \cite{Dicus:1989va}
\be
C_{\tilde G}(m_t) = \frac{g^3_s(m_t)}{16 (4\pi)^3} h(x_t)\times {\rm Im} (Y_t)\,,\qquad h(x_t) = 4x_t^2 \int_0^1 dx \int_0^1 du \frac{u^3 x^3(1-x)}{(x_t\,x(1-ux)+(1-u)(1-x))^2}\,,
\ee
where $h(x_t) \simeq 0.88$. The Weinberg operator should be evolved to lower energy scales. The QCD anomalous dimensions have been computed to three-loop order \cite{deVries:2019nsu} and give rise to a suppressed operator $C_{\tilde G}(1\,\mathrm{GeV}) \simeq 0.40\,C_{\tilde G}(m_t) $. The Weinberg operator then contributes to the neutron EDM through \cite{Haisch:2019bml,Yamanaka:2020kjo}
\begin{equation}
d_n = (25 \pm 12)\,\mathrm{MeV}\,e\,C_{\tilde G}(1\,\mathrm{GeV})\,,
\end{equation}
giving a bound $|v^2 {\rm Im} (Y_t)| < 0.1$, which is far weaker than the electron EDM limit. 
Of course, one can allow for further cancellations to also suppress the neutron EDM in which case EDMs of, for example, ${}^{199}$Hg or ${}^{129}$Xe should be considered. 

In similar fashion, the other CP-violating Yukawa interactions can be constrained. For bottom and charm quarks, care must be taken to consider the scale separation between the quark and Higgs thresholds when generating the Weinberg operator \cite{Brod:2023wsh}. For first-generation quarks and charged leptons the Barr-Zee diagrams involve a virtual top loop. This leads to up and down quark EDMs and chromo-EDMs for the imaginary quark Yukawa couplings and to the electron EDM for the imaginary electron Yukawa. The resulting limits are again very tight \cite{Chien:2015xha}. 

Let us consider one more example where CP violation is mitigated through effective Higgs-gauge couplings. Such interactions have been considered as a CPV source of EWBG in Ref.~\cite{Dine:1990fj}. As an example, we take
\begin{equation}
\mathcal L_{\varphi\varphi WW} = -g^2 C_{\varphi\tilde W} (\varphi^\dagger \varphi)\,W^i_{\mu\nu}\tilde W_i^{\mu\nu}\,.
\label{WWcoupling}
\end{equation}
After EWSB the piece without explicit Higgs bosons can be removed from the Lagrangian without consequences \cite{FileviezPerez:2014xju}, but the part linear and quadratic in $h$ remains. The part linear in $h$ induces EDMs of charged leptons and quarks at the one-loop level (see the right diagram of Fig.~\ref{fig:EDM1}) through the effective $ h F_{\mu\nu} \tilde F^{\mu\nu}$ and  $ h F_{\mu\nu} \tilde Z^{\mu\nu}$ components of the operator. The EDMs are proportional to the fermion Yukawa couplings and thus suppressed for the electron and first-generation quarks, but the limits are still severe. The HfF$^+$ limit leads to $v^2 C_{\varphi\tilde W} <10^{-6}$ while the neutron and diamagnetic EDMs are roughly three orders of magnitude weaker \cite{Cirigliano:2019vfc}. For comparison, prospects for the high-luminosity LHC indicate a sensitivity to $v^2 C_{\varphi\tilde W} \sim \mathcal O(0.1)$ \cite{Bernlochner:2018opw} and are thus far away from EDM limits.

Finally, we consider an example of a four-fermion operator. These are typically less interesting from the point of view of EWBG as they do not involve the Higgs field and are then not directly related to physics at the bubble wall. Nevertheless, four-fermion operators can be induced as a side effect in general BSM scenarios with extra CP violation. A good example are left-right symmetric models where CP violation in the extended scalar section of the model leads, at lower energies, to the SMEFT operator \cite{Dekens:2014jka}
\begin{equation}
\mathcal L_{LR} = \Xi\,(i \tilde \varphi^\dagger D_\mu \varphi)\,\bar u_R \gamma^u d_R + \mathrm{h.c.}\,,
\end{equation}
where $\Xi \sim 1/M_R^2$ where $M_R$ is the scale of heavy right-handed gauge bosons. The gauge-covariant part of the Higgs derivative is proportional to the SM $W$ boson. Integrating this out leads to a CPV four-quark operator
\begin{equation}
\mathcal L_{LR} = i \,\mathrm{Im} \,\Xi\,\left(\bar u_R \gamma^\mu d_R\,\bar d_L \gamma_\mu u_L -\bar d_R \gamma^\mu u_R\,\bar u_L \gamma_\mu d_L\right)\,,
\end{equation}
which, in turn, induces hadronic EDMs. 

The last step of computing hadronic EDMs is not straightforward. 
It is very difficult to compute EDMs of nucleons, nuclei, or diamagnetic atoms in terms of $\mathrm{Im}\,\Xi$ directly. It has proven useful to use chiral perturbation theory to translate the CP-violating four-quark operator (or other quark-gluon interactions) in terms of CP-violating hadronic interactions involving pions, nucleons, and heavier mesons. Pionic interactions are particularly important because of the relative lightness of pions due to their pseudo-Goldstone nature. The construction of the CP-violating chiral Lagrangian has been carried out in Refs.~\cite{deVries:2012ab, Bsaisou:2014oka}. For this particular case the most important hadronic operator is a CP-violating isospin-breaking pion-nucleon interaction
\begin{equation}
\bar L_{\pi N} = \bar g_1\,\bar N N\,\pi_0\,,
\end{equation}
where $\bar g_1$ is a QCD matrix element (often called low-energy constant in the literature) that has to be computed with nonperturbative methods. In this particular case, chiral symmetry can be used to relate $\bar g_1$ to known matrix elements extracted from kaon decays \cite{Cirigliano:2016yhc}, which gives
\begin{equation}
\bar g_1 =(2.7 \pm 1.4)\cdot 10^{-5}\,v^2 \mathrm{Im}\,\Xi\,,
\end{equation}
with a $50\%$ uncertainty.

\begin{figure}[t!]
    \centering
    \includegraphics[width=0.6\textwidth]{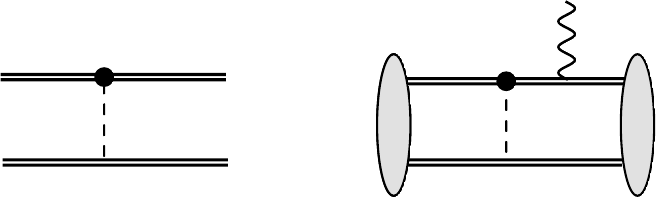}
    \caption{Left: a contribution to the CPV nucleon-nucleon force arising from neutral pion exchange. The double lines denote nucleons and the dashed lines pions. The black circle is again a CPV vertex.  Right: a contribution to the deuteron EDM arising from a CPV nucleon-nucleon force. The gray ovals on the left and right denote the deuteron wave function.}
    \label{fig:EDM2}
\end{figure}

The CP-violating pion-nucleon interaction leads to a neutron EDM at one loop,  but the leading terms vanish and the higher-order corrections are  suppressed by $m_\pi^2/m_N^2$ where $m_N$ is the nucleon mass \cite{Seng:2014pba}. Much more efficient is a tree-level contribution to the CP-violating nucleon-nucleon force shown in the left diagram of Fig.~\ref{fig:EDM2}. Such a force leads to the mixing of $S$ and $P$ waves inside nuclei and, after coupling to an external photon, to nuclear EDMs. The easiest example is the deuteron EDM whose ground state is mainly a ${}^3 S_1$ state. After the CPV pion exchange it obtains a small ${}^3P_1$ component and by coupling to an external photon (see the right diagram of Fig.~\ref{fig:EDM2}) the system can return to the ${}^3S_1$ ground state. The deuteron is simple enough that it can be treated with analytical methods. For example, using an EFT approach for nuclear forces where CP-even pion exchange is treated in perturbation theory \cite{Kaplan:1998tg} leads to the analytic expression \cite{deVries:2011re}
\begin{equation}\label{dEDM}
d_D = d_n + d_p + \frac{e g_A \bar g_1 m_N}{12 \pi F_\pi\,m_\pi }\frac{1+\xi}{(1+2\xi)^2}\,,
\end{equation}
which includes the EDMs of the nucleons and the contribution from pion exchange
in terms of the pion mass, $m_\pi \simeq 135$ MeV, the pion decay constant $F_\pi \simeq 92.2$ MeV, the nucleon axial coupling $g_A \simeq 1.26$, and the ratio of the binding momentum and the pion mass $\xi=\gamma/m_\pi$ where $\gamma \simeq 45$ MeV. For the four-quark operator the term proportional to $\bar g_1$ is significantly larger than the contribution from the nucleon EDMs \cite{Dekens:2014jka} showing that EDMs of nuclei are not simply a sum of nucleon EDMs but can be enhanced. Analytic calculations are not possible for larger nuclei but EDMs of several light nuclei have been obtained with numerical methods and more realistic nuclear forces \cite{deVries:2011an,Bsaisou:2014zwa,Gnech:2019dod,Froese:2021civ}. For the deuteron the results of these calculations are close to \cref{dEDM}.

EDMs of nuclei are currently not directly measured although plans exist for experiments with electromagnetic storage rings. More relevant for present experiments are calculations of nuclear Schiff moments. The Schiff moment is related to the first $q^2$ dependent term in the expansion of the electric dipole form factor $F_A(q^2) = d_A + S_A\,q^2+\dots $\,,
where $d_A$ is the nuclear EDM and $S_A$ the Schiff moment, and $q^2$ the photon momentum. 
Schiff moment calculations of large systems, such as ${}^{136}$Xe, ${}^{199}$Hg, or ${}^{225}$Ra are notoriously difficult because of the many-body problem. An atomic calculation is needed to relate the nuclear Schiff moment to the atomic EDM measured in experiments \cite{Ginges:2003qt,Singh:2014jca}. For such large systems, nuclear models and associated many-body approximations are necessary. Unfortunately, this leads to very uncertain predictions. For example, keeping just the $\bar g_1$ contribution for simplicity, the EDM of the ${}^{199}$Hg atom is expressed as \cite{deJesus:2005nb,Engel:2013lsa, Yanase:2020agg}
\begin{equation}
d_{{}^{199}\mathrm{Hg}} = -\left[(2.1\pm 0.5)\cdot 10^{-4}\right]\times (0.4 \pm 0.8)\,\bar g_1\,e\,\mathrm{fm}\,.
\end{equation}
The overall suppression is arising from Schiff screening and is relatively well understood. However, the evaluation of the Schiff moment in terms of $\bar g_1$ (or other CPV hadronic couplings) is far more uncertain. Not even the sign is known. This hampers the interpretation of the very precise Hg EDM limit in \cref{Hgbound}. Further theoretical studies of nuclear aspects of CP violation are direly needed, see e.g. Ref.~\cite{Zhou:2025jfi} for very recent new computations. 

The examples in this section are far from exhaustive. Nevertheless, we hope they illustrate how EDMs can be used to constrain or detect new sources of CP violation that are necessary for EWBG. Each EWBG model in principle requires a dedicated EDM analysis but by using the EFT methods outlined here, most of the heavy lifting has  been done. 

\subsubsection{Avoiding EDM constraints}
\label{sec:avoid_EDM}

The above discussion indicates that the most simple EWBG scenarios with BSM physics around the TeV scale to generate a first-order electroweak phase transition and new CP violation are stringently constrained by EDM experiments. This is even true in scenarios, such as two-Higgs doublet models, where EDMs are induced at the two-loop level \cite{Dorsch:2016nrg,Biekotter:2025fjx}.
Viable EWBG scenarios require a mechanism to avoid these EDM limits. This can be achieved in several ways:
\begin{itemize}
\item Cancellation mechanisms where large contributions to EDMs are canceled by other contributions. Depending on the source of CP violation the required cancellations can be severe and care must be taken that no other, normally less constraining, EDMs are induced. Cancellation mechanisms have been considered for example in Refs.~\cite{Fuyuto:2019svr,Kanemura:2020ibp,Enomoto:2021dkl}. While technically this provides a way to avoid EDM constraints, it is rather ad hoc and goes against the spirit of EWBG models whose main appeal is their falsifiability.
\item EDMs are flavor-conserving observables and one can therefore consider flavor-changing sources of CP violation. For example, in two Higgs doublet models effective quark-flavor changing Yukawa transitions appear. Viable baryogenesis scenarios include a top-charm CP-violating interaction \cite{Fuyuto:2017ewj,Cline:2021dkf}. As far as we are aware there is no systematic study of these flavor-changing neutral currents in the context of EDMs (but see e.g. Ref.~\cite{Dekens:2018bci} which studied CPV flavor-changing interactions induced in leptoquark models). A rough estimate shows that a top-charm Yukawa can induce a down-quark EDM with a CKM suppression $\sim V_{td} V^*_{cd}$ and a loop suppression $\frac{\alpha_{w}}{4\pi} m_d m_t/m_W^2 $ which renders EDM limits irrelevant. It would be interesting to study if there are more efficient paths to EDMs for CPV flavor-changing neutral currents. More likely, other flavor observables should be considered, see e.g. Refs.~\cite{Cline:2011mm, Kanemura:2023juv}.
\item In non-minimal scenarios with several new fields and interactions, it is possible 
to decouple the source of the strong FOPT from the CP-violating source. For example, by introducing an extra gauge-singlet scalar field that sources the strong FOPT,
EDM constraints on the 2HDM can be softened \cite{Alanne:2016wtx}.
Similar to the cancellation mechanism the downside here is that extra ingredients are added to the model just to avoid testability.
\item In two-step EWBG scenarios a scalar field can obtain a vev in the first phase transition and, in the presence of CP-violating interactions provide an efficient CP-violating source. If in the second step towards the electroweak vacuum, the scalar field  disappears, the source of CP violation is effectively screened from EDM experiments. The CPV coupling can be either to SM degrees of freedom, see e.g. Ref.~\cite{Inoue:2015pza} for an explicit scenario, or to dark matter fields \cite{Cline:2017qpe,Carena:2018cjh}. In the latter case an additional portal coupling is need to transfer the CP asymmetry from the dark to the visible sector. 

\item The additional field in the two-step phase transition is a CP odd field, and CP is spontaneously broken by the (temporary) non-zero vev \cite{Grzadkowski:2018nbc, Huber:2022ndk}. This can be viewed as an explicit implementation of spontaneous baryogenesis \cite{Cohen:1988kt,Cohen:1990py, Cohen:1991iu}.

\end{itemize}

\subsection{Collider probes}

The (minimal) BSM ingredients for successful EWBG are additional sources of CP violation, and an extended scalar sector that induces a strong FOPT. Extra sources of CP violation can in principle be studied at the LHC through CPV collider observables, see e.g. \cite{Han:2009ra, Boudjema:2015nda, Ellis:2015dha, Askew:2015mda, Demartin:2015uha,Bernlochner:2018opw}. However, the projected sensitivity in many cases is much lower than existing, let alone future, EDM constraints. For instance a global analysis of CP-violating Higgs-gauge interactions indicates that EDM constraints are 3-to-6 orders of magnitude more stringent (depending on the gauge fields involved) than (high-luminosity) LHC prospects \cite{Cirigliano:2019vfc}. 
We will therefore not discuss CPV collider observables in detail. Instead, we will focus on direct and indirect collider searches for the extended scalar sector. What is the most promising discovery channel is a rather model-dependent question. An excellent pedagogical review is provided in Ref.~\cite{Ramsey-Musolf:2019lsf} and here we will suffice with some general remarks.

The phenomenology of an extended scalar sector depends sensitively on whether the additional fields have a zero vev -- which can be enforced by $Z_2$ symmetries -- or non-zero vev.  We will refer to these set-ups as $Z_2$ models and $Z_2$-breaking models respectively. For simplicity, we focus on set-ups with one additional field, which we denote by $X$.

The most direct way to probe the scalar sector is to measure the couplings in the Higgs potential, and see if they deviate from their SM values. Unfortunately, the cross section for multiple Higgs production is small, and only weak bounds exist. Defining the deviations from the SM via $\kappa_i =\lambda_i/\lambda_i^{\rm SM}$ for $i=3,4$, the current bound on the cubic Higgs interaction is  $-1.2 \leq \kappa_3 \leq 7.2$ at the 95\% confidence level \cite{ATLAS:2022vkf,CMS:2022dwd,ATLAS:2024ish}. The quartic coupling was only recently probed for the first time, but no meaningful bounds can be put on $\kappa_4$ in the perturbative regime \cite{ATLAS:2024xcs}. We can compare this with the expectations from FOPT models. In $Z_2$ models any change to the SM Higgs potential can only be generated by quantum loop effects involving the new scalars, and deviations from the SM are suppressed, and consequently out of reach of current colliders.

This is not necessarily the case for $Z_2$-breaking models, in which deviations already come at tree-level. As argued in \cref{sec:SMEFT_FOPT}, the SMEFT is not a good framework for the FOPT, as the necessary separation of scales is lacking. Nevertheless, we can use it as a toy model to estimate the impact on the Higgs couplings. Adding a dim-6 $(\varphi^\dagger \varphi)^3/\Lambda^2$ to the SM scalar potential, this changes the triple and quartic Higgs couplings
\be
\kappa_{3} =1+\frac{16 v^4}{m_h^2 \Lambda^2}
\,,\qquad
\kappa_4 = 1 + \frac{96 v^4}{ m_h^2 \Lambda^2}\,.
\ee
Taking $\Lambda = 900\,{\rm GeV}$, a typical cutoff scale for a FOPT, gives $\kappa_3\simeq 6$ and $\kappa_4 \simeq 29$.
The deviation from the  SM are parametrically the same for the cubic and quartic coupling. Given the much stronger experimental constraint on $\kappa_3$, this will always give the dominant bound. What is interesting, is that the value of the triple Higgs coupling is of the order of the current constraint. We would of course like to understand whether such large deviations can actually arise in a concrete model implementation. In \cite{Ramsey-Musolf:2019lsf} a lower bound $\kappa_3 \gtrsim 1+0.01 T_n/(2v)$ was derived in Higgs-singlet models with a FOPT, but this bound comes with caveats, and it does not necessarily say anything about the typical values. It might be more informative then to look at actual parameter points; for the benchmark points in \cite{Huang:2017jws} the range is $\kappa_3 =1.4-3.5$, which stems more optimistic.

Although multi-Higgs production gives most direct information on the Higgs potential, it is not necessarily the observable with the largest discovery potential. The additional field $X$ in $Z_2$-breaking models has to be either a singlet or an electroweak doublet with the same quantum numbers as the Higgs doublet, all other options are already ruled out by
electroweak precision observables, and particularly the $\rho$-parameter. 
The scalar mass matrix is no longer diagonal, leading to mixing 
between the mass and flavor eigenstates.  The couplings $g$  to the fermions and gauge fields of the mostly-Higgs mass eigenstate is altered from the SM values as $g= \cos \theta g_{\rm SM}$, whereas the mostly-$X$  eigenstate couples with $g_X= \sin \theta g_{\rm SM}$, with $\theta$ the mixing angle (in the two Higgs doublet model of \cref{sec:2HDM} we used the notation $\theta =\beta -\alpha$, with $\beta$ the ratio of the doublet vevs). Non-zero mixing affects all SM rates, which can be bounded by LHC data. For example, Higgs decay data give the constraint $\sin^2 \theta \leq 0.12$ at the 95\% C.I. \cite{ATLAS:2015ciy}. Additionally, one can search for the additional $X$ field directly, via $pp \to X \to xx$, with $x$ SM fields. This process gives the strongest bounds for $x=W, Z$ gauge bosons; for $m_X \lesssim 900\,$GeV this gives a similar size bound on the mixing \cite{Huang:2017jws,ATLAS:2015pre,ATLAS:2015oxt,CMS:2013vyt,CMS:2015hra}.

The $Z_2$ symmetric models are much harder to constrain, as the interaction to the SM only is via the Higgs portal coupling $\L = \kappa_{\phi X}|X|^2|\phi|^2$. If the $X$ field is light enough, it might be probed via Higgs decay $h\to XX$ or via direct production; alternatively, one can look for changes in $Zh$-production or the triple Higgs coupling as discussed above. But for a heavier singlet $X$ this set-up can lead to the `nightmare scenario' where it cannot be probed by current and upcoming colliders \cite{Curtin:2014jma}.

\subsection{Gravitational waves} \label{sec:GWs}

The first-order phase transition, required for EWBG, causes a vacuum energy release, distributed inhomogeneously in the nucleated bubbles. This energy can partially be converted into a  stochastic gravitational wave (GW) background, for reviews see e.g. \cite{Caprini:2019egz, Hindmarsh:2020hop, Athron:2023xlk, Croon:2024mde}. A GW signal sourced at the EW scale would have a peak sensitivity in the mHz regime \cite{Grojean:2006bp}, and could therefore be observable with space-based GW detectors such as the Laser Interferometer Space Antenna (LISA) \cite{LISA:2017pwj}, Taiji \cite{10.1093/ptep/ptaa083} or next-generation (and significantly more speculative!) detectors such as Big Bang Explorer \cite{Harry:2006fi} and DECIGO \cite{Kawamura:2006up}. 

The phase transition can cause GWs via three different sources:
\begin{itemize}
	\item{Gradient energy in the scalar field,}
	\item{Sound waves in the plasma surrounding the bubbles,}
	\item{Magneto-hydrodynamic, vortical, or acoustic turbulence.}
\end{itemize}
The corresponding GW spectra have different spectral shapes, which can be predicted by a combination of analytical and numerical approaches; for GWs sourced by the scalar field energy see \cite{Kosowsky:1992rz, Kosowsky:1992vn, Huber:2008hg, Child:2012qg, Jinno:2016vai, Weir:2016tov, Jinno:2017fby, Konstandin:2017sat, Cutting:2018tjt, Jinno:2019bxw,Lewicki:2019gmv, Cutting:2020nla, Lewicki:2020jiv}, for GWs from sound waves \cite{Hindmarsh:2013xza,Hindmarsh:2015qta,Hindmarsh:2017gnf,Cutting:2019zws, Jinno:2020eqg,Jinno:2022mie, Caprini:2024gyk, Correia:2025qif}, and for turbulent contributions \cite{Kosowsky:2001xp, Caprini:2009fx, RoperPol:2019wvy, Kahniashvili:2020jgm, Auclair:2022jod, Dahl:2024eup}. 
The total GW spectrum is the sum of all different contributions, although, depending on the PT parameters, either sound waves or the scalar field energy give the dominant contribution.

The phase transition strength parameter $\alpha$, which largely determines the amplitude of the GW spectrum, is defined as\footnote{
See \cite{Giese:2020rtr, Giese:2020znk} for an alternative definition of the phase transition strength, relevant in the case where the sound speed deviates significantly from the conformal value $c_s = 1/\sqrt 3$.
}
\begin{equation}
	\alpha = \frac{4\Delta \theta}{3w}, \qquad \theta = e - 3p,
\end{equation}
where $\Delta \theta$ denotes the difference in the trace of the energy-momentum tensor between the two phases, $p$ the pressure, $e = T dp/dT - p$ the energy density, and $w = e+p$ the enthalpy. 
The pressure is obtained as $p=-V_T(\phi_{{\rm min},a}(T), T)$, where $\phi_{{\rm min},a}(T)$ with $a = \{{\rm sym, brok}\}$ are the field configurations of the symmetric and broken phase; and $V_T$ should include the field-independent part as well (proportional to $T^4$, up to sub-leading corrections).
$\Delta \theta$ and $w$ should be evaluated at the percolation temperature; for EWBG models this is typically very close to $T_n$.
Other phase transition parameters that affect the spectrum are the inverse phase transition duration (normalized by the Hubble parameter) $\beta/H_*$, \cref{eq:betaoH}, the wall velocity $v_w$, the percolation temperature $T_*$ and the speed of sound $c_s$.
Here and in the following, the subscript $*$ is used to indicate the moment of GW generation. 

We will briefly review predictions for the different contributions to the GW spectrum. 
As successful EWBG requires wall velocities smaller than the speed of light, 
typically phase transitions with moderate values of $(\alpha \lesssim 1)$ are preferred. 
In this regime, the sound wave contribution to GWs dominates \cite{Hindmarsh:2013xza,Caprini:2019egz}, and, while we will discuss all three sources, we put more emphasis on sound wave contributions.

\subsubsection{Gravitational waves from scalar gradient energy}
When the phase transition is very strong, i.e. $\alpha \gg 1$, the dominant contribution to the GW spectrum is from the gradient energy in the scalar field itself.
This source is relevant for strongly supercooled phase transitions, \emph{cf.} \cref{sec:conformal} and \cref{sec:composite}.

Originally, the source was modeled in the envelope approximation, which assumed that only the uncollided regions contributed to the GW signal \cite{Kosowsky:1992rz, Kosowsky:1992vn}. 
Numerical \cite{Huber:2008hg} and semi-analytical approaches \cite{Jinno:2016vai} found a broken power-law spectrum with an IR-tail proportional to $f^3$ and a UV-tail proportional to $f^{-1}$, see e.g. \cite{Huber:2008hg} for a numerical fit. The amplitude (up to redshift and overall numerical factors) is proportional to $K_{\nabla} (H_*/\beta)^2$, with $K_{\nabla}$ the energy budget in the gradient energy. For very strong phase transitions $K_\nabla \sim 1$.

The assumption that the collided regions do not contribute to the GW spectrum turns out to be an over-simplification \cite{Weir:2016tov, Cutting:2018tjt}, and including these regions changes the scaling of the UV and IR-tails. Lattice simulations \cite{Weir:2016tov, Cutting:2018tjt, Cutting:2020nla} and a combination of analytical modeling and simplified simulations \cite{Jinno:2017fby, Konstandin:2017sat, Jinno:2019bxw}, however, do not agree on the exact scalings.
Because of their good prospects for detectability with LISA, phase transitions with $\alpha \gg 1$ and correspondingly $\gamma_w \gg 1$ are particularly interesting. These are also most difficult to study on the lattice, and the largest value obtained is $\gamma_w \sim 8$ \cite{Cutting:2018tjt}.
To better probe the regime of strongly supercooled PTs, Refs.~\cite{Lewicki:2019gmv, Lewicki:2020jiv, Lewicki:2022pdb} combined lattice simulations of 2 bubbles with a many-bubble simulation in the thin-wall limit. 
Ref.~\cite{Caprini:2024hue} provides a template for the GW spectrum for strong phase transitions based on \cite{Lewicki:2022pdb}.

\subsubsection{Gravitational waves from sound waves}
We will now discuss the GW spectrum from sound waves. In the following, we will discuss a relatively simple fit for the GW spectrum, which is expected to be valid for rather weak phase transitions. 
This fit was obtained from hydrodynamic lattice simulations of the scalar field coupled to the plasma. 
Although these lattice simulations are the most faithful way to study the coupled scalar field-plasma system, they are also numerically expensive. 
A major challenge is the scale separation between the bubble profile, the size of the moving fluid shells and the size of the bubbles. 
Simulating a sufficient amount of bubbles requires large simulation boxes,
and resolving the different length scales requires a sufficiently small lattice spacing,
 limiting the range of parameters that can be simulated.
Below, we will discuss alternative computation schemes that are numerically more efficient -- at the cost of simplifying assumptions -- and that allow to scan over larger ranges of $\alpha$ and $v_w$. 

Refs.~\cite{Hindmarsh:2017gnf, Caprini:2019egz} provide the following fitting formula\footnote{There are some slight differences between the formulae in the Erratum of \cite{Hindmarsh:2017gnf} and the one in \cite{Caprini:2019egz}. We follow  \cite{Hindmarsh:2017gnf}, but like \cite{Caprini:2019egz}, we include the suppression factor due to shock formation.} for the GW power spectrum as a function of frequency $f$, based on simulations performed in \cite{Hindmarsh:2013xza,Hindmarsh:2015qta,Hindmarsh:2017gnf}
\begin{equation}
	\frac{d\Omega_{\rm{gw}}}{d\ln{(f)}} = 2.061 \,F_{\rm{gw},0}\, K_s^2 (H_* R_*)(H_* \tau_s)\, \tilde \Omega_{\rm{gw}}\, C\left(\frac{f}{f_{p,0}} \right)\,.\label{eq:LISAfit}
\end{equation}
Here $F_{{\rm gw},0}$ describes the redshift factor from the time of production to today
\begin{equation}
	F_{{\rm gw},0} = (3.57 \pm 0.05)\times 10^{-5}\left( \frac{100}{g_*}\right)^{\frac{1}{3}},
\end{equation}
with $g_*$ the effective number of relativistic degrees of freedom right after the phase transition. It was assumed that the entropic degrees of freedom equals the effective number of relativistic degrees of freedom, $g_*(T) =g_{*s}(T)$, which holds in the SM before electron decoupling $T>0.1\,$MeV. 
$\tilde \Omega_{\rm gw} \sim 0.012$ denotes a numerical constant, extracted from the lattice simulations, and $C$ is a function describing the spectral shape as a (single) broken power law.
$K_s$ is the energy budget available for GWs from sound waves, which is typically estimated by solving the hydrodynamic equations for a single bubble \cite{Kamionkowski:1993fg,Espinosa:2010hh}. In principle, $K_s$ depends on the details of the full particle physics model, but to a good approximation, this model-dependence is captured by the bubble wall velocity and phase transition strength \cite{Espinosa:2010hh}, and the speed of sound $c_s$ \cite{Giese:2020rtr, Giese:2020znk}. 
In the limit of $c_s^2 \sim 1/3$, $K_s$ can be obtained from the fitting formula provided by \cite{Espinosa:2010hh}. For deviations from $c_s^2 \sim 1/3$, the code snippet of \cite{Giese:2020rtr, Giese:2020znk} can be used.
$H_* R_*$ denotes the typical size of the bubbles at the moment they collide  as a fraction of the Hubble radius.
It can be estimated as
\begin{equation}
	H_* R_* = \frac{(8\pi)^{1/3}v_w}{\beta/H_* },
\end{equation}
with $\beta/H_*$ defined in \cref{eq:betaoH}.
The factor $H_* \tau_s = {\rm min} (H_* \tau_{\rm sh}, 1)$ describes the duration of the gravitational wave source.

For relatively weak phase transitions, the source can persist for a Hubble time, but for stronger phase transitions, the fluid will develop shocks on a time scale $\tau_{\rm sh} \sim R_* /K_s^{1/2}$, effectively shortening the GW sourcing time and therefore suppressing the signal. 
This behavior happens on a time scale \emph{beyond} the simulation times of \cite{Hindmarsh:2013xza,Hindmarsh:2015qta,Hindmarsh:2017gnf},
but recent simulations have started to capture this phase more accurately (see below). 
Lastly, the peak frequency $f_{p,0}$ is set by
\begin{equation}
	f_{p,0} = 26 \left(\frac{1}{H_* R_*} \right) \frac{z_p}{10} \frac{T_*}{100 \,{\rm GeV}} \left( \frac{g_*}{100} \right)^{1/6} \, \mu{\rm Hz },\label{eq:fpeak}
\end{equation}
where $z_p \sim 10$  a numerical constant that is
determined from simulations.
Compared to the fits for the GW signal from gradient energy, the spectrum from sound waves is enhanced by the factor $\beta \tau_s$, since the sound waves keep sourcing the GW spectrum, even after completion of the phase transition.
The energy budget $K_s$ is however significantly smaller than 1, and the spectrum falls off more steeply in the UV, as $f^{-3}$.  

Since \cite{Hindmarsh:2017gnf, Caprini:2019egz}, simulations and understanding of the sound wave source has developed further,
and realistic GW predictions typically require going beyond \cref{eq:LISAfit}.
Lattice simulations for stronger phase transitions \cite{Cutting:2019zws, Correia:2025qif} demonstrate that the signal gets suppressed compared to \cref{eq:LISAfit}, especially for smaller wall velocities, due to a slowdown of the bubbles before completion of the PT. 
Ref.~\cite{Correia:2025qif} evolved two parameter points of \cite{Cutting:2019zws} on larger grids and longer time scales, allowing them to observe the development of nonlinearities (turbulence). 
Modeling these effects beyond the two studied benchmarks will require further study.

Refs.~\cite{Jinno:2020eqg,Jinno:2022mie} developed an alternative computation scheme, the \emph{Higgsless simulation}. In this scheme, the scalar field is not tracked in the simulation, but the evolution of the bubbles (at fixed $v_w$) is imposed, and only the evolution of the plasma is tracked.
This simplifying assumption significantly reduces the numerical cost, and therefore allows to study a larger set of parameter points.
Moreover, as the Higgsless simulation scheme does not need to resolve the scalar field profile, it can more easily resolve the length scale of the hydrodynamics fluid profile, resulting in a doubly-broken, rather than single broken power law gravitational wave spectrum. 
A drawback of the method is that backreaction of the fluid to the scalar field motion, which was found to be relevant in some cases in \cite{Cutting:2019zws}, can not be captured.
In \cite{Caprini:2024gyk}, the development of non-linearities is also observed in the Higgsless simulation. The resulting gravitational wave spectrum is described well when the energy budget decays with a power law.
The authors provide a set of templates for the GW spectrum, that take the decay of the source into account. 
{\tt CosmoGW} \cite{RoperPol_CosmoGW_2025} is a Python package for the prediction of the GW spectrum based on these templates.

A semi-analytical computation of the GW spectrum can be obtained in the so-called Sound Shell Model \cite{Hindmarsh:2016lnk,Hindmarsh:2019phv,Guo:2020grp,Cai:2023guc,RoperPol:2023dzg}, which predicts an even more intricate spectral shape than the double power law. 
The advantage of the Sound Shell Model is that the resolution of the spectrum is not limited by the lattice parameters; a drawback is that it assumes that the sound waves are in the linear regime, and this does not hold for stronger phase transitions.
The public code {\tt PTTools} \cite{Hindmarsh_PTtools_2025} computes the sound shell spectrum for user-defined phase transition parameters. 
It takes into account the suppression effect due to slow-down of the walls observed in \cite{Cutting:2019zws} using an interpolation developed in \cite{Gowling:2021gcy}.

Finally, we remark that, even though the maximum phase transition strength attained in simulations has grown significantly over the years, very strong phase transitions $\alpha \gg 1$ are not within reach. 
The authors of \cite{Lewicki:2022pdb} have argued, based on the simplified simulation scheme of \cite{Lewicki:2019gmv, Lewicki:2020jiv} that for $\alpha \gg 1$, the GW spectrum from sound waves and gradient energy becomes similar. This motivates the use of the template of \cite{Caprini:2024hue} for GW predictions from sound waves in very strong phase transitions.

\begin{figure}[t]
    \centering
    \includegraphics[scale=0.8,trim=0 0 0 0]{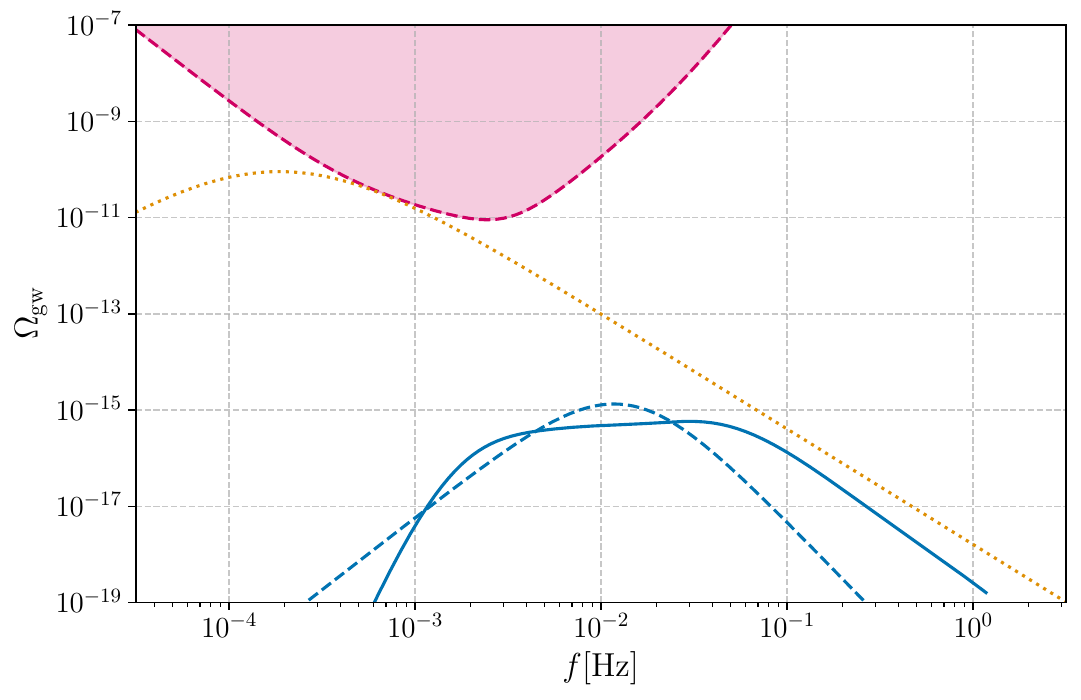}
    \caption{Examples of the GW spectrum. The blue lines are for an xSM BM point from \cite{Ellis:2022lft}. The solid line was computed in the Sound Shell  model using {\tt PTtools}, the dashed line was obtained with {\tt PTPlot} \cite{weir_2023_8220720}, and corresponds to \cref{eq:LISAfit}. The yellow line is a datapoint of the composite Higgs model \cite{Bruggisser:2022rdm} and obtained with the template from \cite{Caprini:2024hue} for relativistic fluid shells. The magenta dashed curve demonstrates the LISA sensitivity (curve from {\tt PTPlot}).}
    \label{fig:GWspectrum}
\end{figure}

\Cref{fig:GWspectrum} shows three example GW spectra. The LISA sensitivity curve is shown in magenta.
The blue lines correspond to a data point from \cite{Ellis:2022lft}, in the singlet-extension of the Standard Model (see also \cref{sec:GWEWBG}).
The dashed line is obtained with the formula \cref{eq:LISAfit}, using {\tt PTPlot} \cite{weir_2023_8220720, Caprini:2019egz} and the solid line is computed in the Sound Shell Model \cite{Hindmarsh:2016lnk, Hindmarsh:2019phv}, using {\tt PTTools} \cite{Hindmarsh_PTtools_2025}. The GW signal is much too weak to detect with LISA.
The yellow dotted line shows a much stronger GW signal from a composite Higgs model with a glueball-like dilaton \cite{Bruggisser:2022rdm}. The graph was obtained with the template from \cite{Caprini:2024hue} for ultrarelativistic fluid shells. This signal would likely be detectable.

\subsubsection{Gravitational waves from turbulence}
The contribution from turbulence to the GW spectrum is the least understood.
Even though a lot of progress has been made in understanding the shape of the GW spectrum from vortical and acoustic turbulence \cite{Pen:2015qta, Brandenburg:2017neh, Dahl:2021wyk,Auclair:2022jod, Dahl:2024eup}, the input of these simulations is \emph{not} a plasma undergoing a first-order phase transition.
This leaves an uncertainty in the overall amplitude of the signal, but also in the amplitude of the sound wave signal, which stops being sourced once the sound waves decay into turbulence.
Significantly longer simulation times could remedy this problem, and determine simultaneously the spectrum from sound waves and the resulting turbulence.
Indeed, the simulations \cite{Caprini:2024gyk, Correia:2025qif} observe the onset of the turbulent phase. 
However, the simulation time used in e.g. \cite{Correia:2025qif} corresponds only to $\mathcal O(1)$ shock formation time (relevant for acoustic turbulence), and less than an eddy turnover time (relevant for vortical turbulence), and therefore the characteristic spectra do not have enough time to develop. 
Longer simulation times come at significantly higher computational costs, and a better description of this phase of GW production is thus still a work in progress.

If magnetic fields are present or sourced during the phase transition, magneto-hydrodynamic (MHD) turbulence could get sourced due to the coupling between the magnetic field and the motion of the plasma.
Numerical simulations of GW formation due to MHD turbulence were performed in \cite{RoperPol:2019wvy, Kahniashvili:2020jgm}.
A template for GWs from MHD is provided in \cite{Caprini:2024hue}.
So far, no lattice simulations of a FOPT in the presence of a non-zero magnetic field have been performed.

\subsubsection{Gravitational waves and EWBG}\label{sec:GWEWBG}
For some time it was believed that observable GWs and successful EWBG were mutually exclusive, as the GW observability requires large $\alpha$ and $v_w$, whereas the baryon asymmetry was thought to be zero for bubble wall velocities exceeding the speed of sound $v_w > c_s$. Models predicting the correct baryon asymmetry therefore typically gave rise to a GW signal that is too weak to be observable with LISA.
As shown by \cite{Cline:2020jre} and confirmed by \cite{Dorsch:2021ubz} however, the baryon asymmetry does \emph{not} vanish for supersonic velocities $v_w > c_s$, although the baryon asymmetry is still a decreasing function of $v_w$. This has been discussed in some detail in \cref{sec:supersonic}. In principle, this opens up the possibility for successful EWBG \emph{and} observable GWs. 

In \cite{Cline:2021iff, Lewicki:2021pgr,Ellis:2022lft} the baryon asymmetry, wall velocity, and gravitational wave spectrum are computed for a singlet extension of the Standard Model, undergoing a two-step phase transition, using the updated fluid equations of \cite{Cline:2020jre} for the semi-classical force term. Refs. \cite{Cline:2021iff, Lewicki:2021pgr} assume a ${Z}_2$-symmetric potential for the singlet, while \cite{Ellis:2022lft} allows for ${Z}_2$-breaking terms. In all three studies, the CP-violating operator is
\begin{equation}
    \mathcal{L}_{\rm CPV} = -y_t \bar Q \tilde \Phi t_R \left( \frac{i s}{\Lambda_{\rm CPV}} \right) + {\rm h.c.},
\end{equation}
where $s$ is the singlet field. The conclusion of \cite{Cline:2021iff, Lewicki:2021pgr,Ellis:2022lft} is unanimous: even though the model under study \emph{can} give rise to the observed value of the baryon asymmetry, in that part of the parameter space the GW signal would be unobservable.
This is illustrated in \cref{fig:GWspectrum}, which shows the GW spectrum of the benchmark point of \cite{Ellis:2022lft} with successful baryogenesis with the strongest phase transition.
The GW signal is far too weak for detection with LISA.
Vice versa, in regions of the parameter space that source an observable GW spectrum, the generated baryon asymmetry is much smaller than the observed value.
It is still an open question whether the same conclusion holds for other models for EW baryogenesis.
Especially composite Higgs models are promising sources of GWs, as illustrated in \cref{fig:GWspectrum}, but it has to be confirmed that EWBG can be successful for the corresponding large values of $v_w$.

    \newpage
	\section{Models for Electroweak Baryogenesis}\label{sec:models}
        In the early days, model building for electroweak baryogenesis was mostly focussed on supersymmetric theories \cite{Cohen:1991iu,Nelson:1991ab,Dine:1990fj}  and two Higgs doublet models (2HDMs) \cite{Turok:1990zg,McLerran:1990zh,Dine:1990fj}, as these set-ups could naturally provide both the necessary first-order phase transition dynamics {\it and} the additional CP violation.  This lore was challenged when the LHC came online. The non-observation of supersymmetric particles and the measurement of the  $m_h \approx 125\,$GeV Higgs mass rules out a FOPT in the Minimal Supersymmetric SM (MSSM) \cite{Giudice:1992hh,Espinosa:1993yi,Carena:1996wj,Laine:2012jy,Liebler:2015ddv}, although it is still possible in extensions such as e.g. the NMSSM  \cite{Cheung:2012pg,Bian:2017wfv,Akula:2017yfr}. Moreover, a radiatively generated FOPT, such as generically occurs in 2HDMs, requires large couplings close to the non-perturbativity bound \cite{Kainulainen:2019kyp}. 

In addition, no beyond-the-SM CP violation  was observed, and together with 
the tremendous improvements of the EDM constraints in the last decade discussed in \cref{sec:EDM}, this severely bounds the amount of CP violation that can be added to the SM. All together, this puts the EWBG scenario under considerable duress, with researchers asking whether the scenario is dead \cite{Cline:2017jvp} (but, of course, see \cite{dead}). The vanilla scenario -- a radiatively generated FOPT and CPV in the top quark Yukawa sector -- seems ruled out. CPV operators involving the other fermions are less strongly constrained, but at the cost of a less efficient BAU production. 

Much of the model building efforts in recent years have thus been aimed at obtaining a FOPT with perturbative physics and especially on `hiding' the CPV from EDM measurements (a list of ways to avoid the EDM bounds is given in \cref{sec:avoid_EDM}).  In this section we highlight some of these approaches; we do not attempt to provide a complete overview and the list is far from complete.

\paragraph{CPV in the lepton sector}

The electron EDM measurements strongly constrain the scale of new CPV corrections to the top Yukawa coupling, but still allows for new CPV physics at the TeV scale in the lepton sector. An additional advantage is that leptons propagate more easily into the symmetric phase and are not affected by washout by strong sphaloron transitions \cite{Joyce:1994bi,Guo:2016ixx,DeVries:2018aul}. The resulting baryon asymmetry was calculated using the now excluded VIA method in \cite{DeVries:2018aul,Fuchs:2020uoc}. Recent work suggests that also with the WKB approach the BAU may be reproduced \cite{Athron:2025iew}, although it is unclear to us how this avoids the more pessimistic conclusions of \cite{Cline:2021dkf}.
The electron EDM only constrains one combination of CPV phases, which can be tuned small; the orthogonal directions can have much larger CPV \cite{Fuchs:2019ore}, which can be used for EWBG. Under specific flavor assumptions, such as minimal flavor violation, stronger indirect limits can be set on the CP-odd $\tau$ Yukawa coupling \cite{Alonso-Gonzalez:2021jsa}. 

\paragraph{CPV in flavor-changing interactions}
As discussed in \cref{sec:avoid_EDM} it is possible to avoid EDM constrants to a large extent by only considering CP violation in flavor-changing effects. For example, one can imagine an effective CPV top-charm Yukawa interaction. Such couplings can be realized in two Higgs doublet models \cite{Fuyuto:2017ewj,Kanemura:2023juv}. However, the extended Yukawa structures in the models have to be set by hand to avoid CP-violating effects in flavor-diagonal Yukawa interactions that would be strongly constrained by EDMs.

\paragraph{Two-step Phase Transition}
In (a limited) part of parameter space the EWPT can occur via a multistep process. The simplest possibility is that the Higgs potential is extended with a single field, which allows for a two-step PT, as discussed in \cref{sec:xSM}. In the first step the additional field -- be it a singlet \cite{Choi:1993cv,Espinosa:2011ax,Espinosa:2011eu, Kurup:2017dzf, Lewicki:2024xan, Niemi:2024vzw}, a doublet \cite{Fromme:2006cm,Dorsch:2013wja,Cline:1995dg}, or a triplet \cite{FileviezPerez:2008bj,Patel:2012pi, Niemi:2020hto} -- obtains a vev while the Higgs remains at the origin, while in the second step the Higgs vev becomes non-zero while the extra scalar field returns to the origin. 
One of these transitions can be first order. For the first transition the barrier can either be generated radiatively or at tree level, as the mass and couplings of the extra degree of freedom are less constrained than for the Higgs field. However, the requirement that the additional field has zero vev in the global minimum at zero temperature complicates this approach, and most attention has been focussed on a FOPT at the second step. For the second transition there can be a tree-level barrier for certain parameter values, which helps to get a strong  FOPT.  CP violation can reside in complex couplings to the additional field. Since this field has zero vev today, EDM signals are at least loop-suppressed. In \cite{Cline:2017qpe}, the singlet-extension was analyzed in a concrete UV completion and with a self-consistent calculation of the bubble wall velocity and profile; also \cite{Carena:2018cjh} describes a concrete UV realization that allows to transfer the CP violation to the SM sector. 

Variations to this theme exist.  For example, the CP-violating coupling can be to a dark sector instead of a SM field~\cite{Cline:2017qpe,Carena:2018cjh}, or the additional scalar can be CP-odd and CP gets broken spontaneously by its non-zero vev \cite{Huber:2022ndk}.

\paragraph{Composite Higgs models}
In composite Higgs Models  the SM Higgs field emerges from a strongly coupled sector that confines in the infrared. In Ref.~\cite{Espinosa:2011eu} this yields an effective Higgs-singlet potential, and EWBG baryogenesis occurs as in the two-step scenario. In the more recent works
\cite{vonHarling:2016vhf, Bruggisser:2018mus,Bruggisser:2018mrt} the EWPT happens together with the confining phase transition, and as the latter is naturally first-order, so is the former. The phase transition dynamics is described by a two-field Higgs-dilaton potential. The dilaton is the pseudo-Nambu-Goldstone boson of a broken conformal invariance, and is a light field that needs to be included in the effective field theory description at low energies. The CPV comes from complex Yukawa couplings, which arise from mixing with heavy composite states. The hierarchy in fermion masses is a result of the  renormalization group (RG) evolution.  The RG running also leads to varying Yukawa couplings during the EWPT \cite{Bruggisser:2017lhc}, which may change considerably as the dilaton obtains a vev. This way, the Yukawa couplings and the CP violation can be large enough during the EWPT to produce the observed BAU, while they are small enough in today's universe to satisfy the strong EDM bounds. The set-up is built on generic features of strong theories, all the required properties of the  potential and the quark sector can in principle be obtained, however the strong gauge dynamics hampers the formulation of an explicit and specific UV embedding. Moreover, the very strong phase transition of \cite{vonHarling:2016vhf, Bruggisser:2018mus,Bruggisser:2018mrt} may give rise to very large bubble wall velocities. It remains to be seen if the baryon asymmetry can be efficiently generated under such circumstances.

\paragraph{Non-standard EWPT}
If the EWPT has occurred at a scale $\Lambda \gg 100\;$GeV much above the EW scale, there would be no sphaleron washout.
Any new CP-violating interactions active at a scale $\Lambda \sim 100-1000\,$TeV will have negligible impact on the low energy phenomenology and bypass e.g. the EDM constraints \cite{Glioti:2018roy}.  In the SM the thermal effects of particularly the top quark restore the EW symmetry at $T \sim 160\,$GeV. Hence, new degrees of freedom  that couple to the Higgs and that live at the EW scale are needed to push the transition to much higher energies. Arguably the simplest option is to add a set of $N$ scalar singlets that couple bilinearly to the Higgs, although a large number of them is required $N>10^2$. In \cite{Baldes:2018nel} more modest values of the $\Lambda$-scale were proposed, here in combination with varying CPV Yukawa couplings that are larger during the EWPT than they are today, to circumvent EDM bounds.  

Ref. \cite{Ellis:2019flb} considers the effect of dimension-5 SMEFT operators that lead to dynamical gauge couplings during the phase transition. This may enhance (suppress) the electroweak (QCD) sphaleron rate, and bring models that are otherwise excluded back into the viable window.

\paragraph{Baryogenesis with relativistic bubbles}
More drastic departures from the vanilla EWBG scenario 
with subsonic bubbles, have been proposed in \cite{Azatov:2021irb} (see also \cite{Baldes:2021vyz} where the mechanism is applied to a PT that is not the EWPT).
With (moderate) supercooling the FOPT can be strong and the bubbles can expand with ultrarelativistic velocities, i.e., with large gamma-factors $\gamma_w \gg 1$. The phase transition is still the cause of the non-equilibrium dynamics, although it manifests very differently. Plasma particles that enter the bubble may obtain a large mass; this can be seen in the plasma frame where the radiation has energy $\sim \gamma_w T$. In addition, heavy plasma particles that only get a small correction to their mass when entering the bubble can also be produced, from mixing with light quanta that are swept up by the bubble. In either way, the production and subsequent decay is out-of-equilibrium for masses $M \gg T$; if either of these processes violate CP a baryon asymmetry can be generated. The heavy particle may be a right-handed neutrino  \cite{Azatov:2021irb}, which gives an implementation of (phase-transition induced) leptogenesis.  With CPV in the neutrino sector, all observational bounds can be satisfied.  
The ultrarelativistic bubble expansion generically generates a strong spectrum of gravitational waves that is potentially observable in the future.

\paragraph{Baryogenesis from a dark phase transition} 
A simple way to get baryogenesis from a dark phase transition is to copy the EWBG scenario \cite{Hall:2019ank} to the dark sector. That is,  introduce a dark ${\rm SU(2)}'$ gauge group that is broken via a FOPT, and with CPV couplings to the dark Higgs field(s). The asymmetry generated in the dark sector can be transferred to the visible sector via a portal coupling; this requires a generalized notion of baryon number. Although this mechanism shares all the features of EWBG, it can probably not be called as such.

	\newpage
    \section{Summary and outlook}\label{sec:sum}
        It has been almost sixty years since the appearance of Sakharov's seminal paper \cite{Sakharov:1967dj} and we still do not understand how our universe came to be dominated by matter. Electroweak baryogenesis is an elegant possibility that makes use of  interesting features of the Standard Model such as the anomalous violation of baryon and lepton number and the occurrence of an electroweak phase transition. However, EWBG does not work in the SM because the electroweak phase transition is not first order and, even if it was, the lack of a sufficiently effective mechanism of CP violation. Successful EWBG thus requires BSM physics which, because the asymmetry is generated around the electroweak temperatures,  must be active not too far from the electroweak scale. It is not surprising then that EWBG scenarios have gained a lot of attention. The chance to solve a major open problem in particle physics and cosmology in a way that can be tested with present-day experiments is simply too good to pass up. 

        While only a single number, computing the baryon asymmetry $Y_b$ in \cref{Yb} in a specific BSM model and testing its validity is a daunting task that involves a broad range of physics at vastly separated energy scales. The practitioner has to study how a first-order phase transition is induced and compute the corresponding bubble profile including the bubble wall velocity that should not be too large nor too small. It is then necessary to understand how a CP asymmetry is generated in front of the bubble wall. This step requires rather involved out-of-equilibrium quantum field theory calculations and solving a complicated set of transport equations. The CP asymmetry must then be converted into a baryon asymmetry through electroweak sphaleron transitions that must be sufficiently suppressed inside the bubbles to avoid washout. Finally, once all of this is done, to test the specific model our practitioner has to compute the phenomenological signatures at
        high-energy particle colliders, electric dipole moments experiments involving leptons, neutrons, atoms, and molecules, and gravitational wave observatories. This review is mainly meant as a guide to navigate all these steps. We have tried to include sufficient detail and intermediate steps in derivations such that a newcomer can actually compute $Y_b$.

        A second goal has been to highlight recent developments in the calculational framework of EWBG and areas where improvements are welcome. In particular, great progress has been made in the systematic computation of electroweak phase transitions and the resulting bubble properties and gravitational wave spectrum. Automated tools have been developed that can facilitate the study of large classes of EWBG models. 
        The EWPT has been studied on the lattice, both in the SM and in some BSM models. For certain BSM models with heavy new physics the lattice results can be readily applied; other models still have to resort to perturbative studies, but also perturbative methods have greatly improved. The incorporation of these developments into EWBG computations is an important next step.
        The bubble wall velocity computations have also improved and can now be readily computed with the package {\tt WallGo}. 
        Interestingly, it has been realized that EWBG is viable for rather large bubble wall velocities that were originally thought to be inconsistent with EWBG. This opens the door to new models and possible observable gravitational wave signals.

        A major source of confusion in the literature has been the derivation of the CP-violating source term that drives the system out of equilibrium. The so-called semi-classical and VIA source terms predict very different values of $Y_b$ within the same model. It is now understood that the VIA source, as used in the literature, actually vanishes at leading order in the gradient expansion once all terms are systematically included. This also means that the phenomenological viability of various scenarios studied in the literature that were based on the VIA must be reconsidered. Whether a `VIA-like' source exists at higher order in gradients is still an open question.
        In models with a misalignment of mass and interaction eigenstates, the flavor source could give an additional contribution to the chiral asymmetry. This source appears at lower order in gradients than the semi-classical one and could thus be the dominant source of CP violation. Since the CP-violating effect depends on an interplay between flavor-diagonal and off-diagonal distribution functions, the computation is more cumbersome, and only few studies with a flavor source exist.

        The last 10 to 20 years have witnessed incredible progress in all experimental aspects of EWBG. The Higgs boson has been discovered with a mass that firmly excludes a SM FOPT. Furthermore, as far as we can presently tell, the Higgs interacts as the SM dictates. That being said, many Higgs couplings (in particular the self-coupling) have not or not precisely been measured and BSM physics can still be hidden.  Similarly, EDM searches have seen marvelous progress with further improvements expected in the near future. The experimental developments have ruled out the most minimal EWBG scenarios, but, of course, model builders are creative and there are various ideas on how to avoid the stringent LHC and EDM constraints. Gravitational waves have finally been detected, opening the door to observe a signal from a potential FOPT. Unfortunately, currently studied EWBG scenarios tend to predict signals out of reach for prospected gravitational wave observatories. 
     
        If EWBG is realized in nature it should ultimately lead to an experimental signature. Of course, it will be difficult to confirm EWBG from such an observed signature but, with sufficient evidence, a convincing claim can probably be made. At the same time, while EWBG can never be fully ruled out, further experimental tests, for example at a new high-energy collider combined with more precise EDM searches, can strongly reduce the appeal of the framework. More and more bells and whistles must be added to avoid the constraints while leading to the observed value of $Y_b$ and Occam's razor would suggest to move on to alternative answers to the question of how our universe came to be matter dominated. However, right now this conclusion is premature and EWBG remains a promising and rich framework to generate the world we see around us with more matter than anti-matter.

\section*{Acknowledgements}
	It is a great pleasure to thank   
    Wen-Yuan Ai,
    James Cline,
    Andreas Ekstedt,
    Bj\"orn Garbrecht,
    Oliver Gould,
    Joonas Hirvonen,
    Ryusuke Jinno,
    Kimmo Kainulainen,
    Maciej Kierkla,
    Thomas Konstandin,
    Benoit Laurent,
    Lauri Niemi,
    Martijn Pellegrom,
    Tomislav Prokopec,
    Philipp Schicho,
    Julian Schoenmakers,
    Bogumi{\l}a {\'S}wie{\.z}ewska,
    Carlos Tamarit,
    Tuomas Tenkanen,
    David Weir,
    and
    Graham White
    for collaboration and/or discussions on the topics discussed in this review.
    We are grateful to Benoit Laurent, Marco Merchand and Marek Lewicki for providing us phase transition parameters used in figures \ref{fig:vwLW} and \ref{fig:GWspectrum},
    and we thank Jaakko Annala for providing us with \cref{eq:fitSph}.
    This work was supported by the Dutch Research Council (NWO) in the form of a VIDI grant (JdV) under project number VI.Vidi.203.005 and a VENI grant (JvdV) under project number VI.Veni.212.133.

\newpage
\appendix
\renewcommand*{\thesection}{\Alph{section}}

\section{Review of the SM}

\subsection{Lagrangian}
\label{A:SM}

The SM Lagrangian is written as
\begin{align}
\mathcal L_{SM}  &=   (D_{\mu} \vp)^{\dagger} D^{\mu} \vp      +\sum \bar \psi i \slashed{D}\, \psi            
-\frac{1}{4} 
\left(G^a_{\mu \nu} G^{a\, \mu \nu}
+W^{i}_{\mu\nu} W^{i\, \mu \nu}
+B_{\mu \nu} B^{\mu \nu}\right)
 \nonumber \\ 
&              
 +\mu_h \vp^\dagger \vp -\lambda(\vp^\dagger \vp)^2
                   - \left( \bar q_L Y_u \tilde \vp\,u_R -  \bar q_L Y_d \vp \,d_R  
+ \bar l_L Y_e \vp \,e_R  + \textrm{h.c.}\right)
. \label{SM}
\end{align}
The first line gives the kinetic terms for all SM fields: the SU(2) Higgs doublet $\vp$; the  fermions $\psi_i$ consisting of the left-handed quark and lepton doublets  $q_L$ and $l_L$, and right-handed singlets $u_R$, $d_R$, and $e_R$; and the  QCD, weak, and hypercharge gauge fields $G^a, W^i,Y$.
The covariant derivative is given by
\begin{eqnarray}\label{Cov}
D_{\mu} = \partial_{\mu}  - i \frac{g_s}{2}\, G^a_{\mu} \lambda^a 
- i \frac{g}{2}\, W^i_{\mu} \tau^i - i g^{\prime} Y B_{\mu}\,.
\end{eqnarray}
Here $g_s$,
$g$, and $g^{\prime}$ are, respectively, the SU(3)$_c$, SU(2)$_L$, and U(1)$_Y$ coupling constants; and $\lambda^a/2$ and $\tau^i/2$ denote SU(3) and SU(2) generators, 
in the representation of the field on which the derivative acts.
The hypercharge $Y=\{
1/6,2/3,-1/3, -1/2, -1, 1/2\}$ for $\{q_L, u_R, d_R,l_L,e_R,H \}$. The 
field strengths are
\begin{align}
G^a_{\mu \nu} &= \partial_{\mu} G^a_{\nu} - \partial_{\nu} G^a_{\mu}
- g_s f^{a b c} G^b_{\mu} G^c_{\nu}\,, \nn\\
W^i_{\mu \nu} &= \partial_{\mu} W^i_{\nu} - \partial_{\nu} W^i_{\mu} -
g \epsilon^{i j k} W^j_{\mu} W^k_{\nu}\,, \nn \\
  B_{\mu \nu} &= \partial_{\mu} B_{\nu} - \partial_{\nu} B_{\mu}\,,
              \label{Fmunu}
\end{align}
with
$f^{abc}$ and $\epsilon^{ijk}$ 
the SU(3) and SU(2) structure constants. 

The second line in \cref{SM} contains the Higgs potential and the Yukawa interactions. Here we defined $\tilde \vp = (i\sigma^2) \vp$, with $\sigma^2$ the Pauli matrix. 
We parameterize the Higgs doublet
\be
\vp = \frac1{\sqrt{2}} \( \begin{array}{c} \theta_2 + i \theta_3 \\
                          \phi + h +i \theta_1  \end{array} \),
\label{Higgs_doublet}
\ee
with $\phi$ the classical background, and $h$ and $\theta_i$ the Higgs
and Goldstone boson fluctuations.
The  Higgs and Goldstone tree-level masses are
\begin{equation}
    m_h^2(\phi) =  -\mu_h^2 + 3\lambda \phi^2 
    ,\qquad
    m_\chi^2(\phi) = -\mu_h^2 + \lambda \phi^2.
\label{mass_higggs}
\end{equation}
The background potential is minimized at the minimum of the tree-level potential $v\equiv \phi_0= \mu_h/\sqrt{\lambda}=
246\,$GeV, with the subscript $0$ denoting today's value. The vacuum Higgs mass then is $m_h^2 =2\lambda v^2$ from which it follows that
$\lambda  \approx 0.12$.

The classical Higgs background breaks the EW symmetry and gives masses to the EW bosons. The mass eigenstates are
\be
Z= \frac1{\sqrt{g^{'2}+g^2}}(g W_3-g' B), \quad
A_\gamma = \frac1{\sqrt{g^{'2}+g^2}}(g W_3+g' B),\quad
W^\pm= \frac1{\sqrt{2}}(W^1 \mp W^2),
\ee
with masses
\be
\{m_Z,m_W,m_\gamma\} =\{ \sqrt{g^{'2}+g^2} \frac{\phi}{2}, g \frac{\phi}{2},0\}.
\label{W_mass}
\ee
Fermion masses arise from the yukawa interactions. In the Higgs background
\begin{align}
\L &\supset
  -\(\frac{1}{\sqrt{2}}Y_u \phi \bar u_L u_R +\frac{1}{\sqrt{2}}Y_d \phi \bar d_L d_R+\frac{1}{\sqrt{2}}Y_e \phi \bar e_L e_R + \textrm{h.c.} \).
     \label{mass_yukawa}
\end{align}
In the SM there are no right-handed neutrinos and the neutrinos are massless.

\subsection{Charge and parity symmetry}
\label{A:CP}

\paragraph{Bosons} The charge C and parity P transformation of bosonic fields is
\begin{equation}
\chi_i^c(u) = \C \chi_i(u) \C^\dagger=C \chi_i^*(u), \quad \chi_i^p(u) = \P \chi_i(u) \P^\dagger=P \chi_i( \bar u),
\label{phi_CP}
\end{equation}
with $C,P$ arbitrary global phases, and  $\bar u= (u^0,-\vec u)$.  For the bosonic system described in this review, we can set the phase factors to unity, and we will omit them below. For non-trivial representations, the charge conjugation as defined above acts on the component fields labeled by $i$. In the background of a spherical bubble, the system is parity invariant.
The Wightman functions \cref{Wightman_def} transform as
\be
i\Delta^>(u,v) \stackrel{\C}{\to} (i\Delta^<)^T(v,u), \qquad
i\Delta^>(u,v) \stackrel{\C\P}{\to} (i\Delta^<)^T(\bar v,\bar u),
\ee
where the transpose acts in flavor space $(i\Delta^\lambda_{ij})^T = i\Delta^\lambda_{ji}$, and would be absent for a single complex scalar field.
Charge conjugation inverts the relative momenta, which yields a sign for the momenta in Wigner space
\be
i\Delta^>(k,x) \stackrel{\C}{\to} i\Delta^<(-k,x)^T, \qquad
i\Delta^>(k,x) \stackrel{\C\P}{\to} i\Delta^<(-\bar k,\bar x)^T,
\label{CP_boson_delW}
\ee
with $\bar k  = (k^0, - \vec{k})$.

CP violation requires complex couplings in the Lagrangian, for example, in the mass matrix 
$\L \supset - \chi^\dagger M^2 \chi $. Altough the mass matrix is hermitean and has real eigenvalues, in a multi-flavor system the off-diagonal entries in flavor space can be complex; in the mass basis this translates to phases in the mixing matrix. What matters is whether CP is violated in a physical process; physical phases cannot be rotated away by field redefinitions $\chi \to \e^{i\alpha} \chi$. If CP violation is induced by the non-zero Higgs vev, there is no global field redefinition to eliminate the phase in the space-time dependent bubble wall background.

\paragraph{Fermions} The C and P transformations of a  Dirac spinor are
\begin{equation}
\psi^c(u) = \C \psi (u) \C^\dagger=C \psi^*(u ) , \quad 
\psi^p(u) = \P \psi (u) \P^\dagger=P \psi (\bar u).
\end{equation}
Just as for bosons, for non-trivial representations the charge conjugation here acts on the individual component fields. 
The $\gamma$-matrices transform
\be
\C \gamma^{\mu T} \C^\dagger = -\gamma^\mu, \quad \C \gamma^{5} \C^\dagger = \gamma^5,\quad
\P \gamma^{\mu T} \P^\dagger = -\gamma^{\mu\dagger}, \quad \P \gamma^{5} \P^\dagger = -\gamma^5.
\label{CP_fermions}
\ee
In the chiral representation of the $\gamma$-matrices  $C =i\gamma^2 $
 and $P= \gamma^0$, with the overal phase choice a convention.
The Wightman functions \cref{Wightman_def} transform as
\be
S^>(u,v) \stackrel{\C}{\to}  \gamma^0 \gamma^2 S^<(v,u)^T  \gamma^0 \gamma^2, \qquad
S ^>(u,v) \stackrel{\C\P}{\to} -\gamma^2 S^<(\bar v,\bar u)^T \gamma^2,
\ee
which in Wigner space becomes
\be
S^>(k,x) \stackrel{\C}{\to}  \gamma^0 \gamma^2 S^<(-k,x)^T  \gamma^0 \gamma^2, \qquad
S ^>(k,x) \stackrel{\C\P}{\to} -\gamma^2 S^<(-\bar k,\bar x)^T \gamma^2,
\label{S_CP}
\ee
with the transpose acting on the flavor space indices. 

CP violation requires complex couplings in the Lagrangian, such as a complex correction to the Yukawa interaction term \cref{CP_yukawa}, which effectively gives a complex mass term for non-zero Higgs vev. In a constant background, most phases can be eliminated by chiral rotations of the fermions, and in the SM it takes three generations to be left with a physical CP-violating CKM phase.  In the bubble wall background, no global rotation can get rid of the phase, and CP violation can already occur for a single fermion.

\section{CTP formalism and Green's functions}
\label{A:CTP}

The  Green's functions for bosons $\chi$ respectively fermions $\psi$ in the CTP formalism are
\be
i\Delta (u,v)
 = \langle  T_{\cal C} \, \left[ \chi(u) 
 \chi^\dagger (v) \right] \rangle, \qquad
  iS_{\alpha\beta} (u,v)
 = \langle  T_{\cal C} \, \left[ \psi_\alpha(u) \bar
 \psi_\beta (v) \right]\rangle,
\label{Greens}
\ee
with $T_{\cal C}$ indicating time-ordering along the closed time path ${\cal C}$; the contour runs from initial time $t_0 = -\infty$ to time $t$ and back. ${\cal C}$ can be split into the forward and backward running branch, and the fields living on them are labeled by $a=+$ and $a=-$ respectively. 
Following the literature, we use the notation $G^t = G^{++}, \, G^{\bar t} =G^{--},\, G^> = G^{-+}, \, G^< = G^{+-}$; here $G^{ab} =\{\Delta^{ab},S^{ab}\}$ refers to both the bosons and fermion two-point functions. We further use the label $\lambda =\{ >,\, <\}$ for the Wightman functions; 
explicitly they are given by
\begin{align}
&{\rm bosons}: &  iG^>(u,v) &= \langle\chi(u) \chi^\dagger(v) \rangle, &iG^<(u,v) &= \langle \chi^\dagger(v) \chi(u)\rangle, 
& (i\Delta^\lambda(u,v))^\dagger &= i\Delta^\lambda(v,u),\nn \\
&{\rm fermions}: &  iG^>(u,v) &= \langle\psi(u) \bar \psi(v) \rangle, &iG^<(u,v) &= \langle\bar  \psi(v) \psi(u)\rangle, &
(\gamma^0 iS^\lambda(u,v))^\dagger& = \gamma^0 iS^\lambda(v,u),
\label{Wightman_def}
\end{align}
where we have also given the hermiticity properties in the right-most expressions.

All other Green functions can be written in terms of the Wightman functions.
The time-ordered and anti-time-ordered Green's functions are
\begin{align}
G^t(u,v) &= \theta(u,v) G^>(u,v) + \theta(v,u) G^<(u,v), \nn \\
G^{\bar t}(u,v) &= \theta(v,u) G^>(u,v) + \theta(u,v) G^<(u,v).
\end{align}
The retarded and advanced propagators can be expressed in terms of Green's functions \cref{Greens} via
\begin{align}
  G^r &\equiv G^t - G^< = G^> - G^{\bar t} ,& G^a &\equiv G^t - G^>  =  G^< - G^{\bar t}.
\label{Gdefs1}
\end{align}
They can be split into hermitian and
anti-hermitian parts
\be
G^h \equiv \frac 12 (G^r+G^a) =\frac12(G^t-G^{\bar t})
\;\; \; \& \;\; \; G^{\cal A} \equiv \frac {1}{2i} (G^a-G^r) =\frac {i}{2}(G^>-G^<).
\label{Gdefs2}
\ee

The Schwinger-Dyson equations give the full propagator with self-energies resummed; for the CTP contour they are \cite{Riotto:1998zb,Prokopec:2003pj}:
\begin{align}
	D(u)G^{ab}(u,v) &= a  \delta_{ab} \delta^4(u-v) + \sum_c c
                          \int d^4 w \, \Pi^{ac}(u,w) G^{cb}(w,v),
                          \label{eq_SD}
\end{align}
with $a,b,c= \pm$. The differential operator is the Klein-Gordon operator $D(u) = -(\partial^2_u + M_{(0)}(u)^2)$ for bosons, and the Dirac operator $D(u)= (i\slashed{\partial}_u- M_{(0)}(u))$ for fermions. Here we added the subscript ${(0)}$ on the mass to indicate these are the tree-level zero-temperature masses. For a system of many fields (flavors), the above expression can be read as a matrix equation in flavor space. The self-energy corrections are denoted by $\Pi^{ab}$. They satisfy the same relations as the Green functions $G^{ab}$, and the advanced/retarded self-energy can be split into real and imaginary parts.  $\Pi^h$ yields a thermal mass correction and $ \Pi^{\cal A}$ a thermal width. 

In a slowly varying bubble wall background, the Schwinger-Dyson equations can be expanded in derivatives. This is most easily done in Wigner space, where the gradient expansion corresponds to an expansion in the diamond operator. To go to Wigner space, introduce the relative and center-of-mass coordinates $ r = u-v$ and $x= \frac12(u+v)$; the Wigner transform is then defined as the Fourier transform with respect to the relative coordinate
\be
G(k,x) = \int \dd^4 (u-v) \, \e^{ik.(u-v)} G(u,v) =\int \dd^4 r \,
\e^{ik.r} G(x+\frac12 r,x- \frac12 r).
\label{Wigner}
\ee
The Wigner transformed Schwinger-Dyson \cref{eq_SD} can be found as follows  \cite{Berges:2015kfa}.  Consider e.g. the position space expression $A(u)B(u,v)$, which can be expanded as
\begin{align}
  A(u) &= A(x+\frac12r )
         = \e^{\frac12r \partial_x^A }A(x), &
  B(u,v) &= B (x ,r) =\int_p \e^{-i k r}B (x ,k).
\end{align}
The Wigner transform is
\begin{align}
\int_r \e^{ipr} A(u)B(u,v)&=
\int_r \int_k \e^{-i (k-p) r} \e^{\frac12r \partial_x^A }A (x ) B (x ,k) 
=\int_r \int_k \e^{-i (k-p) r} \e^{-i\frac12 \partial^B_k \partial_x^A }A (x ) B (x ,k)
\nn\\
&=\e^{-i\frac12 \partial^B_p \partial_x^A }A (x ) B (x ,p) =\e^{-i\diamond} A (x ) B (x ,p).
\end{align}
To get the third expression we used integration by parts.  
The diamond
operator is defined as
\be
\diamond \big(A(k,x) B(k,x) \big)= \frac12 \big(\partial_x A(k,x) \cdot\partial_k B(k,x) -\partial_k A(k,x) \cdot \partial_x B(k,x)\big).
\label{A:diamond}
\ee
The resulting set of Wigner-transformed Schwinger-Dyson equations can be split into a hermitian and anti-hermitian part, together known as the Kadanoff-Baym (KB) equations. The constraint equation encodes the spectral information, while the kinetic equation describes the evolution of the system.

In the plasma background the Green's functions are temperature dependent. In thermal equilibrium they satisfy the KMS relations 
\be
iG^<(k_0,\vec k) =s \e^{\beta k_0} iG^>(k_0,\vec k),
\label{KMS}
\ee
with $\beta =1/T$ the inverse temperature and  $s=1 \,(-1)$ for bosons (fermions).

\paragraph{KB equations for bosons}

Consider the bosonic system
\be
\L = -\chi_f^\dagger \[\partial^2 + M^2_{(0),f} \]\chi_f + \L_{\rm int}(\chi_f)= -\chi_m^\dagger \[\partial^2 + M_{(0)}^2 + \Sigma^2 +2\Sigma \cdot \partial +(\partial \Sigma) \]\chi_m + \L_{\rm int}(\chi_m),
\ee
where the first expression is in the flavor basis in which the interactions with the thermal bath are diagonal, and the second expression in the mass basis with a diagonal mass matrix $M_{(0)}^2$. Here we defined $\Sigma_\mu = U^\dagger \partial_\mu U$ with $\chi_m = U \chi_f$ the rotation matrix that diagonalizes the mass matrix. We will give the Kadanoff-Baym (KB) equations in the mass basis.

The constraint and kinetic equation are\footnote{The equations are different if one goes first to the mass frame and then performs the Wigner transformation, or whether one first goes to Wigner space and only then transforms to the mass frame. The difference should vanish on the solutions. The diamond operator only acts on $\Sigma$ in the $2ik\cdot \Sigma$-term.} 
(the $[\Sigma,\partial_x \Delta^\lambda]$-term is dropped in \cite{Cirigliano:2011di})
\begin{align}
  (k^2-\frac14 \partial_x^2) \Delta^\lambda &=\frac12\e^{-i\diamond}\bigg( \{M^2_{(0)} +\Pi^h+ \Sigma^2-2i k\cdot \Sigma,\Delta^\lambda\}
+\frac{1}{2} \([ \Sigma,\partial_x \Delta^\lambda] +\{ \Pi^\lambda,\Delta^h\} \)+\C_-\bigg)\, ,
 \label{A:constraint_B}\\
2ik\cdot\partial_x \Delta^\lambda &= \e^{-i\diamond}\bigg( [M^2_{(0)} +\Pi^h+ \Sigma^2-2i k\cdot \Sigma(x),\Delta^\lambda] +\(\{ \Sigma,\partial \Delta^\lambda\} +[ \Pi^\lambda,\Delta^h]\) +\C_+\bigg) \, ,
 \label{A:kinetic_B}
\end{align}
with 
\be
\C_\pm=\frac{1}2\([\Pi^>,\Delta^<]_\pm -[\Pi^<,\Delta^>]_\pm\),
\label{calC}
\ee
and $[..,..]_-=[..,..]$ and $[..,..]_+=\{..,..\}$ the commutator and anti-commutator respectively. 

The Wightman functions are a function of momentum and collective coordinate $\Delta^\lambda =\Delta^\lambda(k,x)$,  $M_{(0)}^2=M_{(0)}^2(x)$ and $\Sigma=\Sigma(x)$ depend on $x$ only, and to first approximation $\Pi=\Pi(k)$ is the equilibrium space-time independent self-energy. For $n$ flavors all these quantities are $n\times n$ matrices in flavor space.
 $\Pi^h$ gives the thermal mass corrections, which we can simply absorb in the mass matrix $M^2 = M_{(0)}^2+\Pi^h$. 
The KB equations in the flavor basis follow from \cref{A:constraint_B,A:kinetic_B} setting $\Sigma =0$ and  replacing $\{M^2,\Delta,\Pi \} \to \{M_f^2,\Delta_f,\Pi_f \}$.

The anti-hermitian constraint equation can be solved to find the spectral properties of the system, while the hermitian kinetic equation describes the time evolution. So far, no approximation have been made, and these equations are still exact. The challenge is to find a controlled approximation scheme to describe the plasma dynamics during the electroweak FOPT accurately.
The gradient expansion consists of an expansion of the diamond operator $\e^{-i\diamond} = 1 -i\diamond+ ...$.

For a single free boson -- vanishing self-energies $\Pi =0$ and constant mass term -- the solution for the Wightman functions is:
\begin{align}
i\Delta^> &= 2\pi \delta(k_0^2-\omega^2) \sgn(k_0) \mathfrak{g}^>(k_0,\vec k ,x) =
2\pi \delta(k_0^2-\omega^2) \[ \theta( k_0) (1 + f(k_0,\vec k ,x))+  \theta(-k_0) \bar f(k_0,-\vec k,x)\], \nn \\
i\Delta^< &= 2\pi \delta(k_0^2-\omega^2) \sgn(k_0) \mathfrak{g}^<(k_0,\vec k ,x) =
2\pi \delta(k_0^2-\omega^2) \[ \theta(k_0) f(k_0,\vec k ,x)+ \theta(-k_0)   (1+\bar f(k_0,-\vec k,x))\],
\label{Grho}
\end{align}
with
\be
\mathfrak{g}^>(k_0,\vec k ,x) =  (1+ \mathfrak{n}(k_0,\vec k ,x)) \;\;\; \& \;\;\; \mathfrak{g}^<(k_0,\vec k ,x) =  \mathfrak{n}(k_0,\vec k ,x).
\label{g_frak}
\ee
To get the rightmost expressions we split into particles  $f(k_0,\vec k,x)=\mathfrak{n}(k_0,\vec k,x)$ with $k_0>0$ and antiparticles $\bar f(k_0,\vec k,x) =\mathfrak{n}(-k_0,\vec k,x)$ with $k_0<0$. For symmetric bubbles $\vec k \to -\vec k$ is a symmetry.
In thermal equilibrium the Wightman functions satisfy the KMS relation \cref{KMS}, which implies $\mathfrak{n}_0 = (\e^{k_0/t} -1)^{-1}$ the Bose-Einstein distribution. 
Here the subscript $0$ on the phase space density indicates the equilibrium solution.
The normalization of the spectral functions is fixed by the sum rule
\be
\int\frac{\dd k_0}{2\pi}\, k_0 2\Delta^\A = \mathds{1},
\label{sumrule_B}
\ee
which follows from the commutation relations.

The transport equations follow from integrating the kinetic equation over $k_0$, using the projections
\be
\int_0^\infty \frac{\dd k_0}{2\pi} (2k_0) (i\Delta^<) = f(\omega,\vec k,x), \qquad
\int^0_{-\infty} \frac{\dd k_0}{2\pi} (-2k_0) (i\Delta^>) = \bar f(-\omega,-\vec k,x). \quad
\label{project_boson}
\ee
The spatial momentum assignment assures the kinetic \cref{kinetic_B} in absence of self-energies and gradient corrections gives $u\cdot \partial f(\vec k,x)=u\cdot \partial\bar  f(\vec k,x)=0$, with $u=(1,\vec k/\omega)$ the 4-velocity.

\paragraph{KB equations for fermions}

The Lagrangian for a single Dirac fermion with complex mass term is given in the main text \cref{L_fermion}; the corresponding KB equation is \cref{KB_fermion_semi}.
The fermion propagator is normalized by the sum rule
\be
\int_{-\infty}^\infty \frac{\dd k_0}{2\pi} \frac14\Tr \gamma^0 2 S^\A 
=\mathds{1},
\label{sumrule_F}
\ee
with $S^- =S^> -S^<= -2iS^\A$ the spectral density.

\newpage
\bibliography{refs.bib}

\end{document}